\documentclass[11pt]{article}
\pdfoutput=1
\usepackage[english]{babel}
\usepackage{amsmath,amssymb,amsbsy,amstext, amsthm, simplewick, amsfonts}
\usepackage{hyperref}
\usepackage{graphicx}
\usepackage{upgreek}
\usepackage{framed}
\usepackage{bbm}
\usepackage{textcomp}
\usepackage{cleveref}
\usepackage{adjustbox}
\usepackage{makecell}
\usepackage{tcolorbox}
\usepackage{physics}
\usepackage{empheq}
\usepackage[normalem]{ulem}
\usepackage{cite}

\numberwithin{equation}{section}

\def\be {\begin{equation}}
\def\ee {\end{equation}}

\def\e{\epsilon}

\def\ba{\begin{eqnarray}}
\def\ea{\end{eqnarray}}

\def \bfx {\textbf{x}}

\def\vpi {\varphi}
\def \bfk {\textbf{k}}

\def \bfS {\textbf{S}}

\def\disc {\text{Disc}}

\newcommand{\ex}[1]{\langle #1 \rangle}

\usepackage[framemethod=default]{mdframed}
\newmdenv[skipabove=7pt,
skipbelow=7pt,
rightline=false,
leftline=false,
topline=false,
bottomline=false,
backgroundcolor=gray!10,
linecolor=gray,
innerleftmargin=5pt,
innerrightmargin=5pt,
innertopmargin=5pt,
innerbottommargin=5pt,
leftmargin=0cm,
rightmargin=0cm,
linewidth=4pt]{eBox}
\newmdenv[skipabove=7pt,
skipbelow=7pt,
rightline=false,
leftline=false,
topline=false,
bottomline=false,
backgroundcolor=gray!10,
linecolor=gray,
innerleftmargin=5pt,
innerrightmargin=5pt,
innertopmargin=-5pt,
innerbottommargin=5pt,
leftmargin=0cm,
rightmargin=0cm,
linewidth=4pt]{eBox2}

\usepackage{array}

\definecolor{blue3}{RGB}{31, 119, 180}
\definecolor{red3}{RGB}{	214, 39, 40}
\definecolor{orange3}{RGB}{255, 127, 14}
\definecolor{green3}{RGB}{44, 160, 44}

\usepackage{colortbl}
\definecolor{lightgreen}{cmyk}{0.2, 0, 0.2, 0.2}
\definecolor{lightgray}{cmyk}{0.1,0.2,0,0.1}
\definecolor{lightgray2}{cmyk}{0.1,0.1,0,0.1}

\makeatletter
\newlength{\apb@width}
\newcommand{\autoparbox}[2][c]{\settowidth{\apb@width}{#2}\parbox[#1]{\apb@width}{#2}}

\makeatother


\def\Mpl{M_{\text{Pl}}}

\def\H{{\cal H}}

\def \bfp {\textbf{p}}

\def\beq{\begin{equation}}
\def\eeq{\end{equation}}

\allowdisplaybreaks[1]
\begin{document}



\begin{titlepage}
\setcounter{page}{1} \baselineskip=15.5pt 
\thispagestyle{empty}

\begin{center}
{\fontsize{18}{18} \bf Cutting Cosmological Correlators \vspace{0.1cm}
\;}\\
\end{center}

\vskip 18pt

\begin{center}
\noindent
{\fontsize{12}{18}\selectfont Harry Goodhew\footnote{\tt hfg23@cam.ac.uk}$^\dagger$, Sadra Jazayeri\footnote{\tt jazayeri@iap.fr}$^\star$, Mang Hei Gordon Lee\footnote{\tt mhgl2@cam.ac.uk}$^\dagger$ and Enrico Pajer\footnote{\tt enrico.pajer@gmail.com}$^\dagger$}
\end{center}

\begin{center}
  \vskip 8pt
$\dagger\,$\textit{Department of Applied Mathematics and Theoretical Physics, University of Cambridge, Wilberforce Road, Cambridge, CB3 0WA, UK}\\
$\star\,$\textit{ Institut d'Astrophysique de Paris, GReCO, UMR 7095 du CNRS et de Sorbonne Universit\'{e},\\ 98bis
boulevard Arago, 75014 Paris, France}

\end{center}

\vspace{1.4cm}

 \noindent The initial conditions of our universe appear to us in the form of a classical probability distribution that we probe with cosmological observations. In the current leading paradigm, this probability distribution arises from a quantum mechanical wavefunction of the universe. Here we ask what the imprint of quantum mechanics is on the late time observables. We show that the requirement of unitary time evolution, colloquially the conservation of probabilities, fixes the analytic structure of the wavefunction and of all the cosmological correlators it encodes. In particular, we derive in perturbation theory an infinite set of \textit{single-cut rules} that generalize the Cosmological Optical Theorem and relate a certain discontinuity of any tree-level $n$-point function to that of lower-point functions. Our rules are closely related to, but distinct from the recently derived Cosmological Cutting Rules. They follow from the choice of the Bunch-Davies vacuum and a simple property of the (bulk-to-bulk) propagator and are astoundingly general: we prove that they are valid for fields with a linear dispersion relation and any mass, any integer spin and arbitrary local interactions with any number of derivatives. They also apply to general FLRW spacetimes admitting a Bunch-Davies vacuum, including de Sitter, slow-roll inflation, power-law cosmologies and even resonant oscillations in axion monodromy. We verify the single-cut rules in a number of non-trivial examples, including four massless scalars exchanging a massive scalar, as relevant for cosmological collider physics, four gravitons exchanging a graviton, and a scalar five-point function.

 \vspace{1.4cm} { \begin{center}
     \textit{This work is dedicated to the memory of Prof. John D. Barrow}
 \end{center}}


\end{titlepage}


\newpage
\setcounter{tocdepth}{2}
\tableofcontents

\newpage


\section{Introduction}

We have convincing evidence that the origin of cosmological perturbations in our universe can be traced back to a primordial phase preceding the hot big bang. According to the leading paradigm, it is the quantum mechanical vacuum that is responsible for seeding the small inhomogeneities that later seed the formation of structures. It is therefore natural to ask whether the footprint of quantum mechanics can be found anywhere in the classical probability distribution that we use as initial conditions to make predictions for cosmological surveys. To answer this question, it is convenient to work with the quantum mechanical \textit{wavefunction} of the universe, from which all late time correlators and the related probability distribution can be extracted. \\

Recently, it has been realized that unitary time evolution during the primordial universe leads to a particular relation at tree-level in perturbation theory between the quartic and cubic wavefunction coefficients for a single scalar field. This has been named the Cosmological Optical Theorem \cite{COT} and can be alternatively understood as a conserved quantity under unitary time evolution \cite{Cespedes:2020xqq}. One expects that, just as it is the case for the optical theorem in particle physics, this relation has avatars to each order in perturbation theory for any set of fields. In this paper we prove that this is indeed the case and extend previous results to an infinite set of single-cut rules for a general tree-level diagram. A related but distinct result are the Cosmological Cutting Rules, which apply also to any loop diagram. Those are discussed in a separate paper \cite{sCOTt}. \\

In the original derivation of the Cosmological Optical Theorem, one starts from the iconic unitarity relation $UU^\dagger =1$, where $U$ is the time-evolution operator, and then evaluates this operator identity inside correlators in perturbation theory. This derivation makes the role of unitarity very explicit, but is also lengthy and cumbersome to extend beyond the lowest order in the perturbative expansion. Here we take instead a different approach that makes generalizations and extensions much more straightforward, but which leaves the role of unitarity somewhat implicit. In particular, we introduce a series of \textit{single-cut rules} that cut any given diagram representing a wavefunction coefficient into the product of two lower-order diagrams. This is qualitatively analogous to Cutkosky's cutting rules in flat spacetime \cite{Cutkosky:1960sp} (see also \cite{tHooft:1973wag,Veltman:1994wz} for alternative derivations) and to the cutting rules in AdS (see e.g. \cite{Aharony:2016dwx,Meltzer:2020qbr}).
\\

The main focus of this work is extending the validity of the Cosmological Optical Theorem beyond the simplest case of a single massless scalar field in de Sitter space, which was studied in \cite{COT}. Indeed, we find that precisely the same Cosmological Optical Theorem applies much more generally. In addition to the requirement that coupling constants are real, our derivation relies on two properties: (\textit{i}) Hermitian analyticity of the (bulk-to-boundary and bulk-to-bulk) propagators, namely $K^\ast(-k^\ast)=K(k)$, and (\textit{ii}) the simple factorization of the imaginary part of the bulk-to-bulk propagator. The latter is guaranteed by the nature of the problem, namely the computation of the wavefunction on some constant-time hypersurface, and reduces to a very analogous factorization of the Feynman propagators in flat spacetime. The first condition is more interesting. Here we prove that it follows under very general assumptions for any set of fields with linear dispersion relation, $E\propto |\bfk|$ (in the infinite past), on any flat FLRW spacetime on which we can impose a Bunch-Davies initial state. This includes many spacetimes that are relevant for cosmology, such as de Sitter and inflation (quasi-de Sitter), as well as (accelerating) power-law cosmologies. It also applies for arbitrary speeds of sound and masses that can be approximated as constant on sub-Hubble time scales. The presence of different species of particles is just a matter of bookkeeping: every spatial momentum in the Cosmological Optical Theorem can be expanded to include additional quantum numbers such as charges and flavours. Likewise, spinning (boson) fields can be treated on the same footing as scalars by considering helicity as yet another quantum number (and leaving polarization tensors to the vertices). \\

When using our single-cut rules the derivation of the Cosmological Optical Theorem becomes very straightforward. However, the relations we obtain are highly non-trivial in the general case. In particular, we demonstrate here that our results hold for the scalar four-point functions generated by the exchange of a massive scalar computed in \cite{Arkani-Hamed:2018kmz}. Verifying this relation is challenging and requires careful treatment of special functions and their branch points/cuts. We also compute a four-graviton diagram from graviton exchange by invoking a particularly simple interaction that is expected to appear in the Effective Field theory of Inflation \cite{EFTofI} (see e.g. \cite{Bordin:2020eui,Cabass:2021iii}).\\

Our results show us how general principles such as unitarity shape the resulting observables, namely the wavefunction or correlators. This is useful both in the context of recent progress on perturbative calculations \cite{Arkani-Hamed:2015bza,Arkani-Hamed:2017fdk,Arkani-Hamed:2018bjr,Benincasa:2018ssx,Benincasa:2019vqr,Sleight:2019mgd,Sleight:2019hfp,Sleight:2020obc,Baumgart:2019clc,Gorbenko:2019rza,Cohen:2020php,Baumgart:2020oby} and in the very promising bootstrap approach \cite{Maldacena:2011nz,Creminelli:2011mw,Kehagias:2012pd,Mata:2012bx,Bzowski:2013sza,Ghosh:2014kba,Kundu:2014gxa,Kundu:2015xta,Pajer:2016ieg,Arkani-Hamed:2018kmz,Bzowski:2019kwd,Baumann:2019oyu,PSS,Isono:2020qew,Green:2020ebl,Baumann:2020dch,BBBB}. For example, the Cosmological Optical Theorem and our single-cut rules were recently used to derive a powerful Manifestly Local Test and partial energy recursion relations \cite{MLT}. In turn, these bootstrap tools give us a transparent and computationally effective way to derive both contact and exchange correlators, without any reference to de Sitter boosts, which are incompatible with large non-Gaussianity in single field inflation \cite{Green:2020ebl}.\\

The rest of this paper is organized as follows. In Section \ref{sec2}, after reviewing the perturbative computation of the wavefunction of the universe and discussing the associated propagators, we introduce single-cut rules in the simplest case of a single scalar field without time-derivative interactions. Then in Section \ref{sec3}, we discuss generalizations to time-derivative interactions, general FLRW spacetimes and spinning fields. Here we also show that a linear dispersion relation is a necessary property to satisfy our single-cut rules, in their current formulation. In Section \ref{sec:examples} we study in detail several non-trivial examples, in particular the scalar four-point functions from the exchange of a conformally coupled and general massive scalar and the graviton four-point function from the exchange of a graviton with $\dot\gamma^3$ interactions. Finally we conclude in Section \ref{conclusions} and leave technical details on massive exchange (Appendix \ref{MassiveAppendix}), resonant mode functions (Appendix \ref{RNG}) and the WKB approximation (Appendix \ref{WKB}) to the appendices.

\paragraph{Notation and conventions} Our conventions for the Fourier transform are
\begin{align}
f(\bfx)&=\int \dfrac{d^3\bfk}{(2\pi)^3}{f}(\bfk)\exp(i\bfk\cdot\bfx)\equiv\int_{\bfk}{f}(\bfk)\exp(i\bfk\cdot\bfx) \,.
\end{align}
We will use bold letters to refer to vectors, e.g. $\bfk$, and non-bold letters for their magnitude $k\equiv |\bfk|$. We will usually call $k$ the “energy" by analogy with on-shell, massless particles in Minkowski. We expand the wavefunction\cite{anninos2015late} as\footnote{Notice that our convention differs by a minus sign from that used in \cite{sCOTt}, $\psi_n^{\text{here}}=-\psi_n^{\text{there}}$. We apologize for this inconvenience.}
\begin{align}
\Psi [ \phi  ] &\propto \exp \left[-
 \sum_{n=2}^{\infty} \int_{\bfk_1,..,\bfk_n}\frac{1}{n!} \psi_{ \bfk_1 ... \bfk_n }  \phi_{\bfk_1} ... \phi_{\bfk_n} \right]\; ,
\end{align}
and use a prime when we omit the momentum-conserving delta function, 
\begin{align}
\psi_n(\bfk_1,\dots ,\bfk_n)& \equiv \psi_n'(\bfk_1,\dots ,\bfk_n)(2\pi)^3 \delta^3\left(\sum \bfk_a\right)\,,\\ \nonumber
\langle {\cal O}(\bfk_1)\dots {\cal O}(\bfk_n)\rangle &\equiv \langle {\cal O}(\bfk_1)\dots {\cal O}(\bfk_n)\rangle' (2\pi)^3\delta^3\left(\sum \bfk_a\right)\,.
\end{align}
When discussing $\psi_4$, we will use the following variables:
\begin{align}
 \bfp_s =  \bfk_1 + \bfk_2  \;\; , \;\; \bfp_t = \bfk_1 + \bfk_3\;\; , \;\; \bfp_u =  \bfk_1 + \bfk_4 \,.
\end{align}
These are not independent since $p_s^2 + p_t^2 + p_u^2 = \sum_{a=1}^4 k_{a}^2$. We also sometimes use the notation 
\begin{align}
    k_{ij}\equiv k_i+k_j\,.
\end{align}
We denote the power spectrum of a $\phi^{(\alpha)}$ as
\begin{align}\label{pk}
    \ex{\phi_\bfk^{(\alpha)}(\eta_0) \phi_{\bfk'}^{(\beta)}(\eta_0)}=(2\pi)^3 \delta(\bfk+\bfk') \delta_{\alpha\beta} P_k^{(\alpha)}(\eta_0)\,,
\end{align}
where $\alpha$ refers collectively to a set of quantum numbers. In general we will omit the $\eta_0$ dependence and, when talking about a single scalar field, we'll omit the quantum numbers $\alpha$. 
When discussing general tree-level diagrams we will use $\bfk_a$ with $a=1,\dots,n$ to denote the $n$ external momenta and $\bfp_{m}$ with $m=1,\dots,I$ to denote the $I$ internal momenta. Since we work at tree level, all the $\bfp_m$ are fixed in terms of the $\bfk_a$ by momentum (but not energy) conservation at every vertex. In particular, for every $\bfp_m$ there is a subset $\{\bfk\}_m$ of external momenta such that $\bfp_m = \sum_{a\in \{\bfk\}_m} \bfk_a$. We denote internal and external energies by $k_a$ and $p_m$, respectively, and use them as variables together with additional rotational invariant contractions of the external momenta
\begin{align}
    \psi_n&=\psi_n(\text{ external energies ; internal energies ; contractions })\\
    &=\psi_n(k_1,\dots,k_n;p_1,\dots,p_I; \bfk_a\cdot \bfk_b, \bfk_a \cdot (\bfk_b \times \bfk_c), \bfk_a \cdot \e(\bfk_b),\dots )\\
    &\equiv \psi_n(\{k\};\{p\};\{\bfk\})\,.
\end{align}

We define the Hermitian analytic image of a function of the energy $k$ and momenta $ \bfk$ to be the complex conjugate of the function evaluated at $-k^*$ and $-{\bfk}$. So, for example, the Hermitian analytic image of the function $F(k,{\bfk})$ is $F^*(-k^*,-{\bfk})$. We will refer to a function that is equal to its Hermitian analytic image as Hermitian analytic. When evaluating an expression for real $k$, we will often drop the complex conjugate in analytically continued energies, $-k^\ast \to -k$. This must be interpreted with the following convention: \textit{negative energies $-k$ should be approached from the lower-half complex plane}, namely $-k = -k -i \e$ with $\e>0$ infinitesimal. 

\section{Cutting diagrams}\label{sec2}


In this section, we derive a single-cut rule for tree-level wavefunction coefficients, namely a simple expression for the discontinuity of an $n$-point wavefunction coefficient $\psi_n$ in terms of discontinuities of lower-point coefficients. This result follows from the Hermitian analyticity of the bulk-to-boundary propagator, $K^\ast_{-k}=K_{k}$ and the fact that the imaginary part of the bulk-to-bulk propagator $G$ factorizes. The Cosmological Optical Theorem (COT) for 4-point exchange diagrams derived in \cite{COT} is found as a special case (see also \cite{Cespedes:2020xqq} and parallel developments on the AdS side in \cite{Aharony:2016dwx,Meltzer:2020qbr}). We start our presentation considering a single canonical massless or conformally coupled scalar field $\phi$ in de Sitter with no time-derivative interactions. Generalizations will be discussed later: time-derivatives are discussed in Section \ref{Deriv}, general FLRW spacetimes in Section \ref{sec:flrw} and  multiple fields with arbitrary spin in Section \ref{sec:spinning}.


\subsection{Diagrammatic representation of the wavefunction coefficients}

Our starting point is a brief review of the formalism to compute the wavefunction coefficients in perturbation theory (for more details see e.g. App. A of \cite{COT} or \cite{Goon:2018fyu}).

A given (connected) contribution to $\psi_n$ is represented as a (connected) diagram with $n$ external lines, each with one end on the boundary at $\eta_0 =0$ and the other end on some vertex at time $\eta_A$. Vertices are connected pairwise by $I$ internal lines with momenta $\bfp_m$ with $m=1,\dots, I$ completely fixed (at tree level) in terms of the external momenta by momentum conservation at every vertex. Every vertex may involve spatial derivatives and therefore factors of the momenta contracted in a rotational invariant way and time derivatives. To simplify the presentation, for the time being we assume there are no time-derivative interactions. Later, in Section \ref{Deriv}, we will relax this assumption and arrive at the same results. To capture spatial derivatives we allow for a vertex function $F(\bfk_a,\eta)$, which contains contractions of the momenta ending on a given vertex with each other or with the (3d) Levi-Civita symbol. Later on, in Section \ref{sec:spinning}, we will allow $F$ to also include  polarization tensors of spinning fields.

Every external line is associated with a bulk-to-boundary propagator $K_k(\eta)$ of momentum $\bfk$ and “energy" $k\equiv |\bfk|$ and every internal line to a bulk-to-bulk propagator $G_p(\eta,\eta')$ with internal energy $p$. These solve the following differential equations
\begin{align}
\mathcal{O}_k ( \eta ) K_k ( \eta )  =0 \; ,\qquad 
\mathcal{O}_p (\eta) G_p ( \eta,\eta') = \delta (\eta-\eta')\,,
\end{align}
subject to the boundary conditions, 
\begin{align}
 &\, \; \lim_{\eta\rightarrow\eta_0} K_k  ( \eta )=1,&  \, \; \lim_{\eta\to -\infty(1-i\epsilon)}K_k   ( \eta )=0\,,\\ 
& \lim_{\eta \rightarrow\eta_0} G_p  (\eta,\eta')=0,&  \lim_{\eta \to -\infty(1-i\epsilon)} G_p   ( \eta,\eta')=0\, , \label{boundaryG}
\end{align}
where $\mathcal{O}_k (\eta) \phi_k$ denotes the linearized equations of motion obtained by demanding that $\phi$ is an extremum of the quadratic action. Notice that because of the boundary conditions $K$ and $G$ depend on $\eta_0$, but we will omit writing this dependence explicitly. Also, we will generally have in mind $\eta_0\to 0^-$. Both propagators can be written in terms of the positive and negative frequency mode functions $\phi^\pm$, which solve the same differential equation as $K$ but have different boundary conditions. Namely $\phi^+$ asymptotes to the positive-frequency Minkowski mode functions at $\eta\rightarrow -\infty$ whilst $\phi^-$ is defined so that the Wronskian of the two solutions, $W(\phi^+,\phi^-)$, is equal to $-i$ and, for real energies, $\phi^-_p(\eta)=\left[\phi^+_p(\eta)\right]^*$. Explicitly this gives
\begin{align}
    K_k(\eta)&=\frac{\phi^+_k(\eta)}{\phi^+_k(\eta_0)}\,.\\
G_p ( \eta, \eta') &= i\left[ \theta(\eta-\eta')\left(\phi^+_p(\eta')\phi^-_p(\eta)-\frac{\phi^-_p(\eta_0)}{\phi^+_p(\eta_0)}\phi^+_p(\eta)\phi^+_p(\eta')\right)+(\eta\leftrightarrow\eta')\right]\\
&=iP_p\left[\theta(\eta-\eta')\frac{\phi^+_p(\eta')}{\phi^+_p(\eta_0)}\left(\frac{\phi^-_p(\eta)}{\phi^-_p(\eta_0)}-\frac{\phi^+_p(\eta)}{\phi^+_p(\eta_0)}\right)+(\eta\leftrightarrow\eta')\right],
\end{align}
where $P_p$ is the power spectrum of $\phi$ defined in \eqref{pk}. When we cut internal lines we restrict the momenta of the cut line to be real and so the restriction of this definition to real $p$ is particularly useful. When $p$ is real we can use the relationship $\left[\phi^+_p(\eta)\right]^*=\phi^-_p(\eta)$ and so this can equivalently be expressed as
\begin{align}
G_p ( \eta, \eta') &=iP_p  \left[  \theta(\eta-\eta') K_p^{\ast}(\eta ) K_p(\eta' )+\theta(\eta'-\eta) K_p^{\ast}(\eta')K_p(\eta)-K_p(\eta )K_p(\eta' )\right] \label{G2}\\&=
 2 P_{p}  \left[  \theta( \eta - \eta') K_{p} ( \eta'  )\text{Im} \, K_p ( \eta  )  + \left( \eta \leftrightarrow \eta' \right)    \right]\; .
 \label{eqn:B2B}
\end{align}
Notice the overall factor of $i$ in our definition of $G$. With this definition\footnote{Another prescription would be to remove the overall $i$ from the definition of $G$ as well as the overall factor of $i$ for every diagram and put an $i$ for every vertex, e.g. $i\lambda$ for $\lambda \phi^3/3!$. At tree level, this is equivalent to our prescription because $V=I+1$. This alternative definition might be more intuitive for some because we are expanding $e^{i S_{cl}}$ in perturbation theory.} every diagram has an overall factor of $-i$ and every vertex has no factor of $i$. For example the vertex corresponding to the interaction $\lambda \phi^3/3!$ is simply $\lambda$. 

In this section, we consider only massless and conformally coupled fields, with mode functions given by:
\begin{align}
    \phi^+_k(t)&=\frac{1}{\sqrt{2c_sk}}e^{ic_skt} & \text{(massless Minkowski)}\,, \label{Minkmf}\\
    \phi^+_k(\eta)&=\frac{H}{\sqrt{2c_s^3k^3}}(1-i c_s k \eta)e^{ic_s k\eta}  & \text{(massless in de Sitter)}\,,\label{masslessdS}   \\
    \phi^+_k(\eta)&=-\frac{iH}{\sqrt{2c_sk}}\eta e^{ ic_sk\eta} & \text{(conformally coupled in de Sitter)}\label{conformaldS} \,.
\end{align}
where we allowed for an arbitrary speed of sound $c_s$ (which for a single scalar can be set to unity and included via dimensional analysis). The corresponding action is given in \eqref{eq:FLRWaction}. Later in Section \ref{MassivedS} we will generalize this discussion to the case of arbitrary masses. The final step in computing $\psi_n$ is to integrate over the conformal time of all vertices $\eta_A$ from $-\infty (1-i\e)$ to $\eta_0 \to 0$, where $\e >0$ is a small real number to be taken to zero at the end of the calculation.


\subsection{Properties of the propagators}

In this section, we discuss some remarkable properties displayed by the propagators $K$ and $G$, which in unitary theories\footnote{More precisely, our technical assumption in the derivation is that the coupling constants are real.} lead to powerful relations among the wavefunction coefficients $\psi_n$. At a general level, this discussion is strongly influenced by the cutting rules in Minkowski (see e.g. \cite{Cutkosky:1960sp,tHooft:1973wag,Veltman:1994wz,Schwartz:2013pla}), but the details are quite different. 

Let's start by discussing the bulk-to-bulk propagator $G$. A first thing to notice is that the first two terms in $G$ as written in \eqref{G2} are precisely the standard Feynman propagator $\Delta_p$, namely (for real p)
\begin{align}
    \Delta_p(\eta,\eta')=\ex{T \left( \phi_\bfp(\eta)\phi_{-\bfp}(\eta) \right)}=\theta(\eta-\eta')\phi^+_p(\eta)\phi^{+\ast}_p(\eta')+\theta(\eta'-\eta)\phi^+_p(\eta')\phi_p^{+\ast}(\eta)\,.
\end{align}
It takes this slightly unusual form because we are writing $\Delta_p$ in Fourier space for the spatial coordinates but in position space for the time coordinate. If we used the Minkowski mode functions in \eqref{Minkmf} and Fourier transformed the two times, $\eta$ and $\eta'$, to energy (frequency) space $E$ we would find an energy-conserving delta function and the familiar form $(E^2-|\bfp|^2)^{-1}$. It is straightforward to see that
\begin{align}\label{Gdelta}
    G_p(\eta,\eta')=i\Delta_p(\eta,\eta')-iP_p K_p(\eta) K_p(\eta')\,,
\end{align}
so, the difference between the bulk-to-bulk propagator and the Feynman propagator is a solution of the homogeneous equation of motion, which reminds us of the presence of a boundary at $\eta_0$ where we want to compute the wavefunction. This term is introduced to cancel out with $\Delta_p$ in the limit $\eta\to \eta_0$ or $\eta'\to \eta_0$, so that the boundary conditions in \eqref{boundaryG} are satisfied. In other words, the last term in \eqref{Gdelta} ensures that the interactions are turned off as we approach the boundary\footnote{It may happen that the interactions diverge at $\eta_0 = 0$ faster than $G$ vanishes and the result is IR divergent. In these cases, we have to evaluate the wavefunction at $\eta_0 \neq 0$, where the interactions are finite and $G$ vanishes as $(\eta\rightarrow\eta_0)$.}.

What complicates the calculation of the cosmological wavefunction is the presence of nested time integrals, which is ultimately due to the lack of time-translation invariance. Any way to circumvent or avoid nested integrals enormously simplifies the calculation. In the following we achieve precisely this by taking advantage of the two following properties: the factorized nature of the imaginary part of the bulk-to-bulk propagator $G$ and the Hermitian analyticity of the bulk-to-boundary propagator $K$.


\paragraph{Factorizing the bulk-to-bulk propagator} The “boundary" term of $ G$ in \eqref{Gdelta} is already promising because the $\eta$ and $\eta'$ dependence are factorized and so they cannot give rise to nested integrals. The difficulty sits in the part involving the Feynman propagator. There we can borrow a simple trick from Minkowski spacetime. The key observation is that the imaginary part of $G$ is factorized,
\begin{align}\label{prop1}
 \text{Im} \, G_{p} ( \eta , \eta' ) = 2 P_{p}   \text{Im}  K_{p} (\eta)  \; \text{Im} \, K_{p} (\eta')\,,
\end{align}
where we assumed $p\in \mathbb{R}$.
The hardest part is to put this observation to good use by finding observable quantities that are computed in terms of $\Im G$ as opposed to the full $G$. This problem was solved in \cite{COT}, albeit in a different language, and relies on a second key observation.


\paragraph{Hermitian analyticity of the bulk-to-boundary propagator} Notice that for all the explicit examples of mode functions given in \eqref{Minkmf}-\eqref{conformaldS}, the following relation, which we call\footnote{This nomenclature echos that used in the study of amplitudes where the 2-to-2 amplitude enjoys the following property $A^\ast(s^\ast)=A(s)$, with $s$ the Mandelstam variable.} “Hermitian analyticity", is satisfied
\begin{align}\label{prop2}
    K_{-k^\ast}^\ast(\eta)=K_k(\eta)\,, 
\end{align}
where we have allowed for a complex “energy" $k$ to account for the fact that a negative value of $k$ should be thought of as belonging to the analytic continuation of $K$ from real momenta and positive $k$. To remain compatible with the choice of a Bunch-Davies vacuum, real and negative values of $k$ should always be approached from the lower-half complex plane, $k\in \mathbb{C}^-$. We will now proceed assuming that this relation holds. Later in Section \ref{sec:flrw}, we will show that \eqref{prop2} follows very generally from the choice of the Bunch-Davies vacuum. Our general proof will also involve an additional weak technical assumption about the linearized equations of motion (which is satisfied by all the models in the literature for which we have tested it). The property \eqref{prop2} implies the following relation for the analytic continuation of the bulk-to-bulk propagator
\begin{align}\label{HAG}
    G_{-p^\ast}^\ast(\eta,\eta')&=G_p(\eta,\eta') \,.
\end{align}
This can be seen straightforwardly for both massless and conformally-coupled scalar fields for which the negative energy solutions, $\phi^-_k(\eta)$, obey the same Hermitian analyticity condition as the positive energy ones. For massive fields, this property still holds, but one has to be careful when analytically continuing the mode functions to complex energies. We leave this technical point for Section \ref{MassivedS}.


\subsection{Single-cut rules}

The single-cut rules for the wavefunction coefficients are now easily derived from the two above properties, \eqref{prop1} and \eqref{prop2}.

Consider a general tree-level diagram representing a perturbative contribution to $\psi_n$. According to the rules reviewed in the previous subsections, this diagram translates into the following expression
\begin{align}
    \psi_n (\{k\};\{p\};\{\bfk\})=- i \int \left(\prod_A^V d\eta_A F_A(\bfk)\right) \left(\prod_a^n K_{k_a}\right) \left(\prod_m^I G_{p_m}\right)\,.
\end{align}
Here $V$ is the number of vertices, $I=V-1$ the number of internal lines\footnote{Here we work exclusively at tree level. The cutting rules for loops are derived in \cite{sCOTt}.}, $\bfk_a$ and $\bfp_m$ are the external and internal momenta respectively, $\{\bfk \}$ denote additional non-energy variables obtained from rotation invariant contractions of the external momenta, and we left the time dependence implicit. Now consider choosing as variables for $\psi_n$ the norms (“energies") of the external momenta $k_a$, of all the internal momenta $p_m$ that appear in the graph, and finally of any additional scalar products $\bfk_a \cdot \bfk_b$ that might be needed to have a complete set of variables. This should always be possible according to the following counting. Using the graph relation
\begin{align}
    \sum_A^V n_A = 2I +n\,,
\end{align}
where $n_A$ is the valency of each vertex (the number of internal plus external lines ending on it, matching the number of fields in the corresponding vertex). Restricting to cubic or higher interactions\footnote{In particular here we are neglecting “quadratic interactions" which can arise from mixing of fields at linear order. Although such interactions play important roles in many multifield scenarios (e.g. see \cite{Chen:2009zp, Garcia-Saenz:2018ifx}), they bring about a technical difficulty for us: some internal and external energies in their presence become degenerate. We need the distinction between external and internal energies when flipping the sign of some while keeping unchanged the rest. Nevertheless, our results might be naturally generalized to quadratic interactions too if we think of each one of them as the soft limit of a cubic vertex that involves an auxiliary soft conformally coupled field with an independent (external) energy \cite{quadratic}.}, $n_A \geq 3$, we find the upper bound $I\leq n-3$. Therefore, the $n$ external plus $I$ internal energies account for 
\begin{align}
    n+I \leq n+(n-3)=2n-3
\end{align}
variables. For any $n\geq 3$, this is always less than or equal to the $3n-6 $ variables that are needed to describe a general $\psi_n$ (accounting for rotation and translation invariance). 

We want to consider the following analytic continuation of $\psi_n^\ast$ in all external and internal energies except for the $\tilde m$-th internal energy, which we denote by $ S \equiv p_{\tilde m}$, combined with a parity inversion of all momenta\footnote{In \cite{COT} only parity invariant scalar theories were considered in which case this parity has no effect and can be neglected. However, both in the presence of parity breaking interactions and spinning fields, this is an essential ingredient to satisfy the COT. This can be easily verified on the parity breaking quartic contact interaction of four different scalars, $\phi_1 \partial_i \phi_2 \partial_j \phi_3 \partial_k \phi_4 \e^{ijk}$.}, $\bfk \to -\bfk$,
\begin{align}
    \psi_n^\ast(\{-k\};-p_1,\dots,S,\dots,-p_I; \{-\bfk \})&=i\int \left(\prod_A^V d\eta_A F_A^\ast(-\bfk,-\bfp)\right) \nonumber \\
    & \qquad \times \left(\prod_a^n K_{-k_a}^\ast\right) \left(\prod_m^{I-1} G_{-p_m}^\ast\right) G_{S}^\ast\,,
\end{align}
where to simplify the notation we drop the complex conjugate on $-k^\ast$ and state the general convention that \textit{all the negative real energies are approached from the lower-half complex plane} \cite{COT}. Notice that by momentum conservation one also has $\bfp_m \to - \bfp_m $ so all the momenta attached to any vertex get a minus sign. Let's first see what happens to the vertices $F_A$. In real space, the vertex must involve contractions of spatial derivatives either with $\delta_{ij}$ (parity even) or with the totally anti-symmetric Levi-Civita symbol $\e_{ijk}$ (parity odd). In both cases $F^\ast(-\bfk,-\bfp)=F(\bfk,\bfp)$ because the Fourier transform gives $\partial_i \to i k_i$ and so the minus sign in $\bfk$ cancels with the minus sign from $i^\ast=-i$.
                                 
Using the invariance of the vertices and the Hermitian analyticity of $K$ and $G$, \eqref{prop2} and \eqref{HAG}, we find 
\begin{align}
    \psi_n(\{k\};p_1,\dots,S, \dots,p_I;\{\bfk \})&+\psi_n^\ast(-\{k\};-p_1,\dots,S, \dots,-p_I;-\{\bfk \})\nonumber \\
    &=-i\int \left(\prod_A^V d\eta_A F_A(\bfk,\bfp)\right) \left(\prod_a^n K_{k_a}\right)\left(\prod_m^{I-1}  G_{q_m} \right)2i \Im G_{S}\,.\label{lincomb}
\end{align}
Now we can use the factorization of $\Im G$ to re-write the above $V$ nested integrals as a product of two separated sets of (fewer) nested integrals. Then we recognize that, denoting the $n$-th external leg of a diagram by $\bfS$ and the remaining $n-1$ legs by $\bfk_a$, we have 
\begin{align}\label{lincomb2}
    \psi_n(k_1,\dots,k_{n-1},S;\{p\};\{\bfk \})&+ \psi_n^\ast(-k_1,\dots,-k_{n-1},S;-\{p\};-\{\bfk \})\\&=-i\int \left(\prod_A^V \eta_A F_A(\bfk,\bfp)\right) \left(\prod_a^{n-1} K_{k_a}\right) \left(\prod_m^{I}  G_{p_m} \right) 2i\Im K_S \nonumber\,,  
\end{align}
where $\{p\}$ denotes the set of all internal energies. The linear combination in the first line above and in \eqref{lincomb} is reminiscent of the calculation of discontinuities. Since we will encounter this combination many times in this work, it is convenience to introduce the following notation
\begin{align}
 \text{Disc}_{k_1 \dots k_j } f(k_1,\dots,k_n;\{\bfk \}) = 
 f(k_1,\dots,k_n;\{\bfk \}) - f^\ast(k_1,\dots,k_j,-k_{j+1},\dots,-k_n;-\{\bfk \}) \,,
\end{align}  
where $f$ is a generic function. Notice that the spatial momenta \textit{always} get a minus in the second term, while only the energies $k_a$ that \textit{do not} appear in the argument of $ \disc$ are analytically continued to $-k_a$. In other words, the energies appearing in the arguments of $\disc$ remain untouched and are simple spectators. 

\begin{figure}[t]
    \centering
    \includegraphics[width=0.99\textwidth]{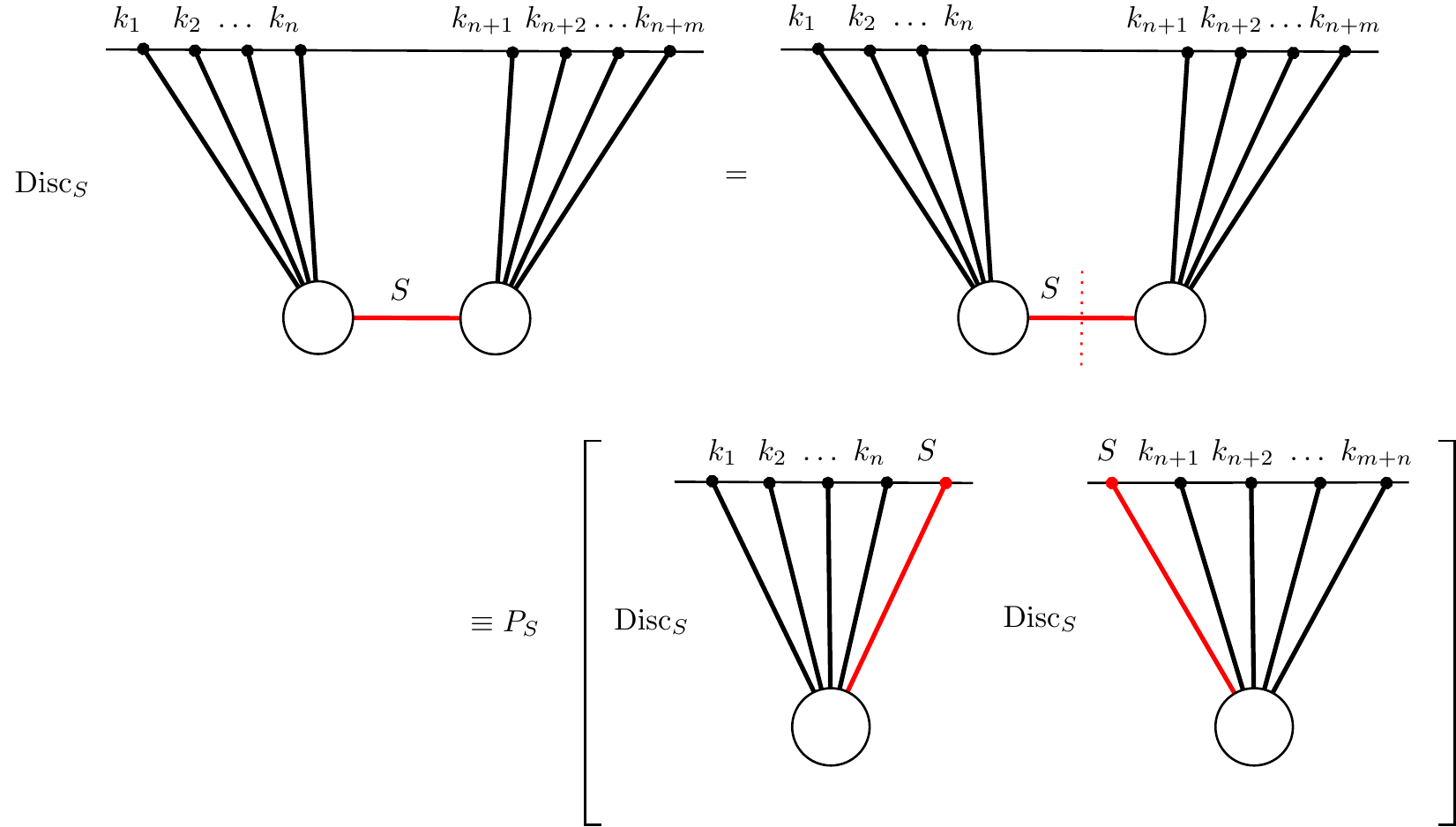}
    \caption{Diagrammatic representation of the single-cut rules defined in \eqref{eq:Cut} demonstrating the interpretation of the right-hand side as the cutting of an internal line in the diagram on the left-hand side. A cut line is pushed to the bounday, i.e. it is substituted by two external lines and a factor of the power spectrum. The discontinuity should be taken of each of the two resulting diagrams. The circles represent an arbitrary tree-level diagram with any number of internal lines.}
    \label{fig:Cut}
\end{figure}

With this notation, we can re-write the left-hand side of \eqref{lincomb} using the factorization \eqref{prop1} of $G$ and \eqref{lincomb2} into the single-cut rule
\begin{align}\label{eq:Cut}
    \disc_S \,i \psi_{n+m}(\{k\};p_1,\dots,S,\dots,p_I;\{\bfk\})& = -i P_S \,\disc_S\, i \psi_{n+1}(k_1,\dots,k_{n},S;\{p\};\{\bfk \}) \nonumber \\
    &\phantom{=} \times  \, \disc_S\, i \psi_{m+1}(k_1,\dots,k_{m},S;\{p\};\{\bfk \})\,.
\end{align}
This relationship is shown diagrammatically in Fig. \ref{fig:Cut} where we demonstrate the interpretation of the right hand side as a cut.

Finally, let's compare these single-cut rules to the recently derived Cosmological Cutting Rules \cite{sCOTt},
\begin{equation}
    i\underset{\substack{\textrm{internal}\\\textrm{lines}}}{\disc}\left[i \psi^{(D)}\right]=\sum_{\textrm{cuts}}\left[\prod_{\substack{\textrm{cut}\\\textrm{momenta}}}\int P\right]\prod_{\textrm{subdiagrams}}(-i)\underset{\substack{\textrm{internal \&}\\\textrm{cut lines}}}{\disc}\left[i\psi^{(\textrm{subdiagram})}\right].
\end{equation}
First, while both set of rules deal with the discontinuities of wavefunction coefficients, here we only consider single-cuts and so the right-hand side of our expressions contain the product of only two discontinuities. Conversely, the Cosmological Cutting Rules require the sum over all possible ways of cutting internal lines, which in general leads to the product of many discontinuities. In particular, in single-cut rules you get to choose where you want to perform the cut, while there you don't. This is particularly useful in some cases, for example in the derivation of the consequences of manifest locality through the Manifestly Local Test \cite{MLT}. Second, in the Cosmological Cutting Rules we never analytically continue internal energies $\{p\}$. This means that there is no need to arrange so that the variables $\{p\}$ appear explicitly in the argument of $\psi_n$. In contrast here we need to access each internal energy independently: the uncut internal energies are analytically continued while the cut energy is not. One consequence of this is that for single-cut rules one generally needs to choose different sets of variables for different channels.
Third, because of the above considerations it seems challenging to extend our single-cut rules to loop diagrams: in that case there are internal energies that are integrated over and it is not clear how one would analytically continue them by manipulating the kinematical variables.

\section{Generalisations}\label{sec3}

For the sake of clarity, the derivation in the previous section was written for the simplest case of a massless or conformally coupled scalar field in a de Sitter background with no time-derivative interactions and Bunch-Davies initial conditions. In this section we show that those results can be greatly generalised to general FLRW spacetimes that admit a Bunch-Davies vacuum for any fields with a linear dispersion relation and arbitrary mass, integer spin and speed of sound.

\subsection{Time Derivatives}
\label{Deriv}
When the interaction is allowed to involve time derivatives this introduces derivatives on the Green's function that potentially alter the analysis as it is no longer immediately possible to exploit the result from \eqref{prop1}. Instead it is necessary to understand the behaviour of terms like
\begin{equation}
 \Im\left[ \partial^N_{\eta}\partial^M_{\eta'}G_p(\eta,\eta')\right]\,.
\end{equation}
This generalisation is simpler than it might seem as the energies are the only complex variables and therefore complex conjugation commutes with the time derivatives and so derivatives of $K$ and $G$ will remain Hermitian analytic,
\begin{equation}
    \left[\partial_\eta^NK_{-k^*}(\eta)\right]^*=\partial_\eta^NK_{-k^*}^*(\eta)=\partial_\eta^NK_k(\eta)\, ,
\end{equation}
\begin{equation}
    \left[\partial_\eta^N\partial_{\eta'}^M G_{-p^*}(\eta,\eta')\right]^*=\partial_\eta^N\partial_{\eta'}^M G^*_{-p^*}(\eta,\eta')=\partial_\eta^N\partial_{\eta'}^MG_p(\eta,\eta')\,.
\end{equation}
Likewise, we can explore the imaginary part of $G$ for real $p$,
\begin{align}
    \Im\left[\partial_\eta^N\partial_{\eta'}^M G_p(\eta,\eta')\right]&=\partial_\eta^N\partial_{\eta'}^M \Im\left[G_p(\eta,\eta')\right]\\&=2 P_{p}   \text{Im}\; \partial_\eta^N K_{p} (\eta)  \; \text{Im} \, \partial_{\eta'}^MK_{p} (\eta')\, .
\end{align}
 From this it is apparent that we can cut lines involving time derivatives in exactly the same way as non-derivative interactions, except each of the diagrams must include the derivatives previously associated with the bulk-to-bulk propagator on the the external lines that are introduced. We can further clarify this by looking at the full wavefunction coefficient,
\begin{align}
    \psi_n (\{k\};\{p\};\{\bfk\})= i \int \left(\prod_A^V d\eta_A F_A(\bfk)\right) \left(\prod_a^n K^{(N_a)}_{k_a}\right) \left(\prod_m^I G^{(N_m,M_m)}_{p_m}\right)\,.
\end{align}
Where we have introduced the notation
\begin{equation}
    K^{(N)}(\eta)=\partial_\eta^{N}K(\eta)
\end{equation}
whilst suppressing the $\eta$ dependence. This has an imaginary discontinuity arising from cutting the internal line with momentum $\textbf{S}$ given by
\begin{align}\nonumber
    &\text{Disc}_{S} \ i\psi_{n+m}(\{k\};p_1,\dots,S,\dots, p_I;\{\bfk\})\\\nonumber&=i \int \left(\prod_A^V d\eta_A F_A(\bfk)\right) \left(\prod_a^{n+m} K^{(N_a)}_{k_a}\right) \left(\prod_{l}^{I} G^{(N_l,M_l)}_{p_l}\right)\left(iG^{(N,M)}_{S}-i\left[G^{(N,M)}_{S}\right]^*\right)\\\nonumber&=i P_{S}\int \left(\prod_A^V d\eta_A F_A(\bfk)\right) \left(\prod_a^{n+m} K^{(N_a)}_{k_a}\right) \left(\prod_{l}^{I} G^{(N_l,M_l)}_{p_l}\right)\left( K_{S}^{(N)}- {K_{S}^{(N)}}^*\right)\left( K_{S}^{(M)}- {K_{S}^{(M)}}^*\right)\\&= -iP_{S}\ \text{Disc}_{S} \left[i\psi_{n+1}\left(k_1,\dots,k_n,S;\{p\};\{\bfk\}\right)\right]\text{Disc}_{S}\left[ i\psi_{m+1}\left(S,k_1,\dots,k_m;\{p\};\{\bfk\}\right)\right]\,.
\end{align}
This is the same as the expression with non time-derivative interactions except the propagators are now allowed to have derivatives. Therefore, the single-cut rules derived in the previous section apply to any derivative interactions as well. The generalization to time derivatives for spinning fields proceeds similarly as the modefunctions remain Hermitian analtyic and the vertex terms are time independent, see Section \ref{sec:spinning} for more details. 


\subsection{General cosmological backgrounds}\label{sec:flrw}

We know that our single-cut rules cannot be completely generic because when we allow for non Bunch-Davies initial conditions, even within the de Sitter case, our derivation breaks down as the pivotal condition $K_k(\eta)=K^*_{-k^*}(\eta)$ is in general no longer valid. Since there are examples in the literature, see e.g. \cite{kofman1997towards,barnaby2012gravity}, of backgrounds that excite negative frequency modes even when they start with Bunch-Davies initial conditions, one might wonder under what conditions our single-cut rules apply to general FLRW spacetimes. 

In this section, we demonstrate that, for a fairly generic FLRW background spacetime, the imposition of Bunch-Davies initial conditions is sufficient to ensure that the propagators are Hermitian analytic. We then also present the mode functions for both a massive field in de Sitter and a massless field in an alternative background, where the background excites negative frequency modes from a Bunch-Davies vacuum, and show that the propagators behave in the same way as for massless and conformally coupled fields. Throughout we focus on linear dispersion relations of the form $E \propto |\bfk|$, which are generic in relativistic theories to leading order in derivatives (two time or space derivatives in the quadratic action). Conversely, we stress that we don't require interactions to be Lorentz invariant.


\subsubsection{Bulk-to-Boundary Propagator}

We will consider here scalar fields with a quadratic action of the form
\begin{align}\label{eq:FLRWaction}
    S=\int d\eta\,d^3\bfx\, a^2(\eta)\,\left(\dfrac{1}{2}\phi'^2-\dfrac{c^2_s(\eta)}{2}(\partial_i \phi)^2-\dfrac{1}{2}a^2(\eta)m^2(\eta)\phi^2\right)\,.
\end{align}
This is the most general quadratic action of a real scalar field to leading (quadratic) order in derivatives. We will discuss higher derivatives and more general dispersion relations in Section \ref{sec:nonlin}. The mode functions $\phi(k,\eta)$ satisfy a second order differential equation of the form
\begin{equation}\label{eq:phiEOM}
\phi''_k(\eta)+p(k,\eta)\phi_k'(\eta)+q(k,\eta)\phi_k(\eta)=0\,,
\end{equation}
where
\begin{align}
    p(k,\eta)=\dfrac{2a'}{a}\,,\qquad q(k,\eta)=c_s^2(\eta)k^2+a^2(\eta)m^2(\eta)\,.
\end{align}
We will assume that $p$ and $q$ are analytic functions of $k$ and $\eta$ over the domain of $\eta$. The ordinary differential equation \eqref{eq:phiEOM} has two linearly independent solutions that we call $\phi^\pm_k(\eta)$ from which we construct $K_k(\eta)= \phi^+_k(\eta)/\phi_k^+(\eta_0)$ where $\eta_0$ is our late time boundary (usually $\eta_0 = 0$). The condition $K_k(\eta)=K^*_{-k^*}(\eta)$ can be expressed in terms of $\phi^+_k(\eta)$,
\begin{equation}\label{eq:HAphi}
    \phi^+_k(\eta)=\frac{\phi^+_k(\eta_0)}{\left[\phi^+_{-k^*}(\eta_0)\right]^*}\left[\phi^+_{-k^*}(\eta)\right]^*\equiv A(k,\eta_0)\left[\phi^+_{-k^*}(\eta)\right]^*.
\end{equation}
Equivalently, in words, these two functions are linearly dependent. It is well known\cite{bocher1901certain} that two analytic functions are linearly dependent if their Wronskian, namely
\begin{equation}
    W(k,\eta)\equiv W\left( \phi^+(k,\eta),\left[\phi^+_{-k^*}(\eta)\right]^*\right)= \phi^+_k(\eta)\ \partial_\eta \left[\phi^+_{-k^*}(\eta)\right]^*- \left[\phi^+_{-k^*}(\eta)\right]^* \partial_\eta \phi^+_k(\eta) ,
\end{equation}
vanishes everywhere. Furthermore, if two functions both satisfy the same differential equation of the form in (\ref{eq:phiEOM}) and their Wronskian is zero at some point $\eta_i$ then, because the Wronskian is given by
\begin{equation}
    W(k,\eta)=W(k,\eta_i)e^{-\int_{\eta_i}^{\eta} p(k,\eta')d\eta'},
\end{equation}
it must vanish everywhere (see e.g. \cite{riley2002mathematical}) by virtue of the assumption that $p$ and $q$ are analytic on this domain. To fix $\phi_k^+(\eta)$ we specify Bunch-Davies initial conditions,
\begin{equation}\label{eq:B-D}
    \lim_{\eta\rightarrow -\infty}\phi^+_k(\eta)\propto a^{-1}(\eta)e^{ic_sk\eta}.
\end{equation}
This solution assumes that $c_sk\eta$ diverges in the infinite past,
\begin{equation}
    \lim_{\eta\rightarrow -\infty} c_s k \neq 0.
\end{equation}
By imposing this initial condition we have assumed that \eqref{eq:B-D} becomes a solution to our differential equation in the infinite past. To understand when this occurs we rewrite the differential equation as
\begin{equation}
    (a(\eta)\phi_k(\eta))''+\left(c_s^2(\eta)k^2+a^2(\eta)m^2(\eta)-\frac{a''(\eta)}{a(\eta)}\right)a(\eta)\phi_k(\eta)=0.
\end{equation}
We can first of all see that the Bunch-Davies initial condition has no dependence on $m$ or $a$ (except through the prefactor of $a^{-1}$) and so the last two terms multiplying $a\phi_k$ must be negligible compared to $c_s^2k^2$ in this limit,
\begin{align}\label{eq:cond1}
    \lim_{\eta\rightarrow -\infty} am&\ll c_s k \,, \\\label{eq:cond2}
    \lim_{\eta\rightarrow -\infty}\frac{a''}{a}&\ll c_s^2k^2.
\end{align}
We further need that $c_s$ is approximately constant. To quantify this condition we insert this asymptotic solution into the differential equation which gives
\begin{equation}
    \lim_{\eta\rightarrow -\infty} c_s^2k^2\left(-2\frac{c_s'}{c_s}\eta-\left(\frac{c_s'}{c_s}\eta\right)^2+2i\frac{c_s'}{c_s}\frac{1}{c_s k}+i\frac{c_s''}{c_s}\frac{\eta}{c_sk}\right)e^{ic_sk\eta}=0\,.
\end{equation}
For this to be an asymptotic solution, we generically require that each of the terms in the bracket vanishes individually and so
\begin{align}
    \lim_{\eta\rightarrow -\infty} \frac{d\log(c_s)}{d\log(\eta)}&\ll 1\,,\\
    \lim_{\eta\rightarrow -\infty}\frac{d^2\log(c_s)}{d\log(\eta)^2}&\ll c_s k\eta\,.
\end{align}
Imposing this initial condition also restricts the asymptotic form of $\left[\phi^+_{-k^*}(\eta)\right]^*$
\begin{equation}
    \lim_{\eta\rightarrow -\infty}\left[\phi^+_{-k^*}(\eta)\right]^*\propto \left[a^{-1}(\eta)\right]^*e^{ic_s^*k\eta}\,.
\end{equation}
So, these two functions are asymptotically proportional to each other provided $c_s$ and $a$ are real in which case the Wronskian in the infinite past is
\begin{equation}
    \lim_{\eta\rightarrow -\infty}W(k,\eta)=0.
\end{equation}
Therefore, provided $\phi^+_k(\eta)$ and $\left[\phi^+_{-k^*}(\eta)\right]^*$ satisfy the same differential equation, they are guaranteed to be linearly dependent. To determine what differential equation $\left[\phi^+_{-k^*}(\eta)\right]^*$ satisfies, we can take the complex conjugate of \eqref{eq:phiEOM} and replace $k\rightarrow -k^*$ everywhere to give
\begin{equation}
\partial_\eta^2\left[\phi^+_{-k^*}(\eta)\right]^*+p^*(-k^*,\eta)\partial_\eta \left[\phi^+_{-k^*}(\eta)\right]^*+q^*(-k^*,\eta)\left[\phi^+_{-k^*}(\eta)\right]^*=0\,.
\end{equation}
This coincides with \eqref{eq:phiEOM} if $p$ and $q$ are Hermitian analytic. Therefore, the bulk-to-boundary propagator is Hermitian analytic if:
\begin{itemize}
    \item We impose Bunch-Davies initial conditions, which requires that
    \begin{itemize}
        \item $\displaystyle\lim_{\eta\rightarrow -\infty} c_s k \neq 0$,
        \item $\displaystyle\lim_{\eta\rightarrow -\infty} \frac{d\log(c_s)}{d\log(\eta)}\ll 1$,
        \item $\displaystyle\lim_{\eta\rightarrow -\infty}\frac{d^2\log(c_s)}{d\log(\eta)^2}\ll c_s k\eta$,
        \item $\displaystyle\lim_{\eta\rightarrow -\infty} a m \ll c_s k$,
        \item $\displaystyle \lim_{\eta\rightarrow -\infty}\frac{a''}{a}\ll c_s^2k^2$,
    \end{itemize}
    \item $\displaystyle\frac{a'}{a},\  a(\eta),\ m(\eta)\text{ and } c_s(\eta)$ are real, analytic functions in the domain $\eta\in(-\infty,0]$.
\end{itemize}
Where we have expressed the conditions on $p$ and $q$ in terms of the functions that appear in the quadratic action. One can also show this by making a WKB approximation (see e.g. \cite{riley2002mathematical}) of the mode function in a general flat FLRW spacetime, where we can see generically that $\phi_{\pm}=e^{ik\sigma_{\pm}(k,\eta)}/a$, where $\sigma$ is Hermitian analytic. The details are included in Appendix \ref{WKB}.


\subsubsection{Bulk-to-Bulk Propagator}
In order to extend our results to diagrams with more than one internal line we also need to prove that for a generic background
\begin{equation}
G_{-p^*}^*(\eta,\eta')=G_p(\eta,\eta').
\end{equation}
To do this we need to know the Hermitian analytic properties of $\phi^-_k(\eta)$ in addition to those of $\phi^+_k(\eta)$, which we have already established. To determine these, consider that $\phi^\pm_k(\eta)$ are defined to have a Wronskian
\begin{equation}
W_k(\eta)\equiv a^2(\eta)\left(\phi^+_k(\eta)\partial_\eta\phi_k^-(\eta)-\phi^-_k(\eta)\partial_\eta\phi^+_k(\eta)\right)=-i.
\end{equation}
We can treat this as a differential equation in $\phi_k^-(\eta)$,
\begin{equation}
\partial_\eta\phi_k^-(\eta)-\frac{\partial_\eta \phi_k^+(\eta)}{\phi_k^+(\eta)}\phi_k^-(\eta)=-\frac{i}{a^2(\eta) \phi_k^+(\eta)},
\end{equation}
which can be formally solved:
\begin{equation}
\phi_k^-(\eta)=-\phi_k^+(\eta)\int_{\eta_0}^{\eta}\frac{i}{a^2(\eta')\phi_k^+(\eta')^2}d\eta'+\frac{\phi_k^+(\eta)}{\phi_k^+(\eta_0)}\phi_k^-(\eta_0).
\end{equation}
If the above assumptions are valid we can then exploit the Hermitian analytic properties of $\phi_k^+(\eta)$ to give
\begin{align}
\left[\phi_{-k^*}^-(\eta)\right]^*&=A(k,\eta_0)\phi_k^+(\eta)\int_{\eta_0}^{\eta}\frac{i}{a^2(\eta')\phi_k^+(\eta')^2}d\eta'+\frac{\phi_k^+(\eta)}{\phi_k^+(\eta_0)}\left[\phi_{-k^*}^-(\eta_0)\right]^*\\&=A(k,\eta_0)\frac{\phi_k^+(\eta)}{\phi_k^+(\eta_0)}\phi_k^-(\eta_0)-A(k,\eta_0)\phi_k^-(\eta)+\frac{\phi_k^+(\eta)}{\phi_k^+(\eta_0)}\left[\phi_{-k^*}^-(\eta_0)\right]^*,\label{eq:HAphi-}
\end{align}
where $A$ was defined in \eqref{eq:HAphi}. The bulk-to-bulk propagator is
\begin{equation}
G_p(\eta,\eta')=i \theta(\eta-\eta')\left(\phi_p^+(\eta')\phi_p^-(\eta)-\frac{\phi_p^-(\eta_0)}{\phi_p^+(\eta_0)}\phi_p^+(\eta)\phi^+ (\eta')\right)+\eta\leftrightarrow \eta',
\end{equation}
and its Hermitian analytic image is
\begin{equation}
    G_{-p^*}^*(\eta,\eta')=-i \theta(\eta-\eta')\left[\phi_{-p^*}^+(\eta')\right]^*\left(\left[\phi_{-p^*}^-(\eta)\right]^*-\frac{\left[\phi_{-p^*}^-(\eta_0)\right]^*}{\left[\phi_{-p^*}^+(\eta_0)\right]^*}\left[\phi_{-p^*}^+(\eta)\right]^*\right)+\eta\leftrightarrow \eta'\, .
\end{equation}
Using the relationships in \eqref{eq:HAphi} and \eqref{eq:HAphi-} we find this is equal to $G_p(\eta,\eta')$ and so the bulk-to-bulk propagator is Hermitian analytic whenever the bulk-to-boundary propagator is.



\subsubsection{Massive Fields in de Sitter}\label{MassivedS}
As a concrete example of these results consider the case of a single scalar field of mass $m$ with a constant speed of sound, which we set to $1$, in de Sitter. With these assumptions $p(k,\eta)=-\frac{2}{\eta}$ and $q(k,\eta)=m^2+k^2$ which are both  analytic functions for all time and, furthermore, are both Hermitian analytic by inspection. Hence in this theory $K_k(\eta)=K^*_{-k^*}(\eta)$. This was shown in \cite{COT}. We can also see that the bulk-to-bulk propagator is Hermitian analytic by introducing $\phi^\pm_k$ for a massive scalar field,
\begin{align}
    \phi^+_k(\eta)&=+ i e^{- i\frac{\pi}{2}\left(\nu+\frac{1}{2}\right)}\sqrt{\pi}\frac{H}{2}(-\eta)^\frac{3}{2}H^{(2)}_\nu(-k\eta)\,,\\
    \phi^-_k(\eta)&=-i e^{+ i\frac{\pi}{2}\left(\nu+\frac{1}{2}\right)}\sqrt{\pi}\frac{H}{2}(-\eta)^\frac{3}{2}H^{(1)}_\nu(-k\eta)\,.
\end{align}
So,
\begin{align}
    \left[\phi_{-k^*}^{+}(\eta)\right]^*&=-ie^{+ i\frac{\pi}{2}\left( \nu^*+\frac{1}{2}\right)}\sqrt{\pi}\frac{H}{2}(-\eta)^{\frac{3}{2}}H_{\nu^*}^{(1)}(k\eta)\,,\\\left[\phi_{-k^*}^{-}(\eta)\right]^*&=+ ie^{- i\frac{\pi}{2}\left( \nu^*+\frac{1}{2}\right)}\sqrt{\pi}\frac{H}{2}(-\eta)^{\frac{3}{2}}H_{\nu^*}^{(2)}(k\eta)\,.
\end{align}
When $\nu$ is real we will recover the original Hankel functions but when $\nu$ is imaginary we pick up a minus sign, this cancels with the sign change in the exponential factor as
\begin{align}
    H_{-\nu}^{(1)}(z)&=e^{\pm i\pi\nu}H_{\nu}^{(1)}(z)\,,\\H_{-\nu}^{(2)}(z)&=e^{- i\pi\nu}H_{\nu}^{(2)}(z)\,,
\end{align}
and so we can just drop the complex conjugates on $\nu$. We wish to express this in terms of the original mode functions and so we replace $k\eta=e^{i\pi} (-k\eta)$ where this particular choice of argument for $-1$ is enforced by the fact that $\Im(k)<0$. We then use the analytic continuation from Section 10.11 of \cite{NIST:DLMF},
\begin{align}
    H_\nu^{(1)}(e^{i\pi}z)&=-e^{-i\pi\nu}H_\nu^{(2)}(z),\\
    H_\nu^{(2)}(e^{i\pi}z)&=e^{i\pi\nu}H_\nu^{(1)}(z)+2\cos(\pi\nu)H_\nu^{(2)}(z),
\end{align}
to give
\begin{align}
   \left[\phi_{-k^*}^{+}(\eta)\right]^*&=i\phi_k^+(\eta),\label{eq:massiveconjp}\\
    \left[\phi_{-k^*}^-(\eta)\right]^*&=i\phi_k^-(\eta)+2\cos(\pi\nu)\phi_k^+(\eta).\label{eq:massiveconjm}
\end{align}
The Hermitian analytic image of the Green's function is
\begin{align}\nonumber
    G_{-p^*}^*(\eta,\eta')&=-i \theta(\eta-\eta')\left[\phi_{-p^*}^+(\eta')\right]^*\left[\phi_{-p^*}^+(\eta)\right]^*\left(\frac{\phi_{-p^*}^-(\eta)}{\phi_{-p^*}^+(\eta)}-\frac{\phi_{-p^*}^-(\eta_0)}{\phi_{-p^*}^+(\eta_0)}\right)^*+\eta\leftrightarrow \eta'\\\nonumber
   &=i \theta(\eta-\eta')\phi^+_p(\eta')\phi^+_p(\eta)\left(\frac{\phi_p^-(\eta)}{\phi_p^+(\eta)}-2i\cos(\pi\nu)-\frac{\phi_p^-(\eta_0)}{\phi_p^+(\eta_0)}+ 2i\cos(\pi\nu)\right)+\eta\leftrightarrow \eta' \\&=G_p(\eta,\eta').
\end{align}
Therefore, the Hermitian analyticity of both propagators is manifest in this theory. 


\subsubsection{Resonant Non-Gaussianity}
One might be worried that due to the evolution in the bulk, negative modes may be excited and ruin the Hermitian analyticity of the propagators, even if we start with Bunch-Davies initial conditions. An example to consider is the resonant production of particles in axion monodromy inflation \cite{McAllister:2008hb,Flauger:2009ab} and the associated non-Gaussianity \cite{Chen:2008wn,flauger2011resonant,Leblond:2010yq}. Here the inflaton potential is modulated by non-perturbative effects
\begin{equation}
    V(\phi)=V_0(\phi)+\Lambda^4\cos\left(\frac{\phi}{f}\right),
\end{equation}
which in turn induce small oscillations in the background. Solving for the slow-roll parameters gives
\begin{align}
    \epsilon&\equiv -\frac{\dot H}{H^2}\simeq \epsilon_*-3b_* f\sqrt{2\epsilon_*}\cos\left(\frac{\phi_0}{f}\right), & \delta&\equiv \frac{\ddot H}{2H\dot H}\simeq  \epsilon_*-\eta_*-3b_*\sin\left(\frac{\phi_0}{f}\right).
\end{align}
where (see \cite{flauger2011resonant} for more details)
\begin{equation}
    \phi_0=\phi_*+\sqrt{2\epsilon_*}\ln(-k_*\eta)\,.
\end{equation}
Putting these slow-roll parameters into the Mukhanov-Sasaki equation gives us the following coefficients for the differential equations:
\begin{align}
    p(k,\eta)&=-\frac{2}{\eta}\left[1+3\epsilon_*-\eta_*-6b_*f\sqrt{2\epsilon_*}\cos\left(\frac{\phi_0}{f}\right)-3b_*\sin\left(\frac{\phi_0}{f}\right)\right],\\
    q(k,\eta)&=k^2.
\end{align}
Notice that both $p(k,\eta)$ and $q(k,\eta)$ are Hermitian analytic (the conditions on the behaviour of the scale factor follow from the fact that this inflationary spacetime is close to de Sitter and $m=0$, $c_s=1$ everywhere). We therefore expect to find that $K_k(\eta)$ is Hermitian analytic. This is not immediately apparent from the results in the paper, however to see that it is, in fact, true we compute the bulk-to-boundary propagator from the modefunctions given in (2.21) of \cite{flauger2011resonant}\footnote{In this paper they define the modefunctions, which they label as $\mathcal{R}_k$, to be
\begin{equation}
    \mathcal{R}_k(-k\eta)=\mathcal{R}_{k,0}^{(o)}\left[i\sqrt{\frac{\pi}{2}}(k\eta)^{\frac{3}{2}}H_{3/2}(-k\eta)-c_k^{(-)}i \sqrt{\frac{\pi}{2}}(-k\eta)^{\frac{3}{2}}H_{3/2}^{(2)}(-k\eta)\right]\propto \phi^-_k(\eta)
\end{equation}
therefore, $c_k(-k\eta)=\left[c_k^{(-)}\right]^*$. }
\begin{equation}
K_k(\eta)=\frac{(1-ik\eta)e^{ik\eta}+c_k(-k\eta)(1+ik\eta)e^{-ik\eta}}{1+c_k(-k\eta_0)}.
\end{equation}
The Hermitian analytic image of this propagator is given by
\begin{equation}
    K^*_{-k^*}(\eta)=\frac{(1-ik\eta)e^{ik\eta}+c_{k^*}^{*}(k^*\eta)(1+ik\eta)e^{-ik\eta}}{1+c_{k^*}^{*}(k^*\eta_0)}.
\end{equation}
Therefore, we conclude that $K_k(\eta)=K^*_{-k^*}(\eta)$ if (and only if)
\begin{equation}
    \left[c_{k^*}^{*}(k^*\eta)\right]^*=c_k(-k\eta).
\end{equation}
Using the definition of $c_k$ from \cite{flauger2011resonant},
\begin{multline}
    c_k=-\frac{3}{2} b \frac{f}{\sqrt{2\epsilon_*}}\left\{e^{\frac{\sqrt{2\epsilon_*}}{f}\left(\frac{\pi}{2}-i\ln 2\right)}e^{i\frac{\phi_k}{f}}\Gamma\left[1+i\frac{\sqrt{2\epsilon_*}}{f},-2ik\eta\right]\right.\\\left.+e^{-\frac{\sqrt{2\epsilon_*}}{f}\left(\frac{\pi}{2}-i\ln 2\right)}e^{-i\frac{\phi_k}{f}}\Gamma\left[1-i\frac{\sqrt{2\epsilon_*}}{f},-2ik\eta\right]\right\}
\end{multline}
we find that, in order for $c_k$ to be Hermitian analytic, it is necessary to keep a term that is suppressed by a factor of $e^{-\frac{\pi\sqrt{2\epsilon_*}}{2f}}$ where $\frac{f}{\sqrt{2\epsilon_*}}$ is assumed small. Therefore, we have recalculated $c_k$ in Appendix \ref{RNG} without making any assumptions on the size of this quantity and we find that it remains true that $c_k$ is Hermitian analytic. Furthermore, we showed earlier in this section that the Hermitian analyticity of the bulk-to-bulk propagator follows from that of the bulk-to-boundary propagator, therefore, even though this background excites negative energy modes the single-cut rules and the Cosmological Optical Theorem still apply exactly as expected. 


\subsection{Non-linear dispersion relations: a non Hermitian-analytic example}\label{sec:nonlin}

So far we showed that Hermitian analyticity is very general for linear dispersion relations, as derived from the leading-derivative quadratic action in \eqref{eq:FLRWaction}. Here we want to point out that this is not the case for general non-linear dispersion relations.

As a toy model, consider the following quadratic action for a real scalar field
\begin{equation}
    S=\int d\eta d^3\bfx \frac{a^4}{2} \left(  \frac{\phi'^2}{a^2}- \frac{c_n^2(\eta)}{a^{2n}} (\partial_{i_1\dots i_n}\phi)^2\right),
\end{equation}
where $c_n$ is some real, possibly time-dependent parameter, $n$ is a positive integer and for concreteness we take $a$ to be the scale factor in de Sitter\footnote{We expect a similar discussion to hold for general accelerated FLRW spacetimes.}. This action has equations of motion that are second order in time, but involves an arbitrary number, $2n$, of spatial derivatives. For $n=1$, this theory reduces to what we studied previously, \eqref{eq:FLRWaction}, while for $n=2$, it reduces to the ghost condensate studied in \cite{ArkaniHamed:2003uy}. For any $n \neq 1$ we have a non-linear dispersion relation $E^2=c_n^2 k^{2n}$.

The equation of motion for $\phi$ is
\begin{equation}
    \phi''_k(\eta)+\frac{2a'}{a}\phi'_k(\eta)+c_n^2a^{2-2n}k^{2n}\phi_k(\eta)=0,
\end{equation}
which satisfies the requirement that $p$ and $q$ are Hermitian analytic for real $a$ and $c_2$. The positive energy initial conditions must now be implemented as
\begin{equation}\label{eq:initialghost}
    \lim_{\eta\rightarrow -\infty}\phi_k^+(\eta)\propto a^{\frac{n-3}{2}} e^{-i\frac{c_n k^n(-\eta)^n}n}.
\end{equation}
To ensure that this is finite in the infinite past we require that $\Im(k^n)<0$. If $n$ is even, then the initial condition for the Hermitian analytic image of $\phi_k^+$ is
\begin{equation}\label{eq:initialghost2}
    \lim_{\eta\rightarrow -\infty}\left[\phi_{-k^*}^+(\eta)\right]^*\propto a^{\frac{n-3}{2}} e^{i\frac{c_n k^n(-\eta)^n}{n}}.
\end{equation}
We see that \eqref{eq:initialghost} and \eqref{eq:initialghost2} are genuinely different solutions, and so the Wronskian will not vanish in the infinite past. Therefore these functions are linearly independent and the bulk-to-boundary propagator will not be Hermitian analytic. This can also be seen explicitly by considering the concrete example of $n=2$, namely the ghost condensate. The mode function and bulk-to-boundary propagator are found to be
\begin{align}
    \phi^+ &=\sqrt{\frac{\pi}{8}}H(-\eta)^{3/2} H^{(2)}_{3/4} \left( \frac{H c_2 k^2 \eta^2}{2} \right)\,, \\
    K_k(\eta)&=\frac{(-\eta)^{3/2}H^{(2)}_{3/4}\left(\frac{1}{2}Hc_2k^2\eta^2\right)}{(-\eta_0)^{3/2}H^{(2)}_{3/4}\left(\frac{1}{2}Hc_2k^2\eta_0^2\right)}.
\end{align}
We can then compute the Hermitian analytic image of $K$:
\begin{equation}
    K_{-k^*}^*(\eta)=\frac{(-\eta)^{3/2}H^{(1)}_{3/4}\left(\frac{1}{2}Hc_2k^2\eta^2\right)}{(-\eta_0)^{3/2}H^{(1)}_{3/4}\left(\frac{1}{2}Hc_2k^2\eta_0^2\right)}\neq K_k(\eta).
\end{equation}
This means that the single-cut rules as written in this paper to do not apply to this case. Because the dispersion relation for the ghost condensate is still a simple monomial, $E \sim k^2$, we can find an ad hoc modification of our single-cut rules that does work. In particular, consider the transformation $k\to \bar{k}=\pm ik^*$. Using this to define a modified Hermitian analytic image, we find 
\begin{equation}
    K_{\bar{k}}^*(\eta)=K_k(\eta).
\end{equation}
 Therefore, in this a theory, we expect it to be possible to derive similar results to those presented in the rest of the paper, with the replacement $-k^* \to \bar{k}$. We don't pursue this further here. Instead, we notice that, for more general, non-monomial dispersion relations, even this modified Hermitian analyticity does not exist. For example consider the theory
\begin{equation}
    S=\int d\eta d^3\bfx \,\frac{a^2}{2} \left(
     \phi'^2-  c_s^2 \partial_i\phi\partial_i\phi-\frac{c_2^2}{a^2} \partial_{ij}\phi\partial_{ij}\phi\right),
\end{equation}
for constant, real $c_s,\, c_2$. In this case we require
\begin{equation}
    q(\bar{k},\eta)=c_s^2\bar{k}^2+\frac{c_s^2\bar{k}^4}{a^2}=q^*(k,\eta),
\end{equation}
which has six solutions,
\begin{align}\label{eq:kcond1}
    \bar{k}&=\pm k^*\,,\\\label{eq:kcond2}
    \bar{k}&=\pm i \sqrt{\frac{a^2c_s^2}{2c_2^2}\left(1\pm\sqrt{1+4{k^*}^2+\frac{4c_2^4}{a^4c_s^4}{k^*}^4}\right)}\,.
\end{align}
The higher derivative term will come to dominate in the infinite past, so, the positive energy initial condition is 
\begin{equation}
    \lim_{\eta\rightarrow -\infty}\phi_k^+(\eta)\propto a^{-\frac{1}{2}} e^{-i\frac{c_2 k^2\eta^2}2}.
\end{equation}
This is equal to the same limit of $\left[\phi_{\bar{k}}^+(\eta)\right]^*$ if
\begin{equation}
    \lim_{\eta\rightarrow -\infty}\bar{k}=\pm ik^*.
\end{equation}
We therefore cannot use the either transformation from \eqref{eq:kcond1}. Provided we take the positive sign for the second $\pm$ in \eqref{eq:kcond2} we find the correct asymptotics. Then, taking
\begin{equation}\label{eq:bark}
    \bar{k}=\pm i\sqrt{\frac{a^2c_s^2}{2c_2^2}\left(1+\sqrt{1+4{k^*}^2+\frac{4c_2^4}{a^4c_s^4}{k^*}^4}\right)},
\end{equation}
will ensure that
\begin{equation}
    K_{\bar{k}}^*(\eta)=K_k(\eta).
\end{equation}
The problem with \eqref{eq:bark} is that now $\bar k$ depends on time. If we defined a $\disc$ with this Hermitian analytic image, the $\disc$ would not commute with the time integrals appearing in the Feynman rules for the calculation of the wavefunction and our derivation would not work. We don't pursue this further here, but notice that because of this it seems unlikely that single-cut or general cutting rules can be derived for general non-linear dispersion relations.

\subsection{Spinning fields}\label{sec:spinning}

In this section we discuss the generalization of our single-cut rules to integer spin fields. For concreteness we focus on the very general class of free theories for such fields that was developed in \cite{Bordin2018}
\begin{align}
    S=\int d^3x dt \,a^3\, \frac{1}{2s!} \left[(\dot \Phi^{i_1 \dots i_s})^2 - \frac{c_s^2}{a^2} (\partial_j \Phi^{i_1 \dots i_s})^2-\frac{\delta c_s^2}{a^2} (\partial_j \Phi^{j i_2 \dots i_s})^2 -m^2 ( \Phi^{i_1 \dots i_s})^2 \right]\,,
\end{align}
where $\Phi^{i_1 \dots i_s}$ is a totally-symmetric, traceless tensor with only spatial indices, $i_{1}=1,2,3$. This theory arises in generic models of inflation where the background of the inflaton selects a preferred time foliation of spacetime into spatial hypersurfaces. The above expression can be written in a covariant way by using the Goldstone boson $\pi$ of time translations to upgrade the spatial tensor $ \Phi^{i_1 \dots i_s} $ to a covariant spacetime tensor. The coupling of $ \Phi^{i_1 \dots i_s} $ to $\pi$ is also dictated by this constructions but we will not need this here. Notice that $ \Phi^{i_1 \dots i_s} $ has $(2s+1)$ components, which each create states (“particles") with helicities $ 0,\pm 1, \dots, \pm s$, respectively.


\subsubsection{Hermitian anaylicity of the propagators}

The equation of motion for the field $\Phi_{i_1\dots i_s}$ is given by:
\begin{equation}
    \Phi_{i_1\dots i_s}''+2\frac{a'}{a}\Phi_{i_1\dots i_s}'-c_s^2\partial^2\Phi_{i_1\dots i_s}-\delta c_s^2 \partial_{i_1}\partial_{j}\Phi_{j\dots i_s}+m^2 a^2\Phi_{i_1\dots i_s}=0.\label{spineq}
\end{equation}
The field can be separated into two parts:
\begin{equation}
    \Phi_{i_1\dots i_s}=\Phi_{i_1\dots i_s}^T+\Phi_{i_1\dots i_s}^{R},
\end{equation}
where $\Phi^T_{i_1\dots i_s}$ is the transverse part of the field, obeying
\begin{equation}
    \partial_j\Phi_{j\dots i_s}^T=0,
\end{equation}
and $\Phi_{i_1\dots i_s}^{R}$ is the remainder. It is straightforward to see that $\Phi^T$ has $2$ degrees of freedom and represents the components with helicity $ \pm s$, while $\Phi^R$ has $2s-1$ components with lower helicities.

For $\Phi^T$, the penultimate term in (\ref{spineq}) vanishes, and the equation of motion becomes:
\begin{equation}
        \Phi_{i_1\dots i_s}^{T''}+2\frac{a'}{a}\Phi_{i_1\dots i_s}^{T'}-c_s^2\partial^2\Phi^T_{i_1\dots i_s}+m^2 a^2\Phi^T_{i_1\dots i_s}=0.
\end{equation}
This equation is in the same form as \eqref{eq:phiEOM}, therefore we can directly apply the analysis in Section \ref{sec:flrw} to show that the propagators of $\Phi^T$ are Hermitian analytic. For $\Phi^R$ we can take the divergence of (\ref{spineq}), which gives us:
\begin{equation}
    (\partial_{j}\Phi_{j\dots i_s}^R)''+2\frac{a'}{a}(\partial_{j}\Phi_{j\dots i_s}^R)'-(c_s^2+\delta c_s^2)\partial^2(\partial_{j}\Phi_{j\dots i_s}^R)+m^2 a^2(\partial_{j}\Phi_{j\dots i_s}^R)=0.
\end{equation}
Once again the equation is in the same form as \eqref{eq:phiEOM}, but with $c_s^2$ replaced with $c_s^2+\delta c_s^2$. We can again directly apply the analysis in Section \ref{sec:flrw}. Working in Fourier space, this tells us that $ik_j\Phi_{j\dots i_s}^R$ is Hermitian analytic. Since $i\bfk$ is Hermitian analytic, and $ik_j\Phi_{j\dots i_s}^R$ has exactly $2s-1$ degrees of freedom, we deduce that the propagators of $\Phi^R$ are also Hermitian analytic. We conclude that the propagator of the full field $\Phi$ must be Hermitian analytic, which establishes the crucial property of our derivation of single-cut rules for free fields of any integer spin (in the spontaneously boost-breaking theories of \cite{Bordin2018}).


\subsubsection{Helicity basis and the diagonalization of propagators}

For practical calculations, we would like to work in a basis where the propagators have a simple form. This can be achieved by looking at the helicity basis of the fields. These are irreps of ISO$(3)$, the isometry group of a flat FLRW spacetime. As we show below, fields of different helicities decouple from each other and the corresponding propagators in this basis become diagonal.

The only non-diagonal term in the action is
\begin{align}
    k_j\Phi_{ji_2\dots i_s}(\bfk)k_l\Phi_{li_2\dots i_s}(-\bfk)&=\Phi_{i_1\dots i_s}k_{i_1}k_{j_1}\delta_{i_1j_1}\dots\delta_{i_s j_s} \Phi_{j_1\dots j_s}\\&\equiv\Phi_{i_1\dots i_s}(\bfk)M_{i_1\dots i_sj_1\dots j_s}(\bfk) \Phi_{j_1\dots j_s}(-\bfk).
\end{align}
We are guaranteed to be able to diagonalise this term because it is real and symmetric in $i$'s and $j$'s. To understand this diagonalisation procedure, let's start by looking at the vector case, $s=1$, for which the tensor equation can be understood as a matrix multiplication, and so is diagonalised by finding the eigenvalues, $\lambda^h$, and eigenvectors, $\epsilon^h$, of $M$,
\begin{align}
    M_{ij}(\bfk)\epsilon^h_j(\bfk)&=\lambda^h(\bfk) \epsilon^h_i(\bfk), &\text{(no sum on $h$)}.
\end{align}
The eigenvalues of $M$ are
\begin{align}
    \lambda^\pm&=0, &    \lambda^0&=k^2.
\end{align}
We define the eigenvectors so that they satisfy the inversion relationship $\epsilon^h(-\bfk)=\epsilon^h(\bfk)^*$\footnote{Naively it may appear that this is sufficient to ensure that the $\disc$ commutes with the vertex contributions. However, as is discussed in Section \ref{SpinVertex}, the explicit dependence of $\epsilon^\pm$ on the energies will ruin this relationship, because not all energies are analytically continued.},
\begin{align}
    \epsilon^\pm_i&=\left(\bfk\times\left(\bfk\times \hat{n}\right) \pm i k \bfk\times \hat{n}\right)_i \,, \\
    \epsilon^0_i&=ik_i
\end{align}
for some normal vector $\hat{n}$ that is perpendicular to $\bfk$. These eigenvectors are orthogonal to each other,
\begin{align}
\left[\epsilon_i^h(\bfk)\right]^*\epsilon_i^{h'}(\bfk)&=C^h(k^2)\delta^{hh'},&\text{(no sum on h)}, 
\end{align}
where $C^h(k^2)$ is a polynomial in $k^2$ (so is guaranteed to be Hermitian analytic) that comes from the normalisation of the eigenvectors. We can therefore express $M$ and the identity in terms of these eigenvectors as
\begin{align}
    M_{ij}=k_ik_j&=\epsilon^h_i(\bfk)\frac{1}{C^h(k^2)}\lambda^{h}(\bfk)\left[\epsilon^{h}_j(\bfk)\right]^*\\
    \delta_{ij}&=\epsilon^h_i(\bfk) \frac{1}{C^h(k^2)}\left[\epsilon^{h}_j(\bfk)\right]^*
\end{align}
We can then see that in the so called "helicity basis",
\begin{equation}
    \Phi^h(\bfk)\equiv\Phi_i(\bfk)\epsilon^h_i(\bfk),
\end{equation}
the action diagonalises,
\begin{equation}\label{eq:helicityaction}
    S_2=\frac{1}{2}\int \frac{d^3k}{(2\pi)^3} d\eta a^2 \, \sum_{h=\pm,0} \left\{ (\Phi^h(\bfk))'   \partial_\tau -\Phi^h(\bfk)\left[c_s^2k^2 -m^2 +\delta c_s^2 \lambda^{h}\right]\right\}\frac{1}{C^h(k^2)} \Phi^h(-\bfk) .
\end{equation}
This also makes it clear why we imposed that $\epsilon^h(-\bfk)=\epsilon^h(\bfk)^*$ as it ensures that
\begin{equation}
    \Phi^h(-\bfk)=\Phi_i(-\bfk)\left[\epsilon^h_i(\bfk)\right]^*.
\end{equation}

Now that we have our eigenvectors for the spin-1 case this procedure can be generalised to arbitrary spin. To keep the symmetries of our field manifest we define a symmetric, traceless basis containing $2s+1$ tensors which are constructed from the symmetrised direct product of $s$ copies of the vector $\epsilon^h$,
\begin{align}
    \epsilon^s_{i_1\dots i_s}&=\epsilon^+_{i_1}\dots \epsilon^+_{i_s},\\
    \epsilon^{s-1}_{i_1\dots i_s}&=\epsilon^+_{(i_1}\dots \epsilon^+_{i_{s-1}}\epsilon^0_{i_s)},\\
    \epsilon^{s-2}_{i_1\dots i_s}&=\epsilon^+_{(i_1}\dots \epsilon^+_{i_{s-2}}\epsilon^0_{i_{s-1}}\epsilon^0_{i_s)}+\frac{1}{3}\epsilon^+_{(i_1}\dots \epsilon^+_{i_{s-2}}\delta_{i_{s-1}i_s)},\\\nonumber
    &\ \, \vdots\\
    \epsilon^0_{i_1\dots i_s}&=\epsilon^0_{i_1}\dots\epsilon^0_{i_s}+\frac{s!}{6}\epsilon^0_{(i_1}\dots\epsilon^0_{i_{s-2}}\delta_{i_{s-1}i_s)}\\\nonumber
    &\ \, \vdots\\
    \epsilon^{-s}_{i_1\dots i_s}&=\epsilon^-_{i_1}\dots \epsilon^-_{i_s}.
\end{align}
The tracelessness of these terms relies on the relationship $\epsilon^+={\epsilon^-}^*$ which ensures that any contractions like $\epsilon^\pm_i\epsilon^\pm_i$ vanish by orthogonality. It can be shown that these tensors inherit the orthogonality of the vectors so
\begin{align}
    M_{i_1\dots i_s j_1\dots j_s}&=\epsilon^h_{i_1\dots i_s}(\bfk)\frac{1}{C^h(k^2)}\lambda^{h}(\bfk)\left[\epsilon^{h}_{j_1\dots j_s}(\bfk)\right]^*,\\
    \delta_{i_1 j_1}\dots \delta_{i_sj_s}&=\epsilon^h_{i_1\dots i_s}(\bfk)\frac{1}{C^h(k^2)}\left[\epsilon^{h}_{j_1\dots j_s}(\bfk)\right]^*.
\end{align}
where $C^h$ is given, as for the vector case, by
\begin{align}\label{eq:Ch}
    \left[\epsilon^h_{i_1\dots i_s}(\bfk)\right]^*\epsilon^{h'}_{i_1\dots i_s}(\bfk)&=C^h\delta^{h h'}, &\text{(no sum on h)}.
\end{align}
The action is therefore exactly that given in \eqref{eq:helicityaction} but with $h$ running from $-s$ to $s$. As this is diagonal, it gives a separate differential equation for each helicity mode\footnote{The contribution from $C^h(k^2)$ factorises out.},
\begin{equation}\label{eq:helicityEOM}
     (a^2 {\Phi^h}'(\bfk))'-a^2\Phi^h\left[c_s^2k^2 -m^2 +\delta c_s^2 \lambda^{h}\right]=0,
\end{equation}
which ensures that the propagators are diagonal in this basis,
\begin{align}
    K^{h h'}_k(\eta)&=\delta^{h h'}K^h_k(\eta)&\text{(no sum on $h$)},\\ 
    G^{h h'}_p(\eta)&=\delta^{h h'}C^h(k^2)G^h_p(\eta)&\text{(no sum on $h$)} .
\end{align}
Here, $K_k^h$ and $G_p^h$ are constructed from the positive energy modefunctions that satisfy \eqref{eq:helicityEOM} subject to the Bunch-Davies initial condition \eqref{eq:B-D} with the substitution
\begin{equation}
    c_s\rightarrow\sqrt{c_s^2+\frac{\lambda^h}{k^2}\delta c_s^2}.
\end{equation}
This is $k$ independent because $\lambda^h\propto k^2$ for all $h$. The proof that these propagators are Hermitian analytic therefore follows similarly to the scalar case. 


\subsubsection{Interaction vertices}\label{SpinVertex}
In order to derive the single-cut rules, we need the interaction vertices to be Hermitian analytic. This is indeed the case if we follow the prescription that the all polarization tensors commute with the $\disc$ operation. In the helicity basis, a generic interaction vertex has the following form:
\begin{equation}\label{prescription}
    S_n=g_n\sum_{\{h_a\}}\int d\eta a^{4}(\eta)\int \prod_{a=1}^{n}\left[\frac{d^3 k_a}{(2\pi)^3}\right]\left(\prod_{a=1}^{n}\Phi^{h_a}_{k_a}(\eta)\right)F^{\{h_a\}}(\{\textbf{k}_a\}).
\end{equation}
The interaction vertex is constructed by taking various contractions of $\epsilon^{h_a}_{i_1\dots i_s}(\bfk_a)$ and $i\bfk_a$ (the latter comes from a spatial derivative). Clearly $i\bfk$ is Hermitian analytic since $(i(-\bfk))^\ast=i\bfk$. Next, we look at the Hermitian analytic image $\left[\epsilon^{h}_{i_1\dots i_s}(-\bfk)\right]^\ast$. Naively one would like to use the property of polarization tensors $ \left[\epsilon^{h}_{i_1\dots i_s}(-\bfk)\right]^\ast = \epsilon^{h}_{i_1\dots i_s}(\bfk)$, but this is subtle because the $\disc$ actually analytically continues some energies (all uncut lines) and not others. To avoid confusion, we provide a clear cut prescription for which our single cut rules are valid: all polarization tensors $\epsilon^{h}_{i_1\dots i_s}$ should be factorized outside all the $\disc$'s. In other words, we should substitute $\left[\epsilon^{h}_{i_1\dots i_s}(-\bfk)\right]^\ast$ in the second term inside each $\disc$ with $\epsilon^{h}_{i_1\dots i_s}(\bfk)$\footnote{To see how this is compatible with our explicit expression for $\epsilon^\pm$ note that we are free to specify how to send $\bfk\rightarrow -\bfk$ under the $\disc$. This is distinct from in the inversion relationship where we must rotate $\bfk$ and its normal vectors with it. We therefore choose to invert $\bfk$ by reflecting it in a plane perpendicular to itself which leaves any vectors perpendicular to it unchanged. This leads to the desired result for all uncut lines whilst for the cut line the helicities are reversed. This is not an issue as all helicities are summed over for internal lines and there is a symmetry between the plus and minus helicities that ensures their propagators are the same.}, which is precisely what appears in the first term. With this prescription, we conclude that the vertex function is Hermitian analytic. 

Due to the form of the propagator, polarization tensors associated with bulk-to-bulk propagators must also come with a sum over helicities. Aside from this, the proof of the single-cut rules proceeds in the same way as the scalar case. Therefore, we have the following single-cut rules:
\begin{align}\nonumber
        \text{Disc}_{S}\,i\psi_{n+m}^{\{h_a\}\{h_b\}}(\{k\};p_1,\dots,S,\dots p_I;\{\textbf{k}\})=
    \sum_{h}&-iP_{\Phi}^h(S) \text{Disc}_{S}\, i\psi_{n+1}^{\{h_a\},h}(\{k_a\},S;\{p\};\{\textbf{k}_a\})\\ \times&\text{Disc}_{S}\, i\psi_{m+1}^{\{h_b\},h}(\{k_b\},S;\{p\};\{\textbf{k}_b\})\,.
\end{align}
Here $P_{\Phi}^h$ is the power spectrum of the exchanged field,
\begin{equation}
    P_{\Phi}^h(S)=C^h\langle \Phi^{h}(\textbf{S})\Phi^{h}(-\textbf{S})\rangle',
\end{equation}
where $C^h$ is defined in \eqref{eq:Ch}.


\subsubsection{Explicit examples: general relativity and massive gravity}

As a demonstration of the various general properties discussed above, let us look at explicit examples involving massive gravity and general relativity. 

The simplest case is that of general relativity. In this theory the (massless) graviton has the same (positive energy) mode functions as a scalar field, \eqref{masslessdS},
\begin{align}
\gamma_{ij} (k)&= \sum_h \epsilon_{ij}^h(\bfk) \frac{H}{\Mpl k^{3/2}}(1-i k \eta)e^{i k\eta}\,,
\end{align}
where now $h=\pm2$ since the lower-helicty modes are removed by diff invariance. As we have seen for the scalar field, the propagators corresponding to this mode function must be Hermitian analytic. 

As an example of an interaction, consider the cubic graviton interaction induced by the spatial Ricci scalar $R^{(3)}$ \cite{Maldacena:2011nz}:
\begin{multline}
    F^{\lambda_1\lambda_2\lambda_3}(k_1,k_2,k_3)= i^2 \left(k_{2\, i} k_{2\, j} \epsilon^{\lambda_1}_{ij}(\bfk_1)\right)\epsilon^{\lambda_2}_{lm}(\bfk_2)\epsilon^{\lambda_3}_{lm}(\bfk_3) \\ -2i^2\left(\epsilon^{\lambda_1}_{ij}(\bfk_1)k_{3\, l}\right)\epsilon^{\lambda_2}_{li}(\bfk_2)\left(k_{2\, m}\epsilon^{\lambda_3}_{jm}(\bfk_3)\right)+(\text{cyclic})  \,.\label{R3vertex}
\end{multline}
The two spatial derivatives are Hermitian analytic thanks to factor of $i^2$ in front, and the polarization tensors can be taken to obey Hermitian analyticity because of the prescription outlined in Section \ref{SpinVertex}. 

As a more interesting example, we also look at the propagators in a theory of massive gravity (see \cite{Goon:2018fyu} for more details):
\begin{multline}
    S=\frac{1}{4}\int d^4x \left[-\nabla_{\rho}h_{\mu\nu}\nabla^{\rho}h^{\mu\nu}-(m^2+2H^2)h_{\mu\nu}h^{\mu\nu}+\nabla^{\rho}h_{\rho\mu}\nabla_{\nu}h^{\nu\mu}\right.\\\left.-\nabla_{\mu}h\nabla_{\nu}h^{\mu\nu}+\frac{1}{2}\nabla_{\mu}h\nabla^{\mu}h+\frac{1}{2}(m^2-H^2(d-2)^2 h^2)\right],
\end{multline}

As shown in the paper, the mode function for the helicity mode $+2$ and $+1$ are found to be:
\begin{align}
    \gamma^{+2}_k(\eta)&=(-k\eta)^{3/2}H^{(2)}_{i\mu}(-k\eta),\\
    \gamma^{+1}_k(\eta)&=\eta(-k\eta)^{1/2}\left(2k\eta H^{(2)}_{i\mu-1}(-k\eta)-(3+2i\mu)H^{(2)}_{i\mu}(-k\eta)\right).
\end{align}
We immediately notice that the $+2$ helicity has a Hermitian analytic propagator. For the $+1$ helicity mode, we use the recurrence relation of Hankel function to rearrange the mode function as:
\begin{equation}
    \gamma^{+1}_k(\eta)=\eta(-k\eta)^{1/2}\left(-3+2\eta\frac{d}{d\eta}\right)H^{(2)}_{i\mu}(-k\eta),
\end{equation}
from which we see that this mode function will give rise to a Hermitian analytic propagator.

\section{Explicit examples} \label{sec:examples}

In this section we show how the single-cut rules are satisfied is a few non-trivial examples. Due to the similarity of the results to those discussed in \cite{COT} we will not reproduce the checks performed there but instead discuss more complex cases. First, we consider the four point function for $4$ external conformally coupled scalars for non-derivative cubic interactions with both a conformally coupled and a massive exchanged field. Then we discuss a novel case involving $4$ gravitons with a graviton exchange. Finally, we explore the $5$-point function with derivative interactions to demonstrate how our results generalise for higher point derivative diagrams.

\subsection{Conformally coupled exchange}
\begin{figure}
    \centering
    \includegraphics{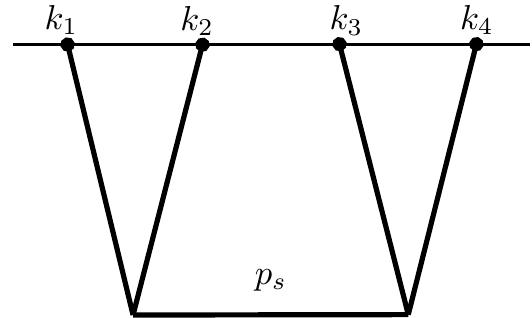}
    \caption{The diagram shows the $s$-channel for the interaction of 4 conformally coupled scalars exchanging a conformally coupled scalar. A similar diagram describes the massive exchange case.}
    \label{fig:scalar4}
\end{figure}

We consider the case of a conformally coupled field with a cubic polynomial interaction $\lambda \phi^3$, as discussed in \cite{Arkani-Hamed:2015bza} (see also \cite{Hillman:2019wgh}). There the in-in correlator was computed, but here we are interested in the wavefunction coefficient so we redo the calculation. It is necessary  to keep the dependence on the time at which the future boundary is taken because the propagators diverge as this is taken to zero. For this interaction the wavefunction coefficients are given by
\begin{align}
    \psi_2&=-ia^2(\eta_0)K'_k(\eta_0),\\
    \psi_3&=-6i\lambda\int_{-\infty}^{\eta_0} d\eta a^4(\eta)K_{k_1}(\eta)K_{k_2}(\eta)K_{k_3}(\eta),\\
    \psi_4&=-36i\lambda^2\int_{-\infty}^{\eta_0} d\eta d\eta' a^4(\eta)a^4(\eta')K_{k_1}(\eta)K_{k_2}(\eta)K_{k_3}(\eta')K_{k_4}(\eta')G_{p_s}(\eta,\eta')+t+u,
\end{align}
where the propagators are defined to be
\begin{align}
    K_k(\eta)&=\frac{\eta}{\eta_0}e^{ik(\eta-\eta_0)},\\
    G_p(\eta,\eta')&=i\frac{\eta\eta'}{2p}\left(e^{-ip\lvert \eta'-\eta\rvert}-e^{ip(\eta+\eta')}e^{-2ip\eta_0}\right).
\end{align}
For simplicity we have set $H=1$, since it can be recovered in the final result using dimensional analysis. We are specifically looking to relate the $s$ channel of $\psi_4$, shown in Fig. \ref{fig:scalar4}, to $\psi_3$ and so we focus on 
\begin{align}
    \psi_3&=-6i\lambda \int d\eta \frac{1}{\eta_0^3\eta}e^{i(k_1+k_2+k_3)(\eta-\eta_0)}=\frac{6i\lambda}{\eta_0^3}e^{-i(k_1+k_2+k_3)\eta_0}\Gamma(0,-i(k_1+k_2+k_3)\eta_0),\\
    \psi_4^s&=36\lambda^2\int d\eta d\eta'\frac{1}{2p_s\eta_0^4\eta\eta'}e^{ik_{12}(\eta-\eta_0)}e^{ik_{34}(\eta'-\eta_0)}\left(e^{-i{p_s}\lvert \eta'-\eta\rvert}-e^{i{p_s}(\eta+\eta')}e^{-2i{p_s}\eta_0}\right),
\end{align}
where $k_{ij}=k_i+k_j$. This can be approached as in \cite{Arkani-Hamed:2015bza} by rewriting the divergent $\frac{1}{\eta}$ terms as integrals over momentum. Furthermore, as the integrals are finite in the limit $\eta_0\rightarrow 0$ we find
\begin{align}
    \lim_{\eta_0\rightarrow 0}\psi_4^s&=-\lim_{\eta_0\rightarrow 0}\frac{18\lambda^2}{{p_s}\eta_0^4}\int_{k_{12}}^\infty dy\int_{k_{34}}^\infty dx\int_{-\infty}^0 d\eta d\eta'e^{iy\eta}e^{ix\eta'}\left(e^{-i{p_s}\lvert \eta'-\eta\rvert} -e^{i{p_s}(\eta+\eta')}\right)\\&=\lim_{\eta_0\rightarrow 0}\frac{18\lambda^2}{{p_s}\eta_0^4}\int_{k_{12}}^\infty dy\int_{k_{34}}^\infty dx\frac{2{p_s}}{({p_s}+x)({p_s}+y)(x+y)}.
\end{align}
This integral can be performed and the resulting logarithms can be combined under the assumption that all energies have negative imaginary parts to give
\begin{equation}
\lim_{\eta_0\rightarrow 0}\psi_4^s=\frac{18\lambda^2}{\eta_0^4 {p_s}}\left[ \frac{\pi^2}{6}-\log\left(\frac{k_{12}+{p_s}}{k_T}\right)\log\left(\frac{k_{34}+{p_s}}{k_T}\right)-\mathrm{Li}_2\left(\frac{k_{12}-{p_s}}{k_T}\right)-\mathrm{Li}_2\left(\frac{k_{34}-{p_s}}{k_T}\right) \right].
\end{equation}
The discontinuity in $\psi_4^s$ is then given by
\begin{equation}
    \lim_{\eta_0\rightarrow 0}\text{Disc}_{{p_s}}\ i\psi_4^s=-i\lim_{\eta_0\rightarrow 0}\frac{18\lambda^2}{\eta_0^4 {p_s}}\log\left(\frac{{p_s}+k_{12}}{k_{12}-s}\right)\log\left(\frac{{p_s}+k_{34}}{k_{34}-s}\right),
\end{equation}
which has been simplified using the fact that the dilogarithm satisfies
\begin{equation}
\mathrm{Li}_2(z)+\mathrm{Li}_2(1-z)=\frac{\pi^2}{6}-\log(z)\log(1-z).
\end{equation}
We wish to compare this to the same limit of the product of the discontinuities in the two three point functions that make up the $s$ channel diagram,
\begin{multline}
\lim_{\eta_0\rightarrow0} P_{p_s}\text{Disc}_{p_s} \left[ i\psi_3'(k_1,k_2,{p_s})\right]\text{Disc}_{p_s}\left[ i\psi_3'(k_3,k_4,{p_s})\right]=\\ \lim_{\eta_0\rightarrow0}\frac{18\lambda^2}{{p_s}\eta_0^4}\left(\Gamma(0,-i(k_{12}+{p_s})\eta_0)-\Gamma(0,i({p_s}-k_{12})\eta_0)\right)\left(\Gamma(0,-i(k_{34}+{p_s})\eta_0)-\Gamma(0,i({p_s}-k_{34})\eta_0)\right)\\=\lim_{\eta_0\rightarrow 0}\frac{18\lambda^2}{{p_s}\eta_0^4}\log\left(\frac{{p_s}+k_{12}}{k_{12}-{p_s}}\right)\log\left(\frac{{p_s}+k_{34}}{k_{34}-{p_s}}\right).
\end{multline}
From this it is clear that, in this limit,
\begin{equation}
    \text{Disc}_{p_s}\ i\psi_4^s=-i P_{p_s}\text{Disc}_{p_s} \left[ i\psi_3'(k_1,k_2,{p_s})\right]\text{Disc}_{p_s}\left[ i\psi_3'(k_3,k_4,{p_s})\right].
\end{equation}
Therefore, our results hold for this interaction involving conformally coupled fields.

\subsection{Massive exchange}\label{Massive}
It has been observed that the four point function of a conformally coupled scalar field serves as a seed solution from which various correlators can be obtained \cite{Arkani-Hamed:2015bza, Arkani-Hamed:2018kmz, Baumann:2019ghk, Baumann:2019oyu, Baumann:2020dch}. Therefore, it will be useful to demonstrate that the single-cut rule is satisfied by this primary building block of the bootstrap program\footnote{We thank Sébastien Renaux-Petel for stimulating discussions about this section.}. In this section, we check the single-cut rule for the 4pt exchange diagram with an intermediate heavy scalar field. In Appendix \ref{MassiveAppendix} we present the case for an arbitrary mass particle as well as some additional details. The three point function arising from the $\vpi^2\,\sigma$ interaction is given by \cite{Arkani-Hamed:2015bza, Arkani-Hamed:2018kmz, Jaz} 
\begin{align}
    \psi^{\varphi\varphi\sigma}(u,{p_s})=-\frac{\lambda}{\eta_0^2\sigma^+_{p_s}(\eta_0)}\sqrt{\frac{2}{{p_s}}}\frac{\pi}{\cos(\pi\nu)}P_{\nu-\frac{1}{2}}\left(\frac{1}{u}\right)\,,
\end{align}
where we have set $H=1$, defined $u=\frac{{p_s}}{k_{12}}$ and introduced the associated Legendre Function of the first kind, see Section 14 of \cite{NIST:DLMF}, which is defined in terms of hypergeometric functions as
\begin{equation}
P_{\nu-\frac{1}{2}}\left(\frac{1}{u}\right)={}_2 F_{1}\left[\frac{1}{2}-\nu,\frac{1}{2}+\nu,1,\frac{u-1}{2u}\right].
\end{equation}
It is also helpful to introduce the power spectrum for the massive field which we will keep implicit as
\begin{equation}
    P_{p_s}^\sigma=\sigma_{p_s}^+(\eta_0)\sigma_{p_s}^-(\eta_0).
\end{equation}
and we note that $p_s$ will be considered real throughout this section in accordance with the procedure for cutting internal lines. The four point correlator for this interaction was computed in \cite{Arkani-Hamed:2018kmz}, however, the corresponding wavefunction coefficient has not been calculated in the literature as far as we are aware. Fortunately, the techniques used in \cite{Arkani-Hamed:2018kmz} can be naturally extended to the wavefunction of the universe coefficients which are defined by
\begin{equation}
    \psi_4=-\frac{4i\lambda^2}{\eta_0^4}\int d\eta d\eta'\frac{1}{\eta\eta'}e^{ik_{12}\eta}e^{ik_{34}\eta'}G_{p_s}(\eta,\eta')+t+u.
\end{equation}
The Green's function here obeys exactly the same differential equation as the Green's functions for the correlators, up to some conventional factors of $i$, we therefore define $F=-\frac{p_s\eta_0^4}{4\lambda^2 }\psi_4$ so that it satisfies the differential equation studied in \cite{Arkani-Hamed:2018kmz},
\begin{equation}
    \left[u^2(1-u^2)\partial_u^2-2u^3\partial_u+\frac{1}{4}-\nu^2\right]F=\frac{u v}{u+v},
\end{equation}
where $v=\frac{p_s}{k_{34}}$. The ansatz in \cite{Arkani-Hamed:2018kmz} uses, as a basis, the functions 
\begin{equation}\label{eq:oldbasis}
    F_\pm(u)=\left(\dfrac{ u}{2\nu}\right)^{\frac{1}{2}\mp \nu} {}_2 F_1 \left[\frac{1}{4}\mp \frac{\nu}{2},\frac{3}{4}\mp \frac{\nu}{2},1\mp\nu,u^2 \right]\,.
\end{equation}
However, the inclusion of $\nu$ in the prefactor of these functions requires coefficients which, when we analytically continue in $u$, lead to complications in the analysis. Furthermore, the Hermitian analytic image of these functions depends on the mass of the particles through the behaviour of the complex conjugate of $\nu$. Therefore, we wish to use an alternative basis. The basis we chose is
\begin{align}
    \tilde{F}_+(u)&=P_{\nu-\frac{1}{2}}\left(\frac{1}{u}\right),\\
    \tilde{F}_-(u)&=\frac{1}{2}\left(Q_{\nu-\frac{1}{2}}\left(\frac{1}{u}\right)+Q_{-\nu-\frac{1}{2}}\left(\frac{1}{u}\right)\right).
\end{align}
Where $P$ and $Q$ are the associated Legendre functions of the first and second type\footnote{These are the associated Legendre functions defined in (14.3.6) and (14.3.7) of \cite{NIST:DLMF} with $\mu=0$ which amounts to choosing the branch cuts as defined below. These are implemented in Mathematica as LegendreP$\left[\nu,0,3,z\right]$ and LegendreQ$\left[\nu,0,3,z\right]$ for $P_\nu(z)$ and $Q_\nu(z)$ respectively.}. We use the Legendre functions rather than generic hypergeometric functions because it simplifies the notation and unifies the two hypergeometric functions that appear in the three and four point correlators. We could have chosen any two linearly independent combinations of these solutions but these particular ones are chosen because
\begin{itemize}
    \item The branch cut in $P_{\nu}(z)$ is along $z\in(-\infty,-1]$, the branch cut in $Q_{\nu}(z)$ is along $z\in(-\infty,1]$ so neither of these solutions have a branch cut with $u$ in its physical range $u\in(0,1)$ and one has no branch cut for $u>0$.
    \item The Wronskian for these two functions is the same as for $F_\pm(u)$ and so the matching condition that arises from the particular integral will take the same form in terms of $\tilde{F}_\pm(u)$ as in \cite{Arkani-Hamed:2018kmz} therefore, the most general solution that is symmetric under exchange of $u,v$ is
    \begin{equation}
        F(u,v)=\begin{cases}\displaystyle \sum_{m,n=0}^\infty c_{mn} u^{2m+1}\left(\frac{u}{v}\right)^n+\frac{\pi}{2\cos(\pi\nu)}\tilde{g}(u,v), \quad \lvert u\rvert \leq \lvert v\rvert \\\displaystyle\sum_{m,n=0}^\infty c_{mn} v^{2m+1}\left(\frac{v}{u}\right)^n+\frac{\pi}{2\cos(\pi\nu)}\tilde{g}(v,u), \quad \lvert v\rvert \leq \lvert u\rvert \end{cases}
    \end{equation}
    where 
    \begin{equation}
    \tilde{g}(u,v)=\beta_+\tilde{F}_+(v)\tilde{F}_+(u)+(\beta_0+1)\tilde{F}_-(v)\tilde{F}_+(u)+\left(\beta_0-1\right)\tilde{F}_+(v)\tilde{F}_-(u)+\beta_-\tilde{F}_-(v)\tilde{F}_-(u),
    \end{equation}
    and $c_{mn}$ are real coefficients that are already completely fixed by the inhomogeneous part of the differential equation.
    \item The Hermitian analytic image of these functions does not depend on whether or not $\nu$ is real,
    \item $F_+$ is finite in the limit $u\rightarrow 1$,
    \begin{align}
        \lim_{u\rightarrow 1}\tilde{F}_+(u)&=1,\\
        \lim_{u\rightarrow1}\tilde{F}_-(u)&=-\frac{1}{2}\log\left(u-1\right).
    \end{align}
    and so in this limit $F(u,v)$ is given by,
    \begin{equation}
        \lim_{u\rightarrow 1}F(u,v)=\lim_{u\rightarrow 1}g(v,u)=  -\frac{1}{2}\left[(1+\beta_0)\tilde{F}_+(v)+\beta_-\tilde{F}_-(v)\right]\log(u-1).
    \end{equation}
    We want this to be finite and so we require that $\beta_-=0$, $\beta_0=-1$. This is slightly simpler than the form in \cite{Arkani-Hamed:2018kmz} where all four possible terms must be kept. 
    \item Both functions are real for $u\in (0,1)$ for both real and imaginary $\nu$.
\end{itemize}
These are all nice properties for our solutions to have but it is possible that some other choice of basis may work. In \cite{Arkani-Hamed:2018kmz} they fix $\beta_0$ using the $u\rightarrow -1$ limit. Here we will instead find a $\beta_0$ that satisfies the single-cut rule and then show that, for real values of $u$ our solution agrees with that in \cite{Arkani-Hamed:2018kmz}. We will start by calculating the discontinuities of the cut diagrams, introducing the notation that
\begin{align}
P_\pm(u)&=P_{\nu-\frac{1}{2}}\left(\pm\frac{1}{u}\right),\\Q_\pm(u)&=Q_{\pm\nu-\frac{1}{2}}\left(\frac{1}{u}\right),
\end{align}
so that
\begin{multline}\label{eq:massCOTRHS}
    -iP_{p_s}^\sigma\text{Disc}_{p_s}\left[ i\psi_3(u,{p_s};\eta_0)\right]\text{Disc}_{p_s}\left[ i\psi_3(v,{p_s};\eta_0)\right]=\\\frac{2i\lambda^2\pi^2}{{p_s}\eta_0^4\cos^2(\pi\nu)} \left(P_-(u)P_+(v)+P_+(u)P_-(v)+\frac{\sigma_{p_s}^+(\eta_0)}{\sigma_{p_s}^-(\eta_0)}P_-(u)P_-(v)+\frac{\sigma_{p_s}^-(\eta_0)}{\sigma_{p_s}^+(\eta_0)}P_+(u)P_+(v)\right).
\end{multline}
To equate this to the left hand side it is easiest to express each term on the left hand side in terms of $P_\pm$. We start by considering the case that $\lvert u\rvert\leq\lvert v\rvert$ and note that $c_{mn}$ are real so the sum is guaranteed to vanish on the left hand side
\footnote{The reality of $c_{mn}$'s is rooted in the fact that the $\sum_{m\,n}\dots$ part of the correlator is equivalent to a sume over an infinite tower of quartic contact diagrams that emerges after integrating out the massive particle $\sigma$. Unitarity (in the form of the COT) and scale invariance (for conformally coupled fields) then demands the reality of each contact diagram, hence the reality of the coefficients $c_{mn}$.}
therefore
\begin{multline}
\text{Disc}_{p_s}\ i\psi_4=  -\frac{4i\lambda^2}{{p_s}\eta_0^4}\left(F(u,v)+F^*(-u^*,-v^*)\right)=\\\frac{2i\pi\lambda^2}{{p_s}\eta_0^4\cos(\pi\nu)}\left(\frac{\pi P_-(u)P_+(v)}{\cos(\pi\nu)}+\frac{\pi P_+(u)P_-(v)}{\cos(\pi\nu)}-(\beta_+^*+\pi i)P_-(v)P_-(u)-(\beta_+-\pi i)P_+(v)P_+(u)\right),\\\ 
\end{multline}
note that the reality of $c_{mn}$ ensures that the sum cancels and this expression is symmetric in $u,v$, ensuring that we will find an identical result for $\lvert u\rvert > \lvert v\rvert$.
By comparison with \eqref{eq:massCOTRHS} it is straightforward to see that the single-cut rule holds if 
\begin{equation}
    \beta_+=i\pi-\frac{\pi}{\cos(\pi \nu)} \frac{\sigma_{p_s}^-(\eta_0)}{\sigma_{p_s}^+(\eta_0)}.
\end{equation}

\subsubsection{Comparison with the Correlator}

We have used the single-cut rule as a condition to fix the wavefunction coefficient. In order to use this to check that this rule holds for the result given in \cite{Arkani-Hamed:2018kmz} we need to compare them. We start by relating the functions $Q_\pm(u)$ to the $F_\pm$ basis \eqref{eq:oldbasis},
\begin{equation}\label{eq:QasF}
Q_\pm(u)=\sqrt{\pi}\frac{u}{2}^{\frac{1}{2}\pm\nu}\frac{\Gamma(\frac{1}{2}\pm\nu)}{\Gamma(1\pm\nu)}\prescript{}{2}{F}_1\left(\frac{3}{4}\pm\frac{\nu}{2},\frac{1}{4}\pm\frac{\nu}{2},1\pm\nu,u^2\right)=-\frac{1}{2\alpha_\mp}F_\mp(u),
\end{equation}
where $\alpha_\pm$ are defined as in \cite{Arkani-Hamed:2018kmz} by
\begin{equation}
    \alpha_\pm\equiv -\left(\frac{1}{2\nu}\right)^{\frac{1}{2}\mp \nu}\frac{\Gamma(1\mp \nu)}{\Gamma\left(\frac{1}{4}\mp \frac{\nu}{2}\right)\Gamma\left(\frac{3}{4}\mp\frac{\nu}{2}\right)}.
\end{equation}
The ambiguity in the definition of these functions discussed under \eqref{eq:oldbasis} can be ignored here because we are only interested in evaluating the correlator on the positive real axis and so the power of $\nu$ in $\alpha_\pm$ will appropriately cancel with that in $F_\pm$. The correlator is related to the wavefunction coefficient by
\begin{equation}
B_4=-2\prod_{a=1}^4\frac{1}{2\Re\psi_2'(k_a)}\left[\Re\psi_4'(u,v)-\frac{\Re\psi'_3(u)\Re\psi'_3(v)}{\Re\psi_2'(s)}-t-u\right].
\end{equation}
As mentioned previously $\tilde{F}_\pm$ are real for $u>0$ and so taking the real part of $\psi_4'$ reduces to taking the real part of the coefficients. We can then express all $\tilde{F}_\pm$ in terms of $Q_\pm$ and use \eqref{eq:QasF} to relate these to $F_\pm$, details in Appendix \ref{MassiveAppendix}, so the $s$ channel of the correlator is
\begin{equation}
    B_4^s=\frac{\eta_0^4\lambda^2}{2p_sk_1k_2k_3k_4}\begin{cases}
    \displaystyle \sum_{m,n=0}^\infty c_{mn} u^{2m+1}\left(\frac{u}{v}\right)^n+\frac{\pi}{2\cos(\pi\nu)}\hat{g}(u,v), \quad \lvert u\rvert \leq \lvert v\rvert \\\displaystyle\sum_{m,n=0}^\infty c_{mn} v^{2m+1}\left(\frac{v}{u}\right)^n+\frac{\pi}{2\cos(\pi\nu)}\hat{g}(v,u), \quad \lvert v\rvert \leq \lvert u\rvert
    \end{cases}
\end{equation}
where we have introduced the function
\begin{equation}
    \hat{g}(u,v)=(\beta-1)F_-(u)\left(F_+(v)-\frac{\alpha_+}{\alpha_-}F_-(v)\right)+(\beta+1)F_+(u)\left(F_-(v)-\frac{\alpha_-}{\alpha_+}F_+(u)\right)
\end{equation}
and defined, for the sake of comparison, $\beta=-\frac{1}{\sin(\pi\nu)}$. This is exactly the result found in \cite{Arkani-Hamed:2018kmz} (except they use $\mu=i\nu$). Therefore, by using the single-cut rule as a condition on the wavefunction of the universe coefficients, we find a result that is consistent with the results in the literature for a four point interaction involving the exchange of a massive scalar.

\subsection{Four graviton exchange}

We consider the case of graviton exchange with a vertex of the form\cite{Bordin:2020eui}:
\begin{align}
    S_{I}&=\frac{1}{3!}\int d^3 xdt\, a^3\dot{\gamma}_{ij}\dot{\gamma}_{jk}\dot{\gamma}_{ki}\\
    &=-\frac{1}{3!}\sum_{\lambda_1,\lambda_2,\lambda_3}\int d\eta d^3 k\, \epsilon_{ij}^{\lambda_1}(\textbf{k}_1)\epsilon_{jl}^{\lambda_2}(\textbf{k}_2)\epsilon_{li}^{\lambda_3}(\textbf{k}_3)\frac{1}{H\eta}\partial_{\eta}\gamma^{\lambda_1}_{k_1}\partial_{\eta}\gamma^{\lambda_2}_{k_2}\partial_{\eta}\gamma^{\lambda_3}_{k_3}.
\end{align}
where we have set the coupling constant to unity to simplify our notation. Setting $\Mpl=1$ for this subsection, the mode function and bulk-to-boundary propagator of the graviton are
\begin{align}
    \gamma^{\lambda}_k(\eta)&=\frac{H}{\sqrt{k^3}}(1-ik\eta)e^{ik\eta}\\
    K^{\lambda\lambda'}_k(\eta)&=\delta^{\lambda\lambda'}K_k(\eta)=\delta^{\lambda\lambda'}(1-ik\eta) e^{ik\eta}.
\end{align}
With this, we can easily obtain the wavefunction coefficient $\psi_3^{\lambda_1\lambda_2\lambda_3}$ as
\begin{equation}
    \psi_3^{\lambda_1\lambda_2\lambda_3}=2\epsilon_{ij}^{\lambda_1}(\bfk_1)\epsilon_{jl}^{\lambda_2}(\bfk_2)\epsilon_{li}^{\lambda_3}(\bfk_3)\frac{k_1^2k_2^2k_3^2}{H(k_1+k_2+k_3)^3}.
\end{equation}
\begin{figure}
    \centering
    \includegraphics{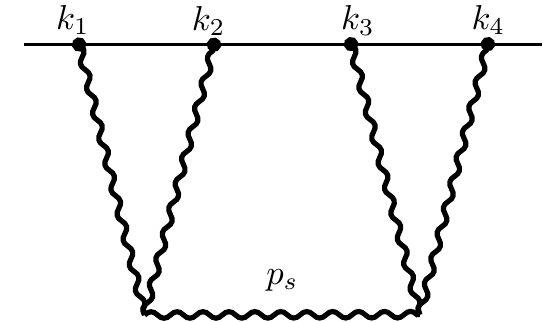}
    \caption{Diagram showing the s-channel for an interaction involving 4 gravitons exchanging a graviton.}
    \label{fig:grav4}
\end{figure}
To calculate the $s$-channel of wavefunction coefficient $\psi_4^{\lambda_1\lambda_2\lambda_3\lambda_4}$, shown in Fig. \ref{fig:grav4}, we need the bulk-to-bulk propagator
\begin{align}
    G^{\lambda\lambda'}_p(\eta,\eta')&=2\delta^{\lambda\lambda'}G_p(\eta,\eta')\\\nonumber
    &=2i\delta^{\lambda\lambda'}\left(\theta(\eta-\eta')\frac{H^2(1+ip\eta)(1-ip\eta')}{p^3}e^{-ip(\eta-\eta')}\right.\\
    &\left.+\theta(\eta'-\eta)\frac{H^2(1+ip\eta')(1-ip\eta)}{p^3}e^{ik(\eta-\eta')}-\frac{H^2(1-ip\eta)(1-ip\eta')}{p^3}e^{ip(\eta+\eta')}\right).
\end{align}
Then we find
\begin{align}\nonumber
    &\psi_4^{\lambda_1\lambda_2\lambda_3\lambda_4}=-i\sum_{\lambda}\int^{0}_{-\infty}d\eta\int^{0}_{-\infty}d\eta'\, \frac{\eta\eta'}{H^2}  2i\e^{\lambda_1}_{ij}(\textbf{k}_1)\e^{\lambda_2}_{jl}(\textbf{k}_2)\e^{*\lambda}_{li}(\textbf{p}_s)\e^{\lambda_3}_{mn}(\textbf{k}_3)\e^{\lambda_4}_{np}(\textbf{k}_4)\e^{\lambda}_{pm}(\textbf{p}_s)\\ \nonumber
    &e^{i(k_1+k_2)\eta}e^{i(k_3+k_4)\eta'}k_1^2k_2^2k_3^2k_4^2 \partial_{\eta}\partial_{\eta'}     \left[\left(\theta(\eta-\eta')\frac{H^2(1+ip_s\eta)(1-ip_s\eta')}{p_s^3}e^{-ip_s(\eta-\eta')}\right)\right.\\ 
    &+\left.\left(\theta(\eta'-\eta)\frac{H^2(1-ip_s\eta)(1+ip_s\eta')}{p_s^3}e^{ip_s(\eta-\eta')}\right)-\left(\frac{H^2(1-ip_s\eta)(1-ip_s\eta')}{p_s^3}e^{ip_s(\eta+\eta')}\right)\right].
\end{align}

Evaluating the integral gives us the following:
\begin{align}\nonumber
    &\psi_4^{\lambda_1\lambda_2\lambda_3\lambda_4}=\sum_{\lambda}\frac{2k_1^2k_2^2k_3^2k_4^2p_s}{k_T^5}\e^{\lambda_1}_{ij}(\textbf{k}_1)\e^{\lambda_2}_{jl}(\textbf{k}_2)\e^{*\lambda}_{li}(\textbf{p}_s)\e^{\lambda_3}_{mn}(\textbf{k}_3)\e^{\lambda_4}_{np}(\textbf{k}_4)\e^{\lambda}_{pm}(\textbf{p}_s)\\
    &\times\left[\left(\frac{24}{E_R}+\frac{12k_T}{E_R^2}+\frac{4k_T^2}{E_R^3}-\frac{24}{p_s}\right)\right.+\left(\frac{24}{E_L}+\frac{12k_T}{E_L^2}+\frac{4k_T^2}{E_L^3}-\frac{24}{p_s}\right)+\left.\left(\frac{4k_T^5}{E_R^3E_L^3}\right)\right].
\end{align}
Here $E_L=k_1+k_2+p_s$, $E_R=k_3+k_4+p_s$, and $k_T=k_1+k_2+k_3+k_4$. Adding the Hermitian analytic image reads:
\begin{align}
   \nonumber\text{Disc}_{p_s}i\psi_4^{\lambda_1\lambda_2\lambda_3\lambda_4}&=\sum_{\lambda}\frac{2ik_1^2k_2^2k_3^2k_4^2p_s}{k_T^5}\e^{\lambda_1}_{ij}(\textbf{k}_1)\e^{\lambda_2}_{jl}(\textbf{k}_2)\e^{*\lambda}_{li}(\textbf{p}_s)\e^{\lambda_3}_{mn}(\textbf{k}_3)\e^{\lambda_4}_{np}(\textbf{k}_4)\e^{\lambda}_{pm}(\textbf{p}_s)\\
   \nonumber &\times\left[\left(\frac{24}{E_R}+\frac{12k_T}{E_R^2}+\frac{4k_T^2}{E_R^3}+\frac{24}{E_L-2p_s}+\frac{12k_T}{(E_L-2p_s)^2}+\frac{4k_T^2}{(E_L-2p_s)^3}\right)\right.\\
   \nonumber &+\left(\frac{24}{E_L}+\frac{12k_T}{E_L^2}+\frac{4k_T^2}{E_L^3}+\frac{24}{E_R-2p_s}+\frac{12k_T}{(E_R-2p_s)^2}+\frac{4k_T^2}{(E_R-2p_s)^3}\right)\\
   &+\left(\frac{4k_T^5}{E_R^3E_L^3}\right)+\left.\left(\frac{4k_T^5}{(E_R-2p_s)^3(E_L-2p_s)^3}\right)\right].
\end{align}
The power spectrum is given by:
\begin{equation}
    P^{\gamma}_{p_s}=2\langle  \gamma^{\lambda}(\bfp_s) \gamma^{\lambda}(-\bfp_s)\rangle'=\frac{2H^2}{p_s^3}.
\end{equation}
Therefore, we have:
\begin{align}\nonumber
    &\sum_{\lambda}-iP^{\gamma}_{p_s}\,\text{Disc}_{p_s}i\psi_3^{\lambda_1\lambda_2\lambda}(k_1,k_2,p_s,\bfk_1,\bfk_2)\,\text{Disc}_{p_s}i\psi_3^{\lambda\lambda_3\lambda_4}(k_3,k_4,p_s,\bfk_3,\bfk_4)\\ \nonumber
    =&\sum_{\lambda}8ik_1^2k_2^2k_3^2k_4^2p_s\e^{\lambda_1}_{ij}(\textbf{k}_1)\e^{\lambda_2}_{jl}(\textbf{k}_2)\e^{*\lambda}_{li}(\textbf{p}_s)\e^{\lambda_3}_{mn}(\textbf{k}_3)\e^{\lambda_4}_{np}(\textbf{k}_4)\e^{\lambda}_{pm}(\textbf{p}_s)\left[\left(\frac{1}{(E_L-2p_s)^3}\frac{1}{E_R^3}\right)\right.\\ &+\left(\frac{1}{E_L^3}\frac{1}{(E_R-2p_s)^3}\right)
    +\left(\frac{1}{E_L^3}\frac{1}{E_R^3}\right)
    \left.+\left(\frac{1}{(E_L-2p_s)^3}\frac{1}{(E_R-2p_s)^3}\right)\right].
\end{align}
With this, it is straightforward to verify the single-cut rule for this interaction.

\subsection{A five-point function}
\begin{figure}
    \centering
    \includegraphics{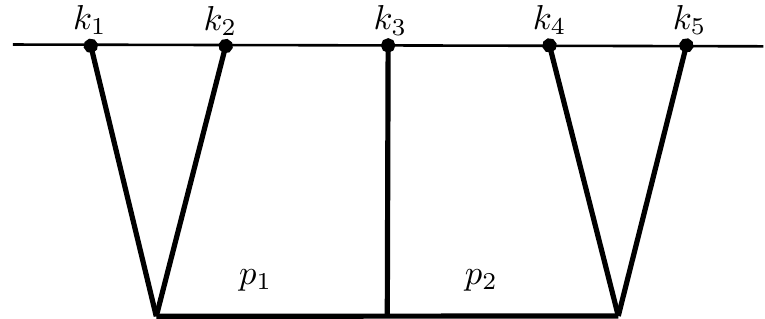}
    \caption{Diagram showing the geometry for the 5 point interaction which is being considered in \eqref{eq:flat5}. There are other diagrams that involve permutations of the labeling of the momenta that must be considered so that the wavefunction coefficient is symmetric in the momenta.}
    \label{fig:5point}
\end{figure}

The simplest case in which we can't remove all derivatives from all the Green's functions by integration by parts is the five point function with a single three particle interaction $\dot{\phi}^3$ so we will consider this interaction here. Unfortunately, the five point function is very complicated and so writing it down here is rather unhelpful. Instead, we include a Mathematica file that demonstrates that the single-cut rules do hold in this case. We do this also for a second five point interaction that involves a combination of $\dot{\phi}^3$ and $\dot{\phi}\partial_i\phi\partial^i\phi$ vertices. Despite the complications with the explicit form of the five point function it was noted by Pajer and Hillman~\cite{EP-Hillman-unpublished} that it is possible to express de Sitter wavefunction coefficients in terms of derivatives of flat space ones. In the case at hand with only $\dot\phi^3$ interactions, this gives
\begin{equation}
    \psi_5(k_1,k_2,k_3,k_4,k_5,p_{1},p_{2})=H^7k_1^2k_2^2k_3^2k_4^2k_5^2\frac{\partial^6}{\partial k_1\partial k_2\partial k_3\partial k_3 \partial k_4\partial k_5} \tilde{\psi}_5^{\textrm{f}}\, .
\end{equation}
Where $\tilde{\psi}_5^{\textrm{f}}$ is a flat space wavefunction coefficient
\begin{equation}\label{eq:flat5}
    \tilde{\psi}_5^{\textrm{f}}=-216i\int dt_1dt_2dt_3K^{\textrm{f}}_{k_1}(t_1)K^{\textrm{f}}_{k_2}(t_1)K^{\textrm{f}}_{k_3}(t_2)K^{\textrm{f}}_{k_4}(t_3)K^{\textrm{f}}_{k_5}(t_3)\tilde{G}^{\textrm{f}}_{p_1}(t_1,t_2)\tilde{G}_{p_2}^{\textrm{f}}(t_2,t_3)\, .
\end{equation}
The geometry for this diagram is shown in Fig. \ref{fig:5point}. Here $K^{\textrm{f}}_k(t)$ is the flat space bulk-to-boundary propagator,
\begin{equation}\label{eq:Kf}
    K^{\textrm{f}}_k(t)=\frac{\left.\phi^{\textrm{f}}_k\right.^+(t)}{\left.\phi^{\textrm{f}}_k\right.^+(t_0)}=e^{-ik(t-t_0)}.
\end{equation}
Whilst $\tilde{G}^{\textrm{f}}$ is related to the typical flat space bulk-to-bulk propagator by
\begin{align}
    \tilde{G}^{\textrm{f}}_p(t_1,t_2)&=p^2 G^{\textrm{f}}_p(t_1,t_2)-\delta(t_1-t_2) \nonumber\\&=\frac{ip}{2}\left(e^{-ip(t_2-t_1)}\theta(t_1-t_2)+e^{ip(t_2-t_1)}\theta(t_2-t_1)-e^{-ip(t_1+t_1)}\right)-\delta(t_1-t_2)\label{eq:Gf}\, .
\end{align}
This integral gives a result that is much more manageable,
\begin{multline}
     \tilde{\psi}_5^{\textrm{f}}=-\frac{216}{(k_{12}+k_{3}+k_{45})(k_{12}+p_{1})(k_{12}+k_{3}+p_{2})(k_{45}+p_{2})(k_3+p_{1}+p_{2})}\\\left[ k_{12}\left(k_{45}(k_3+p_{1})+p_{2}(k_3+k_{45}+p_{1})\right)\left(k_3^2+k_{45}p_{1}+p_{2}\left(k_{45}+p_{1}\right)+k_3\left(k_{45}+2p_{1}+p_{2}\right)\right)\right.\\+\left.p_{1}(k_3+k_{45})(k_3+p_{2})\left(k_{45}\left(k_3+p_{1}\right)+p_{2}\left(k_3+k_{45}+p_{1}\right)\right)\right.\\+\left. k_{12}^2(p_{1}+k_3+k_{45})(k_{45}+p_{2})(k_3+p_{2}+p_{1}) \right].
\end{multline}
Because the $k$'s always appears as pairs in both the prefactor and the derivative the discontinuity commutes with both so the left hand side of the optical theorem becomes
\begin{equation}
    \text{Disc}_{{p}_1}\  i\psi_5=H^7k_1^2k_2^2k_3^2k_4^2k_5^2\frac{\partial^6}{\partial k_{12}\partial k_{12}\partial k_3\partial k_3 \partial k_{45}\partial k_{45}}\text{Disc}_{{p}_1}\ i\tilde{\psi}_5^{\textrm{f}}.
\end{equation}
From this we can see that the structure of the left hand side of the optical theorem depends only on the discontinuity of $\tilde{\psi}_5^{\textrm{f}}$ which can be found straightforwardly (as it is a simple ratio of polynomials) to be
\begin{equation}
   \text{Disc}_{{p}_1}\ i \tilde{\psi}_5^{\textrm{f}}=-i\frac{72p_{1}^3\left(2k_3 k_{45}p_{2}+k_3^2\left(k_{45}+p_{2}\right)-p_{1}^2\left(k_{45}+p_{2}\right)\right)}{\left((k_3+k_{45})^2-p_{1}^2\right)(k_{3}+p_{2}-p_{1})(k_{45}+p_{2})(k_{3}+p_{1}+p_{2})}\frac{6}{(k_{12}^2-p^2_{1})}\, .
\end{equation}
Where the final result has been separated into a term that involves just $k_{12}$ and $p_{1}$ and one that does not include $k_{12}$ as expected from \eqref{eq:Cut}. To relate this to the cut diagram it is convenient to construct an equivalent derivative expression for $\psi_3$ and $\psi_4$. For a generic diagram with only $\dot{\phi}^3$ interactions where each vertex has at least one external momentum, we can construct such an expression by introducing a factor of $k_a^2$ for each external line and introducing two derivatives over the sum of the external momenta at each vertex,
\begin{align}
    \psi_3&=Hk_1^2k_2^2k_3^2\frac{\partial^2}{\partial k_T \partial k_T}\tilde{\psi}_3^{\textrm{f}}\, ,\\
    \psi_4&=H^4k_1^2k_2^2k_3^2k_4^2\frac{\partial^4}{\partial k_{12} \partial k_{12} \partial k_{34}\partial k_{34}}\tilde{\psi}_4^{\textrm{f}}\, .
\end{align}
The flat space wavefunction coefficients, $\tilde{\psi}_{3,4}^{\textrm{f}}$, are constructed from the propagators in \eqref{eq:Kf} and \eqref{eq:Gf},
\begin{align}
    \tilde{\psi}_{3}^{\textrm{f}}&=-6i\int dt_1 K^f_{k_1}(t_1)K^f_{k_2}(t_1)K^f_{k_3}(t_1)=-\frac{6}{k_1+k_2+k_3}\, ,\\
    \tilde{\psi}_{4}^{\textrm{f}}&=-36i\int dt_1dt_2 K^f_{k_1}(t_1)K^f_{k_2}(t_1)K^f_{k_3}(t_2)K^f_{k_4}(t_2)\tilde{G}^f_p(t_1,t_2)=\frac{36(k_{34}p+k_{12}(k_{34}+p))}{(k_{12}+k_{34})(k_{12}+p)(k_{34}+p)}\, .
\end{align}
To calculate the discontinuities of the cut diagrams, first notice that the derivatives over $k_T$ in $\psi_3$ can equivalently be written as derivatives over $k_{12}$ so that the $\psi_3$ discontinuity is
\begin{equation}
     \text{Disc}_{{p}_1}\ i\psi_3(k_1,k_2,p_{1})=Hk_1^2k_2^2p_{1}^2\frac{\partial^2}{\partial k_{12}\partial k_{12}}\text{Disc}_{{p}_1}\ i\tilde{\psi}_3^{\textrm{f}}=iHk_1^2k_2^2p_{1}^2\frac{\partial^2}{\partial k_{12}\partial k_{12}}\frac{12 p_{1}}{p_{1}^2-k_{12}^2}\,.
\end{equation}
Likewise, the $k_{12}$ derivatives in $\psi_4$ can be expressed as $k_2$ derivatives so that
\begin{multline}     
     \text{Disc}_{{p}_1}\ i\psi_4(p_{1},k_3,k_4,k_5,p_{2})=H^4p_{1}^2k_3^2k_4^2k_5^2\frac{\partial^4}{\partial k_{3} \partial k_{3} \partial k_{45}\partial k_{45}}\text{Disc}_{{p}_1}\ i\tilde{\psi}_4^{\textrm{f}}=\\-iH^4p_{1}^2k_3^2k_4^2k_5^2\frac{\partial^4}{\partial k_{3} \partial k_{3} \partial k_{45}\partial k_{45}}\frac{72p_{1}\left(2k_3 k_{45}p_{2}+k_3^2\left(k_{45}+p_{2}\right)-p_{1}^2\left(k_{45}+p_{2}\right)\right)}{\left((k_3+k_{45})^2-p_{1}^2\right)(k_{3}+p_{2}-p_{1})(k_{45}+p_{2})(k_{3}+p_{1}+p_{2})}\,.
\end{multline}
It is then possible to combine all of this together to give the single-cut rule for this 3 site chain,
\begin{equation}
    \text{Disc}_{{p}_1}\ i\psi_5=-iP_{p_1}\text{Disc}_{{p}_1}\left[ i\psi_3(k_1,k_2,p_{1})\right]\text{Disc}_{{p}_1}\left[ i\psi_4(p_{1},k_3,k_4,k_5,p_{2})\right].
\end{equation}
\section{Conclusion} \label{conclusions}

In this work, we have derived \textit{single-cut rules} for the coefficients of the wavefunction of the universe that vastly extend the validity of the Cosmological Optical Theorem \cite{COT}. Our derivation leverages some simple analytic properties of the bulk-to-bulk and bulk-to-boundary propagators. Just like cutting rules in flat space, our results should be regarded as a consequence of unitarity. In particular, our main achievements are summarized as follows:
\begin{itemize}
    \item We generalised the Cosmological Optical Theorem to an arbitrary number of spinning bosonic fields with a linear dispersion relation and arbitrary mass, arbitrary speed of sound and general local interactions at tree level (cutting rules for loops are discussed in \cite{sCOTt}). In particular, we explicitly checked that our relations are obeyed by the four-point scalar correlators from conformally coupled and general massive scalar exchange derived in \cite{Arkani-Hamed:2015bza,Arkani-Hamed:2018bjr}. We also discussed a four-graviton correlator from graviton exchange to demonstrate our treatment of spinning fields.
    \item We proved that the Cosmological Optical Theorem applies to all FLRW spacetimes where a Bunch-Davies initial state can be consistently chosen. This includes most spacetimes relevant for cosmology, such as de Sitter, slow and fast roll inflation and accelerating power-law cosmologies. We also checked that it applies to axion-monodromy inflation, where oscillations in the inflaton potential lead to a resonant particle creation and characteristic non-Gaussianities. 
\end{itemize}
There are several directions for future investigation:
\begin{itemize}
    \item While valid to all orders in perturbation theory at tree level, our results do not give a non-perturbative statement of the Cosmological Optical Theorem. In analogy with flat spacetime, such a non-perturbative formulation would be highly desirable and could provide an important piece of the puzzle to derive positivity constraints on cosmological observables (see \cite{Conjecture,Grall:2020tqc,Positivity2} for developments in that direction) and perhaps numerically bootstrap non-perturbative correlators, along the lines of \cite{Paulos:2016fap,Guerrieri:2021ivu} (see also \cite{Celoria:2021vjw} for recent non-perturbative results). 
    \item It has recently been shown how to bootstrap cosmological correlators for massless scalars and tensors using the Cosmological Optical Theorem and a set of Bootstrap Rules \cite{PSS,BBBB,MLT}. Given our results here, it would be interesting to see if one can extend this derivation to the case of exchanged massive and possibly spinning fields, with potential applications to the cosmological collider phenomenology \cite{Arkani-Hamed:2015bza}.
    \item It would be interesting to investigate what is the holographic interpretation of our single-cut rules in term of a hypothetical boundary field theory. Around de Sitter space one would expect the boundary theory to be a non-unitary CFT \cite{Witten:2001kn,Strominger:2001pn}, but it is not clear what additional property needs to be satisfied to ensure that the bulk time evolution is unitary. 
\end{itemize}
The fundamental and general nature of our results in this work strongly suggests that there are still basic and very general facts about quantum field theory on cosmological spacetimes that are awaiting to be discovered. Because of the ever growing body of cosmological dataset, advancements on the theory side are likely to have important repercussion on the phenomenology and ultimately make a long standing contribution to our understanding of the very early universe.

\section*{Acknowledgements}

We would like to thank Tanguy Grall, Aaron Hillman, Austin Joyce, Jason Joykutty, Scott Melville and Sebastien Renaux-Petel for useful discussions. E.P. has been supported in part by the research program VIDI with Project No. 680-47-535, which is (partly) financed by the Netherlands Organisation for Scientific Research (NWO). SJ is supported by the European Research Council under the European Union’s Horizon 2020 research and innovation programme (grant agreement No 758792, project GEODESI). HG is supported by jointly by the Science and Technology Facilities Council through a postgraduate studentship and the Cambridge Trust Vice Chancellor's Award.

\appendix


\section{Additional details on the massive case}\label{MassiveAppendix}
We present here the details of the calculations involved in the massive case where they expand on the solutions given in the literature.

\subsection{The three-point function}
\label{app:3point}
In \cite{Arkani-Hamed:2015bza} the authors study the three point function, leveraging its invariance under special conformal symmetries. However, for a field of arbitrary mass we cannot exclude the possibility of boundary terms that break this symmetry, so the arguments presented there break down. We are also interested in the proportionality factor that is neglected there. Therefore, in this appendix we calculate the three point function from the explicit bulk time integral,
\begin{equation}
    \psi^{\varphi\varphi\sigma}(k_1,k_2,k_3;\eta_0)=-2i\lambda\int_{-\infty}^{\eta_0}a^4(\eta)d\eta K_{k_1}^\varphi(\eta)K_{k_2}^\varphi(\eta)K_{k_3}^\sigma(\eta).
\end{equation}
As in the main text we set $H=1$ and expand this in terms of the conformally coupled modefunctions defined in \eqref{masslessdS} which gives
\begin{equation}
    \psi^{\varphi\varphi\sigma}(k_1,k_2,k_3;\eta_0)=-\frac{2i\lambda e^{-ik_{12}\eta_0}}{\eta_0^2\sigma^+(k_3\eta_0)}\int_{-\infty}^{\eta_0}d\eta\frac{1}{\eta^2}e^{ik_{12}\eta}\sigma^+(k_3\eta),
\end{equation}
where we have taken out the factor of $k_3^{\frac{3}{2}}$ from $\sigma^+_k(\eta)$ so that it is a function of $k_3\eta$ only. We then define $x=k_3\eta$ so that
\begin{equation}
    \psi^{\varphi\varphi\sigma}(k_{12},k_3;\eta_0)=-\frac{2i\lambda k_3 e^{-i\frac{k_{12}}{k_3}x_0}}{\eta_0^2\sigma^+(x_0)}\int_{-\infty}^{x_0}dx\frac{1}{x^2}e^{i\frac{k_{12}}{k_3}x}\sigma^+(x).
\end{equation}
Now we introduce the variable $U=\frac{k_{12}}{k_3}$ and consider the action of a slightly suggestive differential operator,
\begin{equation}
    \left[(U^2-1)\partial_U^2+2U\partial_U\right]\psi^{\varphi\varphi\sigma}(U,k_3;\eta_0)=-\frac{2i\lambda k_3 e^{-iUx_0}}{\eta_0^2\sigma^+(x_0)}\int_{-\infty}^{x_0}dx\left(1-U^2+2i\frac{U}{x}\right)e^{iUx}\sigma^+(x).
\end{equation}
It is possible to express this in terms of $x$ derivatives of the exponential,
\begin{equation}
    \left[(U^2-1)\partial_U^2+2U\partial_U\right]\psi^{\varphi\varphi\sigma}(U,k_3;\eta_0)=-\frac{2i\lambda k_3 e^{-iUx_0}}{\eta_0^2\sigma^+(x_0)}\int_{-\infty}^{x_0}dx\left(1+\partial_x^2+\frac{2}{x}\partial_x\right)e^{iUx}\sigma^+(x)\,,
\end{equation}
which we integrate by parts
\begin{multline}
    \left[(U^2-1)\partial_U^2+2U\partial_U\right]\psi^{\varphi\varphi\sigma}(U,k_3;\eta_0)=-\frac{2i\lambda k_3 }{\eta_0^2}\left(iU+\frac{2i}{x_0}-\frac{\partial_{x_0}\sigma^+(x_0)}{\sigma^+(x_0)}\right)\\-\frac{2i\lambda k_3 e^{-iUx_0}}{\eta_0^2\sigma^+(x_0)}\int_{-\infty}^{x_0}dxe^{iUx}\left(\partial_x^2\sigma^+(x)-\frac{2}{x}\partial_x\sigma^+(x)+\left(1+\frac{2}{x^2}\right)\sigma^+(x)\right)\,.
\end{multline}
We can exploit the differential equation satisfied by $\sigma^+(x)$,
\begin{equation}
    \partial_x^2\sigma^+(x)-\frac{2}{x}\partial_x\sigma^+(x)+\left(1+\frac{1}{x^2}\left(\frac{9}{4}-\nu^2\right)\right)\sigma^+(x)=0\,,
\end{equation}
to give
\begin{equation}
     \left[(U^2-1)\partial_U^2+2U\partial_U+\left(\frac{1}{4}-\nu^2\right)\right]\psi^{\varphi\varphi\sigma}(U,k_3;\eta_0)=-\frac{2i\lambda  }{\eta_0^2}\left(ik_3U+\frac{2}{\eta_0}-\frac{\partial_{\eta_0}\sigma^+(x_0)}{\sigma^+(x_0)}\right).
\end{equation}
We are interested in the $\eta_0\rightarrow 0$ limit and so we define
\begin{equation}
    \lim_{\eta_0\rightarrow 0}\psi^{\varphi\varphi\sigma}(U,k_3;\eta_0)=\psi^{\varphi\varphi\sigma}(U,k_3)\,,
\end{equation} 
which satisfies
\begin{equation}
     \left[(U^2-1)\partial_U^2+2U\partial_U+\left(\frac{1}{4}-\nu^2\right)\right]\psi^{\varphi\varphi\sigma}(U,k_3)=-\frac{2i\lambda  }{\eta_0^2}\left(\frac{2}{\eta_0}-\frac{\partial_{\eta_0}\sigma^+(x_0)}{\sigma^+(x_0)}\right).
\end{equation}
The homogeneous equation is identical to the one considered in \cite{Arkani-Hamed:2015bza} from the symmetries of the theory and is also the associated Legendre equation whilst the inhomogeneous part is independent of $U$ and so the general solution is
\begin{equation}\label{eq:genpsi3}
    \psi^{\varphi\varphi\sigma}(U,k_3)=AP_{\nu-\frac{1}{2}}(U)+BQ_{\nu-\frac{1}{2}}(U)-\frac{8i\lambda}{\eta_0^2(1-4\nu^2)}\left(\frac{2}{\eta_0}-{K_{k_3}^\sigma}'(\eta_0)\right).
\end{equation}
We want to avoid terms with spurious singularities and so $B=0$. In order to fix $A$ we explore the $U\rightarrow -1$ limit 
\begin{align}\nonumber
    \lim_{p\rightarrow -1}\psi^{\varphi\varphi\sigma}(U,k_3)&=\frac{\lambda }{\eta_0^2\sigma^+_{k_3}(\eta_0)}\sqrt{\frac{\pi}{k_3}}e^{-i\frac{\pi}{2}\left(\nu+\frac{1}{2}\right)}\int_{-x_0}^{\infty}dxe^{ix}(x)^{-\frac{1}{2}}H_\nu^{(2)}(x)\\&=\frac{\lambda }{\eta_0^2\sigma^+_{k_3}(\eta_0)}\sqrt{\frac{2}{k_3}}\log(U+1).
\end{align}
Comparing this to the same limit in \eqref{eq:genpsi3},
\begin{equation}
    \lim_{U\rightarrow -1}\psi^{\varphi\varphi\sigma}(U,k_3)=-A\frac{\cos(\pi\nu)}{\pi}\log(U+1)\Rightarrow A=-\frac{\lambda}{\eta_0^2\sigma^+_{k_3}(\eta_0)}\sqrt{\frac{2}{k_3}}\frac{\pi}{\cos(\pi\nu)}.
\end{equation}
Therefore,
\begin{equation}
   \psi^{\varphi\varphi\sigma}(U,k_3)=-\frac{\lambda}{\eta_0^2\sigma^+_{k_3}(\eta_0)}\sqrt{\frac{2}{k_3}}\frac{\pi}{\cos(\pi\nu)}P_{\nu-\frac{1}{2}}(U)-\frac{8i\lambda}{\eta_0^2(1-4\nu^2)}\left(\frac{2}{\eta_0}-{K_{k_3}^\sigma}'(\eta_0)\right).
\end{equation}
For contact with the four point function we then define $u=\frac{1}{U}$ and take $k_3={p_s}$ so that\footnote{We apologize for the abuse of notation in not taking this as a function of $\psi^{\varphi\varphi\sigma}\left(\frac{1}{u}\right)$, this choice makes later expressions simpler.}
\begin{equation}
    \psi^{\varphi\varphi\sigma}(u,{p_s})=-\frac{\lambda}{\eta_0^2\sigma^+_{p_s}(\eta_0)}\sqrt{\frac{2}{{p_s}}}\frac{\pi}{\cos(\pi\nu)}P_{\nu-\frac{1}{2}}\left(\frac{1}{u}\right)-\frac{8i\lambda}{\eta_0^2(1-4\nu^2)}\left(\frac{2}{\eta_0}-{K_{p_s}^\sigma}'(\eta_0)\right).
\end{equation}
For later convinience we define this as
\begin{equation}
    \psi^{\varphi\varphi\sigma}(u,{p_s})=\psi^H(u)+\psi^I({p_s}).
\end{equation}
This additional, inhomogeneous, term can be ignored if
\begin{equation}
    \lim_{\eta_0\rightarrow 0}\frac{\left(2\sigma^+_{p_s}(\eta_0)-\eta_0\partial_{\eta_0}\sigma^+_{p_s}(\eta_0)\right)}{\eta_0}=\eta_0^{\frac{1}{2}}\left(\eta_0^{-\nu}\alpha(\nu)+\eta_0^\nu\beta(\nu)\right)\ll1,
\end{equation}
where $\alpha$ and $\beta$ are some $\nu$ dependent constants. For imaginary $\nu$ this is always true because the term in brackets is bounded. For real $\nu$ this is true provided
\begin{equation}
    \frac{1}{2}-\nu>0 \rightarrow  m>\sqrt{2},
\end{equation}
where this final condition also includes the case for imaginary $\nu$.


\subsection{General form of the four-point function}
In this appendix we present the details of the calculation of the four-point function for a field of arbitrary mass. Just as for the three point function we will consider the bulk integral representation to ensure that we do not ignore any boundary terms that might be present,
\begin{equation}
    \psi_4=-4i\lambda^2\int_{-\infty}^0 d\eta d\eta' a^4(\eta)a^4(\eta') K_{k_1}^\varphi(\eta) K_{k_2}^\varphi(\eta) K_{k_3}^\varphi(\eta') K_{k_4}^\varphi(\eta') G^\sigma_{p_s}(\eta,\eta')+t+u.
\end{equation}
We can expand $K_\varphi$ to give
\begin{equation}
    \psi_4=-\frac{4i\lambda^2}{\eta_0^4}\int_{-\infty}^0 d\eta d\eta'\frac{1}{\eta^2{\eta'}^2}e^{ik_{12}\eta}e^{ik_{34}\eta'}G_{p_s}^\sigma(\eta,\eta')+t+u.
\end{equation}
We then use the differential equation solved by $G_\sigma$,
\begin{equation}
\frac{1}{\eta^2}{G_p^\sigma}''(\eta,\eta')-\frac{2}{\eta^3}{G_p^\sigma}'(\eta,\eta')+\frac{p^2}{\eta^2}G_p^\sigma(\eta,\eta')+\frac{m^2}{\eta^4}G^p_\sigma(\eta,\eta')=\delta(\eta-\eta'),
\end{equation}
to replace $\frac{G_{p_s}^\sigma}{\eta^2}$ in $\psi_4$,
\begin{multline}
\psi_4=-\frac{4i\lambda^2}{\eta_0^4}\int d\eta d\eta'\frac{1}{\eta'^2}e^{ik_{12}\eta}e^{ik_{34}\eta'}\\\times\left(\frac{\eta^2}{m^2}\delta(\eta-\eta')-\frac{1}{m^2}{G_{p_s}^\sigma}''(\eta,\eta')+\frac{2}{\eta m^2}{G_{p_s}^\sigma}'(\eta,\eta')-\frac{{p_s}^2}{m^2}G_{p_s}^\sigma(\eta,\eta')\right).
\end{multline}
We can integrate this by parts to move all derivatives from $G_{p_s}^\sigma$ onto $e^{ik_{12}}$ where they can be performed so that
\begin{multline}
\psi_4=\frac{4i\lambda^2}{m^2\eta_0^2}\int d\eta'\frac{1}{\eta'^2}e^{ik_{34}\eta'}\frac{\sigma_{p_s}^+(\eta')}{\sigma_{p_s}^+(\eta_0)} - \frac{4\lambda^2}{\eta_0^4m^2k_T} \\- \frac{4i\lambda^2}{\eta_0^4}\int d\eta d\eta'\frac{1}{\eta'^2}e^{ik_{12}\eta}e^{ik_{34}\eta'}\left(\frac{1}{m^2}k^2_{12}-\frac{2}{\eta m^2}\left(ik_{12}-\frac{1}{\eta}\right)-\frac{{p_s}^2}{m^2}\right)G_{p_s}^\sigma(\eta,\eta').
\end{multline}
We can recognise this first term as $\psi_3$ and express each of the remaining terms as derivatives of $\psi_4$ so that
\begin{equation}
(k^2_{12}-{p_s}^2)\partial^2_{k_{12}}{\psi}_4+2k_{12}\partial_{k_{12}}{\psi}_4+\left(\frac{m^2}{H^2}-2\right){\psi}_4=-\frac{4\lambda^2}{k_T\eta_0^4 }-2\lambda\psi^{\varphi\varphi\sigma}(k_3,k_4,{p_s}).
\end{equation}
This is once again the associated Legendre equation however the particular integral arising from the first term on the right hand side is not obvious. We will therefore employ the same tools that were developed in \cite{Arkani-Hamed:2018kmz} by defining $F=-\frac{\eta_0^4s}{4\lambda^2}\psi_4$, $u=\frac{{p_s}}{k_{12}}$ and $v=\frac{{p_s}}{k_{34}}$ so that
\begin{equation}
    \left[\Delta_u+\left(\frac{1}{4}-\nu^2\right)\right]F=\frac{uv}{u+v}+\frac{\eta_0^4 {p_s}}{2\lambda}\psi^{\varphi\varphi\sigma}(k_3,k_4,{p_s}).
\end{equation}
For $\Delta_u$ defined as in \cite{Arkani-Hamed:2018kmz},
\begin{equation}
    \Delta_u=u^2(1-u^2)\partial_u^2-2u^3\partial_u.
\end{equation}
As we identified in Section \ref{Massive}, it is preferable to take the two linearly independent solutions to the homogeneous differential equation to be
\begin{align}
    \tilde{F}_+(u)&=P_{\nu-\frac{1}{2}}\left(\frac{1}{u}\right),\\
    \tilde{F}_-(u)&=\frac{1}{2}\left(Q_{\nu-\frac{1}{2}}\left(\frac{1}{u}\right)+Q_{-\nu-\frac{1}{2}}\left(\frac{1}{u}\right)\right).
\end{align}
We fix the particular integral arising from $\frac{uv}{u+v}$ just as in \cite{Arkani-Hamed:2018kmz} by first considering a series expansion for $\lvert u\rvert <\lvert v\rvert$,
\begin{equation}
    \tilde{F}_<(u,v)=\sum_{m,n=0}^\infty c_{mn}u^{2m+1}\left(\frac{u}{v}\right)^n,
\end{equation}
the series coefficients are
\begin{equation}
    c_{mn}=\frac{(1)^n(n+1)...(n+2m)}{\left[\left(n+\frac{1}{2}\right)^2-\nu^2\right]\left[\left(n+\frac{5}{2}\right)^2-\nu^2\right]...\left[\left(n+\frac{1}{2}+2m\right)^2-\nu^2\right]}.
\end{equation}
Because these solutions have an identical Wronskian to those used in \cite{Arkani-Hamed:2018kmz} the particular integral takes an identical form,
\begin{equation}
    \tilde{F}_<(u,v)=\begin{cases}\displaystyle\sum_{m,n=0}^\infty c_{mn} u^{2m+1}\left(\frac{u}{v}\right)^n  \hphantom{+\ \frac{\pi}{\cos(\pi\nu)}\left(\tilde{F}_+(v)\tilde{F}_-(u)-\tilde{F}_-(v)\tilde{F}_+(u)\right)}\quad \lvert u\rvert\leq\lvert v\rvert,\\\displaystyle
   \sum_{m,n=0}^\infty c_{mn}v^{2m+1}\left(\frac{v}{u}\right)^n+\frac{\pi}{\cos(\pi\nu)}\left(\tilde{F}_+(v)\tilde{F}_-(u)-\tilde{F}_-(v)\tilde{F}_+(u)\right)\quad \lvert u\rvert\geq\lvert v\rvert.\end{cases}
\end{equation}
We also have a second particular integral that arises from the boundary term,
\begin{equation}
    \left[\Delta_u+\left(\frac{1}{4}-\nu^2\right)\right]\tilde{F}_P=\frac{\eta_0^4{p_s}}{2\lambda}\psi^{\varphi\varphi\sigma}(v,{p_s}).
\end{equation}
We expect this to be symmetric in $u,v$ so to find this particular integral we first look at the differential equation solved by $\psi^{\varphi\varphi\sigma}$
\begin{equation}
    \left[\Delta_u+\left(\frac{1}{4}-\nu^2\right)\right]\psi^{\varphi\varphi\sigma}=\frac{2i\lambda}{\eta_0^2}\left(\frac{2}{\eta_0}-{K_{p_s}^\sigma}'(\eta_0)\right).
\end{equation}
So the homogeneous part of $\psi^{\varphi\varphi\sigma}$ is also a homogeneous solution to the differential equation satisfied by $F_P$ therefore, to ensure symmetry in $u,v$ we must add the homogeneous solution in $u$ to this particular integral,
\begin{align}
\tilde{F}_P(u,v,{p_s})&=\frac{2\eta_0^4s}{\lambda(1-4\nu^2)}\left(\psi^H(u)+\psi^H(v)+\psi^I({p_s})\right)\\&=\frac{2\eta_0^2{p_s}}{\left(1-4\nu^2\right)^2}\left(\frac{4\nu^2-1}{\sigma_{p_s}^+(\eta_0)}\sqrt{\frac{2}{{p_s}}}\frac{\pi}{\cos(\pi\nu)}\left(\tilde{F}_+(u)+\tilde{F}_+(v)\right)-8i\left(\frac{2}{\eta_0}-{K_{p_s}^\sigma}'(\eta_0)\right)\right)
\end{align}
This also ensures that it properly satisfies the differential equation in $v$. The only thing that remains is to find the complementary function which is defined so that the total solution is symmetric in $u,v$ and free from unphysical singularities. The first of these conditions is implemented identically to \cite{Arkani-Hamed:2018kmz} so that
\begin{equation}
    F(u,v)=F_P(u,v,{p_s})+\begin{cases}\displaystyle\sum_{m,n=0}^\infty c_{mn} u^{2m+1}\left(\frac{u}{v}\right)^n +\frac{\pi}{2\cos(\pi\nu)}\tilde{g}(u,v)\quad \lvert u\rvert\leq\lvert v\rvert,\\\displaystyle
   \sum_{m,n=0}^\infty c_{mn}v^{2m+1}\left(\frac{v}{u}\right)^n\frac{\pi}{2\cos(\pi\nu)}\tilde{g}(v,u)\quad \lvert u\rvert\geq\lvert v\rvert,\end{cases}
\end{equation}
where
\begin{equation}
    \tilde{g}=\beta_+\tilde{F}_+(v)\tilde{F}_+(u)+(\beta_0+1)\tilde{F}_-(v)\tilde{F}_+(u)+(\beta_0-1)\tilde{F}_+(v)\tilde{F}_-(u)+\beta_-\tilde{F}_-(v)\tilde{F}_-(u).
\end{equation}
However, the removal of unphysical singularities as $u,v\rightarrow 1$ occurs slightly differently. $\psi_3$ is already free from such singularities and the convergence of the sum is ensured by the piece-wise definition. The only term that can diverge is therefore $\tilde{g}$. Physically $u,v<1$ and so we must take the $u\rightarrow 1$ limit with $u>v$, and we need to ensure that
\begin{equation}
    \lim_{u\rightarrow 1} \tilde{g}(v,u)=-\frac{1}{2}\log(u-1)\left[(\beta_0+1)\tilde{F}_+(v)+\beta_-\tilde{F}_-(v)\right]
\end{equation}
is finite. Therefore, we need $\beta_-=0$ and $\beta_0=-1$ so $\tilde{g}(u,v)$ becomes
\begin{equation}
    \tilde{g}(u,v)=\beta_+\tilde{F}_+(u)\tilde{F}_+(v)-2\tilde{F}_+(v)\tilde{F}_-(u).
\end{equation}
We will fix the remaining free coefficient, $\beta_+$ using the single-cut rule and so need to consider the Hermitian analytic image of this function. For this we need to use that
\begin{align}
    P_{\nu^*}(z^*)&=P^*_\nu(z),& Q_{\nu^*}(z^*)&=Q^*_\nu(z),\\
    P_{-\nu}(z)&=P_{\nu-1}(z),& Q_{-\nu-1}(z)&=Q_\nu(z)-\pi\frac{\cos(\pi\nu)}{\sin(\pi\nu)}P_\nu(z)\,.
\end{align}
Using this we find that 
\begin{equation}
\tilde{F}_+^*(-u^*)=P_{\nu-\frac{1}{2}}^*\left(-\frac{1}{u^*}\right)=P_{\nu^*-\frac{1}{2}}\left(-\frac{1}{u}\right)=\begin{cases}P_{\nu-\frac{1}{2}}\left(-\frac{1}{u}\right)&\nu\in\Re,\\
P_{-\left(\nu+\frac{1}{2}\right)}\left(-\frac{1}{u}\right)=P_{\nu-\frac{1}{2}}\left(-\frac{1}{u}\right) &\nu\in\Im.
\end{cases}
\end{equation}
To explore the Hermitian analytic image of $F_-$ we need that, \cite{Wolf:LegendreQ},
\begin{equation}
    Q_\nu(-z)=-e^{-\pi\nu\frac{\sqrt{-z^2}}{z}}Q_\nu(z)=e^{-i\pi\nu}Q_\nu(z),
\end{equation}
where the final equality holds for $\Im(u)>0$, therefore
\begin{equation}
    \tilde{F}_-^*(-u^*)=-i\frac{1}{2}\left(e^{-i\pi\nu}Q_{\nu-\frac{1}{2}}\left(\frac{1}{u}\right)+e^{i\pi\nu}Q_{-\nu-\frac{1}{2}}\left(\frac{1}{u}\right)\right).
\end{equation}
It is convinient at this point to introduce
\begin{align}
    P_\pm(u)&=P_{\nu-\frac{1}{2}}\left(\pm\frac{1}{u}\right),\\
    Q_\pm(u)&=Q_{\pm\nu-\frac{1}{2}}\left(\frac{1}{u}\right).
\end{align}
$\psi_3$ includes only $P_+$ and so to prove that the single-cut rule holds it is convenient to re-express $\tilde{F}_-(u)$ in terms of $P_\pm$ and we find that, for $\Im(u)>0$,
\begin{align}\label{eq:PasQ}
   P_-(u)&=-i\frac{\cos(\pi\nu)}{\pi\sin(\pi\nu)}\left(e^{i\pi\nu}Q_-(u)-e^{-i\pi\nu}Q_+(u)\right),\\
   P_+(u)&=\frac{\cos(\pi\nu)}{\pi\sin(\pi\nu)}\left(Q_-(u)-Q_+(u)\right).
\end{align}
Using this we can express the $\tilde{F}_-$ terms as
\begin{align}
    \tilde{F}_-(u)&=\frac{\pi i}{2}P_+(u)+\frac{\pi}{2\cos(\pi\nu)}P_-(u),\\
    \tilde{F}^*_-(-u^*)&=-\frac{\pi i}{2}P_-(u)+\frac{\pi}{2\cos(\pi\nu)}P_+(u).
\end{align}
The sum term is trivially equal to minus its Hermitian analytic image so, for $\lvert u\rvert\leq \lvert v\rvert$ the discontinuity of the four point wavefunction coefficient is
\begin{multline}
\text{Disc}_{p_s}\ i\psi_4=  -\frac{4i\lambda^2}{{p_s}\eta_0^4}\left(F(u,v)+F^*(-u^*,-v^*)\right)=\frac{64i\lambda^2}{(1-4\nu^2)^2}\frac{1}{\sigma_{p_s}^+(\eta_0)\sigma_{p_s}^-(\eta_0)}+\\\frac{2i\lambda^2}{{p_s}\eta_0^4}\frac{\pi}{\cos(\pi\nu)}\left(\frac{\pi P_-(u)P_+(v)}{\cos(\pi\nu)}-(\beta_+^*+\pi i)P_-(u)P_-(v)-(\beta_+-\pi i)P_+(u)P_+(v)+\frac{\pi P_+(u)P_-(v)}{\cos(\pi\nu)}\right)\\+\frac{8i\lambda^2}{\eta_0^2(1-4\nu^2)}\sqrt{\frac{2}{{p_s}}}\frac{\pi}{\cos(\pi\nu)}\left(\frac{P_+(u)}{\sigma_{p_s}^+(\eta_0)}+\frac{P_+(v)}{\sigma_{p_s}^+(\eta_0)}+\frac{P_-(u)}{\sigma_{p_s}^-(\eta_0)}+\frac{P_-(v)}{\sigma_{p_s}^-(\eta_0)}\right),
\end{multline}
where we have used that for ${p_s}>0$,
\begin{equation}
    \left[\sigma_{p_s}^+(\eta_0)\right]^*=\sigma_{p_s}^-(\eta_0).
\end{equation}
Noting that this is symmetric when we exchange $u$ and $v$ we can conclude that this is also valid for $\lvert v\rvert\leq \lvert u\rvert$ which is required for the single-cut rule to hold. We now need to compare this to the discontinuity of the cut diagram which depends on 
\begin{equation}
    \text{Disc}_{p_s}\ i\psi^{\varphi\varphi\sigma}=-\frac{\lambda i\pi}{\eta_0^2\cos(\pi\nu)}\sqrt{\frac{2}{p_s}}\left(\frac{P_+(u)}{\sigma_{p_s}^+(\eta_0)}+\frac{P_-(u)}{\sigma^-_{p_s}(\eta_0)}\right)-\frac{8i\lambda}{(1-4\nu^2)}\frac{1}{\sigma^+_{p_s}(\eta_0)\sigma^-_{p_s}(\eta_0)}.
\end{equation}
From this we find the right hand side
\begin{multline}
-i P_{p_s}\text{Disc}_{p_s}\left[ i\psi^{\varphi\varphi\sigma}\right] \text{Disc}_{p_s}\left[ i\psi^{\varphi\varphi\sigma}\right]= \frac{64i \lambda^2}{(1-4\nu^2)^2}\frac{1}{\sigma_{p_s}^+(\eta_0)\sigma_{p_s}^-(\eta_0)}+\\\frac{2i\lambda^2}{{p_s}\eta_0^4} \frac{\pi^2}{\cos^2(\pi\nu)}\left(P_-(u)P_+(v)+\frac{\sigma_{p_s}^+(\eta_0)}{\sigma_{p_s}^-(\eta_0)}P_-(u)P_-(v)+\frac{\sigma_{p_s}^-(\eta_0)}{\sigma_{p_s}^+(\eta_0)}P_+(u)P_+(v)+P_+(u)P_-(v)\right)\\+\frac{8i\lambda^2}{\eta_0^2(1-4\nu^2)} \sqrt{\frac{2}{p_s}}\frac{\pi}{\cos(\pi\nu)}\left(\frac{P_+(u)}{\sigma_{p_s}^+(\eta_0)}+\frac{P_+(v)}{\sigma_{p_s}^+(\eta_0)}+\frac{P_-(u)}{\sigma_{p_s}^-(\eta_0)}+\frac{P_-(v)}{\sigma_{p_s}^-(\eta_0)}\right).
\end{multline}
So this satisfies the single-cut rule if 
\begin{equation}
    \beta_+=i\pi-\frac{\pi}{\cos(\pi\nu)}\frac{\sigma_{p_s}^-(\eta_0)}{\sigma_{p_s}^+(\eta_0)}.
\end{equation}


\subsection{Comparison to the correlator}
The results presented in \cite{Arkani-Hamed:2018kmz} were expressed in terms of
\begin{equation}
    \hat{F}_\pm=\left(\frac{u}{2\nu}\right)^{\frac{1}{2}\mp \nu}\prescript{}{2}{F}_1\left(\frac{3}{4}\mp\frac{\nu}{2},\frac{1}{4}\mp\frac{\nu}{2},1\mp\nu,u^2\right).
\end{equation}
This is defined explicitly for real $u$ so that
\begin{equation}
    \nu^{\frac{1}{2}\mp \nu} \hat{F}_\pm=\left(\frac{u}{2}\right)^{\frac{1}{2}\mp \nu}\prescript{}{2}{F}_1\left(\frac{3}{4}\mp\frac{\nu}{2},\frac{1}{4}\mp\frac{\nu}{2},1\mp\nu,u^2\right).
\end{equation}
$Q_\pm$ can also be expressed in terms of hyper geometric functions,
\begin{equation}
Q_\pm(u)=\sqrt{\pi}\left(\frac{u}{2}\right)^{\frac{1}{2}\pm\nu}\frac{\Gamma\left(\frac{1}{2}\pm\nu\right)}{\Gamma(1\pm\nu)}\prescript{}{2}{F}_1\left(\frac{3}{4}\pm\frac{\nu}{2},\frac{1}{4}\pm\frac{\nu}{2},1\pm\nu,u^2\right)=\nu^{\frac{1}{2}\pm \nu}\sqrt{\pi}\frac{\Gamma\left(\frac{1}{2}\pm\nu\right)}{\Gamma(1\pm\nu)}\hat{F}_\mp(u).
\end{equation}
We then introduce
\begin{equation}
    \alpha_\pm=-\left(\frac{1}{2\nu}\right)^{\frac{1}{2}\mp\nu}\frac{\Gamma(1\mp \nu)}{\Gamma\left(\frac{1}{4}\mp\frac{\nu}{2}\right)\Gamma\left(\frac{3}{4}\mp\frac{\nu}{2}\right)}=-\left(\frac{1}{\nu}\right)^{\frac{1}{2}\mp\nu}\frac{\Gamma(1\mp \nu)}{2\sqrt{\pi}\Gamma\left(\frac{1}{2}\mp\nu\right)}.
\end{equation}
Therefore, 
\begin{equation}\label{eq:AQasF}
Q_\pm(u)=-\frac{1}{2\alpha_\mp}\hat{F}_\mp(u),
\end{equation}
so, to compare our results to those in \cite{Arkani-Hamed:2018kmz} we should re-express everything in terms of $Q_\pm$. As we noted in Section \ref{Massive}, $\tilde{F}_\pm$ are real for physical $u,v$ and so we first take the real part and then use \eqref{eq:PasQ} to give
\begin{align}\nonumber
    \frac{\pi}{2\cos(\pi\nu)}\Re\tilde{g}(u,v)&=\frac{\pi}{2\cos(\pi\nu)}\tilde{F}_+(v)\left(\Re\left(\beta_+\right)\tilde{F}_+(u)-\tilde{F}_-(u)\right)\\\nonumber
    &=\frac{Q_-(v)-Q_+(v)}{2\sin(\pi \nu)}\left(\Re\left(\beta_+\right)\frac{\cos(\pi\nu)}{\pi\sin(\pi\nu)}\left(Q_-(u)-Q_+(u)\right)-Q_-(u)-Q_+(u)\right)\\
    &=\frac{Q_-(v)-Q_+(v)}{2\sin(\pi \nu)}\left(\frac{\Re\left(\sigma^-_{p_s}(\eta_0)^2\right)}{\sin(\pi\nu)P_{p_s}^\sigma}\left(Q_+(u)-Q_-(u)\right)-\left(Q_+(u)+Q_-(u)\right)\right).
\end{align}
Similarly, we need to find the real part of the three point function,
\begin{align}\nonumber
   \Re\left(\psi^{\varphi\varphi\sigma}({p_s},s)\right)&=-\frac{\lambda P_+(u)}{P_{p_s}^\sigma\eta_0^2}\sqrt{\frac{2}{{p_s}}}\frac{\pi}{\cos(\pi\nu)}\Re\left( \sigma^-_{p_s}(\eta_0)\right) +\frac{8\lambda}{\eta_0^2(1-4\nu^2)}\Re\left( i {K_{p_s}^\sigma}'(\eta_0)\right)\\&=-\frac{\lambda }{P_{p_s}^\sigma\eta_0^2}\sqrt{\frac{2}{p_s}}\frac{Q_-(u)-Q_+(u)}{\sin(\pi\nu)}\Re\left( \sigma_{p_s}^-(\eta_0)\right) -\frac{4\lambda}{(1-4\nu^2)P_{p_s}^\sigma }.
\end{align}
Where we have used that the Wronskian for the plus and minus solutions is $-i$. We can then calculate the trispectrum from this using that
\begin{equation}
B_4=-2\prod_{a=1}^4\frac{1}{2\Re\psi_2'(k_a)}\left[\Re\psi_4'(u,v)-\frac{\Re\psi'_3(u)\Re\psi'_3(v)}{\Re\psi_2'({p_s})}-t-u\right].
\end{equation}
So, expressing the $s$ channel in terms of $F$ gives
\begin{equation}
    B_4^s=\frac{\eta_0^4\lambda^2}{2{p_s}k_1k_2k_3k_4}\left(\Re(F)+\frac{{p_s}\eta_0^4}{2\lambda^2}P_{p_s}^\sigma\Re\psi'_3(u)\Re\psi'_3(v)\right).
\end{equation}
Therefore, for $\lvert u\rvert \leq \lvert v\rvert$, we find
\begin{multline}
    B_4^s=\frac{\eta_0^4\lambda^2}{2{p_s}k_1k_2k_3k_4}\left(\Re\left(F_P(u,v,{p_s})+\sum_{m,n=0}c_{mn} u^{2m+1}\left(\frac{u}{v}\right)^n+\frac{\pi}{2\cos(\pi\nu)}\tilde{g}(u,v)\right)\right.\\\left.+\frac{{p_s}\eta_0^4}{2\lambda^2}P_{p_s}^\sigma\Re\left(\psi'_3(u)\right)\Re\left(\psi'_3(v)\right)\right).
\end{multline}
We can also see that
\begin{align}\nonumber
\Re(\tilde{F}_P(u,v,{p_s}))&=\frac{2\eta_0^4{p_s}}{\lambda(1-4\nu^2)}\Re\left(\psi^H(u)+\psi^H(v)+\psi^I({p_s})\right)\\&=-\frac{\eta_0^4{p_s}}{2\lambda^2}P_{p_s}^\sigma\Re(\psi^I({p_s}))\Re\left(\psi^H(u)+\psi^H(v)+\psi^I({p_s})\right).
\end{align}
Therefore, this term cancels with the inhomogeneous contribution to the three point function and so 
\begin{multline}
    B_4^s=\frac{\eta_0^4\lambda^2}{2{p_s}k_1k_2k_3k_4}\left(\sum_{m,n=0}c_{mn} u^{2m+1}\left(\frac{u}{v}\right)^n+\right.\\\left.\frac{Q_-(v)-Q_+(v)}{2\sin(\pi \nu)}\left(\frac{\Re\left(\sigma_{p_s}^-(\eta_0)^2\right)-2\Re\left( \sigma^-_{p_s}(\eta_0)\right)^2}{\sin(\pi\nu)P_{p_s}^\sigma}\left(Q_+(u)-Q_-(u)\right)-\left(Q_+(u)+Q_-(u)\right)\right) \right).
\end{multline}
We simplify this using that
\begin{equation}
    \frac{\Re\left(\sigma^-_{p_s}(\eta_0)^2\right)-2\Re\left( \sigma^-_{p_s}(\eta_0)\right)^2}{P_{p_s}^\sigma}=-\frac{\Re(\sigma^-_{p_s}(\eta_0))^2+\Im(\sigma^-_{p_s}(\eta_0)^2}{\Re(\sigma^-_{p_s}(\eta_0))^2+\Im(\sigma^-_{p_s}(\eta_0))^2}=-1,
\end{equation}
so that the s-channel of the trispectrum is
\begin{multline}
    B_4^s=\frac{\eta_0^4\lambda^2}{2{p_s}k_1k_2k_3k_4}\left(\sum_{m,n=0}c_{mn} u^{2m+1}\left(\frac{u}{v}\right)^n+\right.\\\left.\frac{\beta}{2}\left(Q_-(v)-Q_+(v)\right)\left(\left(1-\beta\right)Q_+(u)+\left(\beta+1\right)Q_-(u)\right) \right),
\end{multline}
where we have introduced $\beta=-\frac{1}{\sin(\pi\nu)}$. We now rewrite this in terms of $F_\pm$ using \eqref{eq:AQasF} so that
\begin{multline}\label{eq:smallboundary}
    B_4^s=\frac{\eta_0^4\lambda^2}{4{p_s}k_1k_2k_3k_4}\left(2\sum_{m,n=0}c_{mn} u^{2m+1}\left(\frac{u}{v}\right)^n+\right.\\\left. \frac{\beta}{4\alpha_+\alpha_-}\left(-F_+(v)+F_-(v)\frac{\alpha_+}{\alpha_-}\right)\left(-\left(1-\beta\right)F_-(u)-\left(\beta+1\right)F_+(u)\frac{\alpha_-}{\alpha_+}\right)\right).
\end{multline}
Finally, we simplify this using Euler's reflection formula~\cite{riley2002mathematical},
\begin{equation}
    \frac{1}{\alpha_+\alpha_-}=\frac{4\pi\nu\Gamma\left(\frac{1}{2}-\nu\right)\Gamma\left(\frac{1}{2}+\nu\right)}{\Gamma(1-\nu)\Gamma(1+\nu)}=\frac{-4\pi^2}{\sin\left(\pi\left(\frac{1}{2}-\nu\right)\right)\Gamma(-\nu)\Gamma(1+\nu)}=\frac{4\pi\sin(\pi\nu)}{\cos\left(\pi\nu\right)}=-\frac{4\pi}{\beta\cos(\pi\nu)}.
\end{equation}
Therefore,
\begin{multline}
    B_4^s=\frac{\eta_0^4\lambda^2}{2{p_s}k_1k_2k_3k_4}\left(\sum_{m,n=0}c_{mn} u^{2m+1}\left(\frac{u}{v}\right)^n+\frac{\pi}{2\cos(\pi\nu)}\right.\\\left. \left((\beta-1)\left(F_-(u)F_+(v)-\frac{\alpha_+}{\alpha_-}F_-(u)F_-(v)\right)+(\beta+1)\left(F_-(v)F_+(u)-\frac{\alpha_-}{\alpha_+}F_+(u)F_+(v)\right)\right)\right) .
\end{multline}
The expression on the final line is exactly $\hat{g}(u,v)$ from \cite{Arkani-Hamed:2018kmz}. We can find the expression for $\lvert v\rvert\leq \lvert u\rvert$ by exchanging $u\leftrightarrow v$ and so we have
\begin{equation}
    B_4^s=\frac{\eta_0^4\lambda^2}{2{p_s}k_1k_2k_3k_4}
    \begin{cases} \displaystyle
    \sum_{m,n=0}c_{mn} u^{2m+1}\left(\frac{u}{v}\right)^n+\frac{\pi}{2\cos(\pi\nu)}\hat{g}(u,v),\quad \lvert u\rvert \leq \lvert v\rvert,\\\displaystyle
    \sum_{m,n=0}c_{mn} v^{2m+1}\left(\frac{v}{u}\right)^n+\frac{\pi}{2\cos(\pi\nu)}\hat{g}(v,u),\quad\lvert v\rvert \leq \lvert u\rvert.
    \end{cases}
\end{equation}
This is exactly the same expression for $B_4^s$ as (B.29) from \cite{Arkani-Hamed:2018kmz} and so we find agreement with the results in  \cite{Arkani-Hamed:2018kmz}.


\subsection{The total discontinuity}
We have shown that it is possible to cut an internal massive line in a way that is consistent with the massless case however, our more general results also rely on the ability to leave massive internal lines uncut. To see that it is valid to do this whilst also cutting an internal line would require the calculation of a new, complicated, diagram and so it is convenient to instead consider the total discontinuity of this diagram which, by the Hermitian analyticity of the propagators, is expected to vanish. As before we first restrict to the case where $\lvert u\rvert \leq \lvert v\rvert$,
\begin{multline}
    \text{Disc}\ i\psi_4=-\frac{4i\lambda^2}{{p_s}\eta_0^4}\left(F(u,v)-F^*(u^*,v^*)\right)=-\frac{4i\lambda^2}{{p_s}\eta_0^4}\left(\vphantom{\frac{\pi}{\cos(\pi\nu)}}F_P(u,v,{p_s})-F_P^*(u^*,v^*,-p_s^*)\right.\\\left.+\sum_{m,n=0}u^{2m+1}\left(\frac{u}{v}\right)^n\left(c_{mn}-c_{mm}^*\right)+\frac{\pi}{2\cos(\pi\nu)}\left(\tilde{g}(u,v)-\tilde{g}^*(u^*,v^*)\right)\right).
\end{multline}
To evaluate this we need that
\begin{equation}
    \tilde{F}_{\pm}^*(u^*)=\tilde{F}_{\pm}(u).
\end{equation}
We consider each of the terms in order
\begin{multline}
    F_P(u,v,{p_s})-F_P^*(u^*,v^*,-p_s^*)=\frac{2\eta_0^2s}{\left(1-4\nu^2\right)^2}\left(\vphantom{\sqrt{\frac{2}{{p_s}}}}8i\left({K_{p_s}^{\sigma}}'(\eta_0)-\left[{K_{-p_s^*}^{\sigma}}'(\eta_0)\right]^*\right)\right.\\\left.+\frac{4\nu^2-1}{\sigma_{p_s}^+(\eta_0)}\frac{\pi}{\cos(\pi\nu)}\left(\tilde{F}_+(u)+\tilde{F}_+(v)\right)\left(\sqrt{\frac{2}{{p_s}}}+\sqrt{-\frac{2}{{p_s}}}\frac{\sigma_s^+(\eta_0)}{\left[\sigma_{-p_s^*}^+(\eta_0)\right]^*}\right)\right)=0.
\end{multline}
Where we have used \eqref{eq:massiveconjp} as well as the fact that, for $\Im({p_s})<0$ 
\begin{equation}
    \sqrt{-\frac{2}{{p_s}}}=-i\sqrt{\frac{2}{{p_s}}},
\end{equation}
to cancel the second line. Note that this same relationship also gives the total discontinuity in $\psi^{\varphi\varphi\sigma}$,
\begin{multline}
    \text{Disc}\ i\psi^{\varphi\varphi\sigma}=i\psi^{\varphi\varphi\sigma}(u,{p_s})+i\left[\psi^{\varphi\varphi\sigma}(u^*,-p_s^*)\right]^*=\\\frac{8\lambda}{\eta_0^2(4\nu^2-1)}\left({K_{p_s}^{\sigma}}'(\eta_0)-\left[{K_{-p_s^*}^{\sigma}}'(\eta_0)\right]^*\right)-\frac{i\lambda}{\eta_0^2\sigma_{p_s}^+(\eta_0)}\frac{\pi}{\cos(\pi\nu)}\left(\sqrt{\frac{2}{p_s}}+\sqrt{-\frac{2}{p_s}}\frac{\sigma_{p_s}^+(\eta_0)}{\left[\sigma_{p_s}^+(\eta_0)\right]^*}\right)\tilde{F}_+(u)\\=0,
\end{multline}
which is what has previously been referred to as the contact COT\cite{COT}. The sum trivially cancels as $c_{mn}$ are real and the final term is
\begin{align}
    \tilde{g}(u,v)-\tilde{g}(u^*,v^*)&=\left(\beta_+-\beta_+^*\right)\tilde{F}_+(u)\tilde{F}_+(v)\nonumber\\&=\left(2i\pi-\frac{\pi}{\cos(\pi\nu)}\left(\frac{\sigma_{p_s}^-(\eta_0)}{\sigma_{p_s}^+(\eta_0)}-\left[\frac{\sigma_{p_s}^-(\eta_0)}{\sigma_{p_s}^+(\eta_0)}\right]^*\right)\right)\tilde{F}_+(u)\tilde{F}_+(v)\nonumber\\&=\left(2i\pi-\frac{\pi}{\cos(\pi\nu)}\left(\frac{\sigma_{p_s}^-(\eta_0)}{\sigma_{p_s}^+(\eta_0)}-\frac{\sigma_{p_s}^-(\eta_0)-2i\cos(\pi\nu)\sigma_{p_s}^+(\eta_0)}{\sigma_{p_s}^+(\eta_0)}\right)\right)\tilde{F}_+(u)\tilde{F}_+(v)\nonumber\\&=0,
\end{align}
where the third line we have used the results in Eqs. \ref{eq:massiveconjp} and \ref{eq:massiveconjm}. Combining these results gives us that
\begin{equation}
    \text{Disc}\ i\psi_4=0,
\end{equation}
which is symmetric in $u,v$ and so is valid for all $u,v$. This is exactly the result that was predicted by our consideration of the general case and so we have confirmed that uncut massive lines behave in a way consistent with our cutting rules. 


\section{Resonant non-Gaussianity}
\label{RNG}

In this appendix, we present some technical details of the proof that the propagators of perturbations in axion monodromy inflation are Hermitial analytic even in the presence of background oscillations that lead to the resonant production of perturbations.  

In order to make direct contact with the bulk-to-boundary propagator we will calculate $\phi^+$ (as opposed to $\mathcal{R}_k(-k\eta)\propto\phi^-$ which was considered in \cite{flauger2011resonant}). To this end, we consider the ansatz
\begin{equation}
\phi^+_k(\eta)=C\left[(-k\eta)^{\nu}H^{(2)}_{\nu}(-k\eta)+b_* c_k(-k\eta)(-k\eta)^{\frac{3}{2}}H^{(1)}_{3/2}(-k\eta)\right],
\end{equation}
where $C$ is a $k$-independent constant and the factor of $b_*$ has been taken out of $c_k$ to make the relative size of the terms clear. The second term is of order $\frac{3}{2}$ because $\nu=\frac{3}{2}+2\epsilon_0+\delta_0$ and $\epsilon_0,\ \delta_0=\mathcal{O}(b_*)$ so to linear order it is enough to keep just this term. This is important because we don't have a convenient expression for the Hermitian analytic image of $H_\nu^{(1)}(x)$ for generic $\nu$ and $\Im(x)<0$. The differential equation that $\phi^+$ satisfies is
\begin{equation}
\frac{d^2\phi^+_k}{dx^2}-\frac{2(1+2\epsilon_0+2\epsilon_1(x)+\delta_0+\delta_1(x))}{x}\frac{d\phi^+_k}{dx}+\phi^+_k=0.
\end{equation}
Where $x=-k\eta$ and both $\epsilon_0$ and $\delta_0$ are constant whilst
\begin{align}
    \epsilon_1(x)&=-3b_*f\sqrt{2\epsilon}\cos\left(\frac{\phi_0(\eta)}{f}\right),&    \delta_1&=-3b_*\sin\left(\frac{\phi_0(\eta)}{f}\right).
\end{align}
This then gives an equation for $c_k$ which, to linear order in $b_*$, is
\begin{equation}
6e^{-ix}x\sin\left(\phi_*+\frac{\sqrt{2\epsilon_*}}{f}\ln\frac{x}{x_*}\right)+e^{ix}(2(x(i+x)-1)\frac{d}{dx}c_k(x)+(1-ix)x\frac{d^2}{dx^2}c_k(x)=0.
\end{equation}
If we then define $u(x)=e^{2ix} \frac{d}{dx} c_k(x)$ we find a linear first order differential equation for $u(x)$,
\begin{equation}
6x\sin\left(\phi_*+\frac{\sqrt{2\epsilon_*}}{f}\ln\frac{x}{x_*}\right)-2u(x)+(1-ix)x\frac{d}{dx}u(x)=0,
\end{equation}
which can be solved and then integrated to give
\begin{multline}
c_k(x)=iAe^{-2ix}\left(-\frac{1}{2}-\frac{1}{ix-1}\right)+B+ \\ \frac{3 \left(\left(\frac{\sqrt{2\epsilon_*}}{f}+i\right)\left(2f\sqrt{2\epsilon_*}+i\right)e^{-i\left(\frac{\phi_k}{f}+\frac{\sqrt{2\epsilon_*}}{f}\log(x)\right)}+\left(\frac{\sqrt{2\epsilon_*}}{f}-i\right)\left(2f\sqrt{2\epsilon_*}-i\right)e^{i\left(\frac{\phi_k}{f}+\frac{\sqrt{2\epsilon_*}}{f}\log(x)\right)}\right)}{e^{2i  x}\frac{\sqrt{2\epsilon_*}}{f}\left(1+\frac{2\epsilon_*}{f^2}\right)\left(1-\frac{1}{ix}\right)}\\+\frac{3i}{2\frac{\sqrt{2\epsilon_*}}{f}\left(1+\frac{2\epsilon_*}{f^2}\right)}\left(2^{i\frac{\sqrt{2\epsilon_*}}{f}}\left(\frac{\sqrt{2\epsilon_*}}{f}+i\right)^2\left(2f\sqrt{2\epsilon_*}+i\right)e^{-i\phi_k}e^{-\pi\frac{\sqrt{2\epsilon_*}}{2f}}\Gamma\left(1-i\frac{\sqrt{2\epsilon_*}}{f},2ix\right)\right.\\\left.-2^{-i\frac{\sqrt{2\epsilon_*}}{f}}\left(\frac{\sqrt{2\epsilon_*}}{f}-i\right)^2\left(2f\sqrt{2\epsilon_*}-i\right)e^{i\phi_k}e^{\pi\frac{\sqrt{2\epsilon_*}}{2f}}\Gamma\left(1+i\frac{\sqrt{2\epsilon_*}}{f},2ix\right)\right),
\end{multline}
where $A$ and $B$ are $x$ independent constants. Since $c_k$ must vanish in the infinite past we choose $B=0$. Generically, for $k$ with a negative imaginary part and real $v$ we have that the Hermitian analytic image of the Hankel function is
\begin{equation}
x^{v}H_v^{(2)}(x)\rightarrow (-x)^{v}{H_v^{(2)}}^*(-x^*)=-x^{v}H_v^{(2)}(x).
\end{equation}
However, we don't have an equivalent expression for $H_v^{(1)}$ with arbitrary $v$. Fortunately, we do know that for the special case of $v=3/2$
\begin{equation}
x^{\frac{3}{2}}H_{3/2}^{(1)}(x)=-i\sqrt{\frac{2}{\pi}}(1-ix)e^{ix},
\end{equation}
so its Hermitian analytic image is
\begin{equation}
 i\sqrt{\frac{2}{\pi}}(1-ix)e^{ix}=-x^{\frac{3}{2}}H_{3/2}^{(1)}(x).
\end{equation}
Both Hankel terms transform in the same way, therefore if $c_k(x)$ is Hermitian analytic then we recover the desired relationship. Because $k$ has a negative imaginary part so too does $x$ and so $\log(-x)=\log(e^{i\pi}x)=\log(x)+i\pi$ and 
\begin{equation}
\phi_k=\phi_*-\sqrt{2\epsilon_*}\log\frac{k}{k_*}\rightarrow\phi_{-k^*}^*= \phi_*-\sqrt{2\epsilon_*}\log\left(-\frac{k}{k_*}\right)=\phi_k-i\pi\sqrt{2\epsilon_*}
\end{equation}
Putting this all together the Hermitian analytic image of $c_k(x)$ is then
\begin{multline}
    c_{-k^*}^*(-x^*)=-iA^*e^{-2ix}\left(-\frac{1}{2}-\frac{1}{ix-1}\right)+\\\frac{3 \left(\left(\frac{\sqrt{2\epsilon_*}}{f}-i\right)\left(2f\sqrt{2\epsilon_*}-i\right)e^{i\left(\frac{\phi_k}{f}+\frac{\sqrt{2\epsilon_*}}{f}\log(x)\right)}+\left(\frac{\sqrt{2\epsilon_*}}{f}+i\right)\left(2f\sqrt{2\epsilon_*}+i\right)e^{-i\left(\frac{\phi_k}{f}+\frac{\sqrt{2\epsilon_*}}{f}\log(x)\right)}\right)}{e^{2i  x}\frac{\sqrt{2\epsilon_*}}{f}\left(1+\frac{2\epsilon_*}{f^2}\right)\left(1-\frac{1}{ix}\right)}\\+\frac{3i}{2\frac{\sqrt{2\epsilon_*}}{f}\left(1+\frac{2\epsilon_*}{f^2}\right)}\left(-2^{-i\frac{\sqrt{2\epsilon_*}}{f}}\left(\frac{\sqrt{2\epsilon_*}}{f}-i\right)^2\left(2f\sqrt{2\epsilon_*}-i\right)e^{i\phi_k}e^{\pi\frac{\sqrt{2\epsilon_*}}{2f}}\Gamma\left(1+i\frac{\sqrt{2\epsilon_*}}{f},2ix\right)\right.\\\left.+2^{i\frac{\sqrt{2\epsilon_*}}{f}}\left(\frac{\sqrt{2\epsilon_*}}{f}+i\right)^2\left(2f\sqrt{2\epsilon_*}+i\right)e^{-i\phi_k}e^{-\pi\frac{\sqrt{2\epsilon_*}}{2f}}\Gamma\left(1-i\frac{\sqrt{2\epsilon_*}}{f},2ix\right)\right)
\end{multline}
So, provided $A$ is imaginary this expression shows that $c_k$ is Hermitian analytic; the second and third terms swap whilst the first term is Hermitian analytic by itself. To fix $A$ we consider,
\begin{equation}
-Ae^{-2ix}\left(\frac{i}{2}+\frac{1}{i+x}\right)x^{3/2}H^{(1)}_{3/2}(x)=-A i e^{-ix}\frac{(-i+x)}{\sqrt{2\pi}}=\frac{iA}{2}x^{3/2}H_{3/2}^{(2)}(x).
\end{equation}
Therefore, any contribution from this term can be reabsorbed into the overall constant (once again using that this is at linear order in $b_*$ so the slow roll corrections to the order of the Hankel function can be ignored). The requirement that it is imaginary makes the overall factor between it and the first term real and it can be ignored for the purposes of determining the analytic structure of the function so we conclude that $c_k$ is Hermitian analytic and so is $K_k(\eta)$ for arbitrary $\frac{\sqrt{2\epsilon_*}}{f}$. 


\section{WKB solution to the Klein Gordon equation for flat FLRW spacetime}
\label{WKB}
 As a demonstration of the Hermitian analyticity of the bulk-to-boundary propagator with Bunch-Davies initial conditions, we consider the case $p(k,\eta)=2\frac{a'}{a}$ and $q(k,\eta)=c_s^2(\eta)k^2+m^2$, i.e. the case where $\phi$ satisfy the Klein Gordon equation in an arbitrary flat FLRW spacetime. (One can also carry out the same procedure for the Mukhanov Sasaki equation by replacing the scale factor with $z=\frac{a\bar{\dot{\phi}}}{H}$.) The equation we have is:
\begin{equation}
	\phi''+2\frac{a'}{a}\phi'+(c_s^2k^2+m^2 a^2)\phi=0.
\end{equation}
Re-writing this in terms of $f=a\phi$, we find
\begin{equation}
    f''+\left(c_s^2k^2+m^2 a^2-\frac{a''}{a}\right)f=0.
\end{equation}
For solutions of the form $f=Ce^{ik\sigma(k,\eta)}$ this becomes
\begin{equation}
	(c_s^2-\sigma'^2)+\frac{i}{k}\sigma''+\frac{1}{k^2}\left(m^2 a^2-\frac{a''}{a}\right)=0.\label{ceq}
\end{equation}
Since we are interested in the case where the mode function approaches $e^{ik\eta}$ in the far past, we make the following ansatz:
\begin{equation}
    \sigma(k,\eta)=\pm\sigma_0(\eta)+\frac{1}{k}\sigma_1(\eta)+\frac{1}{k^2}\sigma_2(\eta)+\frac{1}{k^3}\sigma_3(\eta)+\frac{1}{k^4}\sigma_4(\eta)+\dots,
\end{equation}
where
\begin{equation}
    \sigma_0(\eta)=\int^{\eta}_{-\infty}d\bar{\eta}\, c_s(\bar{\eta}).
\end{equation}
We will focus on the solution with $+$ sign for now, though the negative solution can be easily obtained by complex conjugation.

At $\mathcal{O}(k^{-1})$ (\ref{ceq}) tells us that
\begin{equation}
    \frac{2c_s}{k}\sigma'_1=0.
\end{equation}
This means that $\sigma_1$ is constant unless $c_s$ vanishes somewhere in the bulk. This constant can be absorbed into the normalization of the mode function, so we will ignore its contribution.

At $\mathcal{O}(k^{-2})$ we have:
\begin{align}
    	-2c_s(\eta)\sigma'_2+\left(m^2 a^2-\frac{a''}{a}\right)=0,\\
		\sigma_2=\frac{1}{2}\int^{\eta}_{-\infty}d\bar{\eta}\frac{1}{c_s(\bar{\eta})}\left(m^2 a(\bar{\eta})^2-\frac{a''(\bar{\eta})}{a(\bar{\eta})}\right).
\end{align}
Since everything within the integral is real, we expect $\sigma_2$ to be real as well.

At $\mathcal{O}(k^{-3})$ we have:
\begin{align}
    -2c_s\sigma'_3+i\sigma''_2=0,\\
	\sigma_3=\frac{i}{2}\int^{\eta}_{-\infty}d\bar{\eta}\frac{\sigma''_2(\bar{\eta})}{c_s(\bar{\eta})}.
\end{align}
Since everything within the integral is real, $\sigma_3$ is pure imaginary. In general we have the following:
\begin{equation}
    \sigma_r=\frac{1}{2}\int^{\eta}_{-\infty}d\bar{\eta}\frac{1}{c_s(\bar{\eta})}(i\sigma''_{r-1}(\bar{\eta})-\sum_{m+n=r}\sigma'_m(\bar{\eta})\sigma'_n(\bar{\eta})).
\end{equation}
By induction, we see that $\sigma_r$ must be real for even $r$ and pure imaginary for odd $r$. Therefore, as long as this series expansion converges, we conclude that $\sigma(k,\eta)$ is Hermitian analytic.

Since $K(k,\eta)=\frac{\phi^{+}(k,\eta)}{\phi^{+}(k,\eta_0)}$, in terms of the function $f(k,\eta)$ this is simply
\begin{equation}
 K(k,\eta)=\frac{a(\eta_0)e^{ik\sigma_{+}(k,\eta)}}{a(\eta)e^{ik\sigma_{+}(k,\eta_0)}}.   
\end{equation} 
Now since the scale factor is real, and $\sigma$ is Hermitian analytic, the bulk-to-boundary propagator is Hermitian analytic. 

As an example of how this WKB expansion gives us the mode function, let us consider the case of a massless scalar in de Sitter space with $c_s=1$. For de Sitter, $a=\frac{1}{H\eta}$, therefore $\frac{a''}{a}=\frac{2}{\eta^2}$. Since $m^2=0$, we have:
\begin{equation}
	\sigma_2=\frac{1}{2}\int^{\eta}_{-\infty}d\eta'\left(\frac{-2}{\eta'^2}\right)=\frac{1}{\eta}.
\end{equation}
Taking this result, we can obtain the other $\sigma_r$ order by order:
\begin{equation}
	\begin{aligned}
		\sigma_3&=\frac{i}{2}\int^{\eta}_{-\infty}d\eta'\left(\frac{1}{\eta}\right)''=\frac{i}{2}\int^{\eta}_{-\infty}d\eta'\left(\frac{2}{\eta^3}\right)=\frac{-i}{2\eta^2},\\
		\sigma_4&=\frac{1}{2}\int^{\eta}_{-\infty}d\eta'\left[i\frac{-3i}{\eta^4}-\frac{1}{\eta^4}\right]
		=\frac{-1}{3\eta^3}.
	\end{aligned}
\end{equation}
With the help of combinatorics and using induction, we can see that:
\begin{equation}
	\sigma_r=\frac{(-i)^r}{2}\int^{\eta}_{-\infty}d\eta'\left[\frac{r-1}{\eta^{r}}-\frac{r-3}{\eta^{r}}\right]=\frac{-(-i)^r}{(r-1)\eta^{r-1}}.
\end{equation}

Therefore, we have:
\begin{equation}
	\begin{aligned}
		ik\sigma(k,\eta)&=ik\eta-ik\sum_{r=2}^{\infty}\frac{(-i)^r}{(r-1)\eta^{r-1}k^r}\\
		&=ik\eta-\sum_{r=1}^{\infty}\frac{(-i)^r}{r(k\eta)^r}\\
		&=ik\eta+\log({1+\frac{i}{k\eta}}),\\
		f&=Ce^{ik\sigma}=Ce^{ik\eta}\left(1+\frac{i}{k\eta}\right).
	\end{aligned}
\end{equation}
Setting $C=-ik/\sqrt{2k^3}$, and remembering that $\phi=f/a$, we have
\begin{equation}
	\phi=\frac{H}{\sqrt{2k^3}}(1-ik\eta)e^{ik\eta},
\end{equation}
which is the usual de Sitter mode function of a massless scalar.

\bibliographystyle{JHEP}
\bibliography{refs}

\end{document}